\documentclass[useAMS,referee]{biom}
\def\bSig\mathbf{\Sigma}

\usepackage{url}
\usepackage{amsmath, amssymb}
\usepackage{graphicx}
\usepackage{bm}
\usepackage{enumerate, enumitem}
\let\bibhang\relax
\usepackage{natbib}

\usepackage{comment}
\usepackage{booktabs}

\usepackage{multirow, makecell}

\usepackage[dvipsnames]{xcolor}
\usepackage{setspace}
\usepackage[font={footnotesize,stretch=1}]{caption} 
\usepackage{xr-hyper}
\usepackage{hyperref}
\hypersetup{
    colorlinks=true,
    linkcolor=blue,
    filecolor=black,      
    urlcolor=black,
    citecolor=blue
}
\usepackage[capitalise,noabbrev]{cleveref}

\usepackage{float}
\definecolor{ao}{rgb}{0.0, 0.5, 0.0}
\definecolor{brightmaroon}{rgb}{0.76, 0.13, 0.28}
\usepackage{pifont} % for check marks
\newcommand{\cmark}{\ding{51}}%
\bibpunct[, ]{(}{)}{;}{a}{,}{,}

\usepackage{bbold}

\newcommand{\hs}{\hspace{0.1em}}
\newcommand{\iv}{\mathbb{1}}

\definecolor{changed}{rgb}{0,0,0}

\newcommand*{\addFileDependency}[1]{% argument=file name and extension
\typeout{(#1)}% latexmk will find this if $recorder=0
\@addtofilelist{#1}
\IfFileExists{#1}{}{\typeout{No file #1.}}
}\makeatother

\def\ind{\perp\!\!\!\perp}

\title{Nonparametric Motion Control in Functional Connectivity Studies in Children with Autism Spectrum Disorder}

\author{JIALU RAN$^1$, SARAH SHULTZ$^2$, BENJAMIN B. RISK$^1$, AND DAVID BENKESER$^{\ast1}$\\
$^1$ Department of Biostatistics and Bioinformatics, Rollins School of Public Health, Emory University \\
$^2$ Department of Pediatrics, School of Medicine, Emory University\\
\emailx{benkeser@emory.edu}
}

\begin{document}

\pagerange{\pageref{firstpage}--\pageref{lastpage}} 
\volume{}
\pubyear{}
\artmonth{}

% \doi{10.1111/j.1541-0420.2005.00454.x}

\label{firstpage}

\begin{abstract}
Autism Spectrum Disorder (ASD) is a neurodevelopmental condition associated with difficulties with social interactions, communication, and restricted or repetitive behaviors. To characterize ASD, investigators often use functional connectivity derived from resting-state functional magnetic resonance imaging of the brain. However, participants' head motion during the scanning session can induce motion artifacts. Many studies remove participants with excessive motion, and then estimate the effect of diagnosis on functional connectivity using linear regression. However, participant exclusions and linearity assumptions can cause biases. We propose an estimand that quantifies the difference in average functional connectivity in autistic and non-ASD children while standardizing motion relative to the low motion distribution in scans that pass motion quality control. We introduce a nonparametric estimator for motion control, called MoCo, that uses all participants and flexibly models the impacts of motion and other relevant features using an ensemble of machine learning methods. We establish large-sample efficiency and multiple robustness of our proposed estimator. The framework is applied to estimate the difference in functional connectivity between 132 autistic and 245 non-ASD children, of which 34 and 126 pass motion quality control, respectively. MoCo appears to dramatically reduce motion artifacts compared to a standard approach with no participant removal, while more efficiently utilizing participant data and accounting for possible selection biases compared to participant removal. %MoCo appears to dramatically reduce motion artifacts relative to no participant removal {\color{changed}{and the inverse propensity weighting method (IPTW)}}, while more efficiently utilizing participant data and accounting for possible selection biases relative to the na{\"i}ve approach with participant removal, {\color{changed}{and offering greater sensitivity than Nebel’s method}}.
\end{abstract}


\maketitle

\section{Introduction}
\label{sec:intro}
Early studies on neurodevelopment using functional magnetic resonance imaging found that short-range brain connections weakened and long-range brain connections strengthened during development. However, the validity of these findings was undermined by the discovery that motion during imaging can lead to these same patterns \citep{van2012influence,power2014methods}. This discovery led to the widespread adoption of motion quality control via participant removal, which can result in drastic data loss. A recent study removed 60\% of approximately 11,500 children due to excessive motion  \citep{marek2022reproducible}. Removal of these children not only greatly decreases sample size, but also may introduce selection bias \citep{cosgrove2022limits}. This is especially true for studies of neurodevelopmental conditions, such as autism spectrum disorder (ASD). \citet{nebel2022accounting} found that 80\% of autistic children compared to 60\% non-ASD children were removed during quality control, and the removed autistic children had greater social deficits, worse motor control, and lower generalized ability index. The authors concluded that  differential removal of scans may significantly bias results. These studies point to a need to develop efficient statistical methods that can avoid selection bias in order to draw unbiased inferences about brain development.

The current study is motivated by studies of ASD, where investigators often use resting-state functional magnetic resonance imaging (rs-fMRI) to derive measures of \emph{functional connectivity} between regions in the brain. Functional connectivity is commonly defined as the correlation between the blood oxygen level dependent signal of different brain regions across time. Functional connectivity may be atypical in autism \citep{di2014autism}. %\citep{hull2017resting}. 
However, obtaining high-quality rs-fMRI data for functional connectivity analysis is challenging. Participants' head motion during the scanning session can induce \emph{motion artifacts}. The patterns of correlation induced by motion artifacts mimic the connectivity theory of autism, which predicts increased correlations between nearby brain regions and decreased correlations between distant brain areas \citep{deen2012perspective}. Artifact-driven disruptions in brain networks can arise in comparisons of high and low motion rs-fMRI scans \citep{power2014methods}. 

Current guidelines for analyzing rs-fMRI involve four steps. First, rigid body motion correction is used to align fMRI volumes across time. Second, it is generally recommended to remove individuals in which motion is deemed unacceptable, e.g., remove individuals if they have less than five minutes of data free from excessive motion  \citep{power2014methods}. Third, confound regression is applied to the time series, which may include regressing motion alignment parameters, global signal, cerebral spinal fluid signal, and white matter, which may be combined with the removal of high-motion volumes or spike regression \citep{ciric2017benchmarking}. Following the confound regression, a single measure of functional connectivity is derived for each pair of brain regions summarizing the connectivity between the two regions. In the fourth and final step, the association of diagnosis with the connectivity between two regions is estimated via a linear regression. This regression model includes all children who pass motion quality control and regresses the derived functional connectivity outcome onto a group indicator (e.g., ASD vs. no ASD) while adjusting for certain participant-level variables possibly including a measure of average motion during the scanning session.

There are several shortcomings of current guidelines for the analysis of rs-fMRI data. First, participant removal may be inefficient and lead to selection bias \citep{nebel2022accounting}. Second, the reliance on linear models may yield inferences that lack robustness due to model misspecification. %Finally, the current literature lacks appropriate guidance on which participant-level variables should be adjusted for in order to derive appropriate inferences. 
\citet{nebel2022accounting} addressed the issue of selection bias by treating the excluded participants as missing data. They used a doubly robust method from causal inference \citep{benkeser2016highly} that improves upon inverse propensity weighting \citep{petersen2024inverse} to correct for potential bias. However, their approach did not leverage information from the fMRI data in excluded participants. %still relied heavily on the use of linear regression to handle demographic covariates.% and did not address the issue of covariate selection. %\cite{sobel2014causal} formulated a causal framework for task fMRI activation studies that considers the systematic error from motion, which is closely related to the nuisance regression step in resting-state fMRI preprocessing, but it does not address the problem of quality control in resting-state fMRI. 

%Although scan exclusion in the second stage may help alleviate motion artifacts, it can lead to drastic reductions in sample size. For example, in a brain-wide analysis of resting-state functional connectivity,  \citet{marek2022reproducible} removed 60\% of 10,000 children due to excessive motion. A similar study removed 75\%  \citep{nielsen2019evaluating}. Removal of these children not only greatly decreases sample size, but also may introduce selection bias if the excluded children differ meaningfully in terms of characteristics related to functional connectivity. 

 %There is a need to develop a method that can examine functional connectivity group differences in a manner that does not decrease sample size or introduce bias into the analysis.

The objective {\color{changed} in this paper} is to define an estimand (and estimators thereof) that appropriately quantifies the association between functional connectivity and ASD diagnosis that: (i) minimizes the impact of selection bias from motion quality control, (ii) does not rely on correct specification of a linear model, and (iii) provides guidance on which covariates should be adjusted for and how we should adjust for them.
%1) appropriately controls for possibly non-linear impacts of motion, and 2) does not suffer from selection biases due to quality control. We defer to the prevailing view in the fMRI community that motion causes changes in functional connectivity, and these changes are not of biological interest \citep{ciric2017benchmarking,satterthwaite2013improved,power2014methods}. 
Our proposed estimand has connections to direct standardization \citep{rothman2008modern} and certain estimands used in causal inference \citep{diaz2021nonparametric}; however, we do not require counterfactuals to define our estimand. Nevertheless, by connecting the estimand to these areas, we are able to provide insight into whether and how to adjust for certain participant-level characteristics. Our proposed estimators can utilize nonparametric learning approaches (e.g., based on machine learning), while maintaining standard asymptotic behavior. We also show that the estimators enjoy a desirable multiple robustness property. Our approach to motion control, which we call MoCo, is a novel solution to the significant challenges associated with the analysis of rs-fMRI data. We illustrate the MoCo approach via an analysis of functional connectivity between a seed region in the default mode network and other brain regions in children in the Autism Brain Imaging Data Exchange \citep{di2017enhancing}.

\section{Methods}
\label{sec:meth}
%\subsection{Notation}
% write all integrals as being wrt Lebegesue measure
% p_{\delta, G  \mid  a, H, J}(g  \mid  h, j) is the delta-truncated density of G  \mid  A = a, H, J evaluated at (g, h, j)
% leave off delta if not truncated
\emph{Notation}. Let $A \in \{0, 1\}$ denote the diagnosis group, which is equal to 1 if the participant has ASD and 0 otherwise. Let $M \in \mathcal{M}$ denote the motion variable. In our data application, we take $M$ to be mean framewise displacement (FD). FD quantifies head motion between consecutive fMRI frames; each participant’s scan yields approximately 120–180 frames over 5.5–6.5 minutes, and mean FD ($M$) is a commonly used summary measure of motion during resting-state fMRI \citep{di2014autism,power2014methods}. Let $\Delta \in \{0, 1\}$ denote an inclusion indicator, which is equal to 1 if the participant meets a pre-specified set of criteria for inclusion in the study, related to the aggregate amount of movement during a child's scanning session. In our data application, we use the criteria from \cite{power2014methods}, in which $\Delta = 1$ if a child has more than 5 minutes of data after removing frames with FD $>$ 0.2 mm. Let $Y \in \mathcal{Y}$ denote the functional connectivity between two locations in the brain. For clarity, we initially define $Y$ for a pair of regions, but in \cref{sec:simultaneous_conf} we extend to the multivariate case with appropriate family-wise error control. Let $X \in \mathcal{X}$ denote covariates that are putatively related to functional connectivity and are possibly imbalanced across diagnosis groups. Such covariates could include age, sex, and handedness. Let $Z \in \mathcal{Z}$ be variables related to the diagnosis group and the pathophysiology of ASD that could possibly contribute to children moving more/less during a scanning session and that have substantially different distributions with little or no overlapping support in ASD and non-ASD groups. 

The distinction between $X$ and $Z$ is an important scientific decision, as these sets of variables serve distinct roles in defining our estimand and deriving estimators thereof. In general, we wish to include in $X$ any variables that would be balanced across diagnosis groups in an ideal study, but that may be imbalanced due to imperfect recruitment. Thus, variables such as age, sex, and handedness may be important to consider as components of $X$. Each of these variables may biologically relate to connectivity in the brain; however, these biological impacts are not of direct scientific interest and we simply wish to control for any differences in the distribution of these variables across diagnosis groups. %For example, an ideal experiment may recruit age-matched children with and without ASD. However, in practice such designs may be impractical and as a result studies may, for example, recruit older children with ASD compared to not. When analyzing data from such a study, we would not want the chance imbalance in age to be reflected in our quantification of brain connectivity and thus we may look to statistical methods to control for this difference. 
On the other hand, we wish to include in $Z$ any variables that may be related to ASD diagnosis that may also be associated with motion during a scanning session. Such variables could include, for example, a child's score on the autism diagnostic observation schedule (ADOS, a measure of social disability) and/or their  full-scale intelligence quotient score (FIQ, a measure of intelligence). Note that for these variables either (i) we could not balance them by design (e.g., ADOS) or (ii) we would not wish to balance them by design (e.g., FIQ as balancing intelligence across groups may decrease differences between diagnosis groups). In our analysis, we include age, sex, and handedness in $X$ and ADOS score, FIQ score, stimulant medication status, and non-stimulant medication status in $Z$. % However, we recognize that the ultimate decision of which variables should be considered a part of $X$ and which a part of $Z$ requires a discussion within the scientific community. Our goal in this work is not to create a definitive list of covariates that fall within each category, but rather to spark such a discussion. 
Previous analyses have controlled for a limited set of variables in linear regression, e.g., age, FIQ, site, and mean FD \citep{di2014autism}, while we argue in favor of considering a broader collection of variables in order to appropriately control for motion artifacts.

Let $O = (A, M, \Delta, X, Z, Y)$ represent a random variable with distribution $P$. Denote $O_1, \dots, O_n$ as $n$ i.i.d.\ observations of $O$, where $O_i=(A_i, M_i, \Delta_i, X_{i}, Z_{i},Y_i)$. We assume $P \in \mathcal{P}$, where $\mathcal{P}$ is a statistical model for probability distributions on the support of $O$ that is nonparametric up to certain positivity conditions that will be defined below. %We let $Pf = \int f(o) dP(o)$ for a given function $f$, and the corresponding expectation operator is denoted as $E$. 
%We use $P_n$ to denote the empirical distribution of $O_1, \dots, O_n$. 

In our notation, an uppercase letter with no subscript denotes a random variable, an uppercase letter with an index, typically $i$, is an observed value of a random variable, and a lowercase letter indicates a typical realization of the random variable. For example, $E(Y \mid A)$ is a random variable, while $E(Y \mid A=a)$ is {\color{changed} non-random}.%, where $E\{f(O)\}$ denotes expectation under $P$ of any measurable function $f$ of $O$.

Let $P_{M \mid \Delta = 1, A, X}(m \mid a, x)$ denote the probability distribution of $M$ conditional on $\Delta = 1, A, X$ evaluated at value $(m, a, x) \in \mathcal{M} \times \{0, 1\} \times \mathcal{X}$. $P_{M \mid \Delta = 1, A, X}$ is thus the probability distribution of motion given fixed diagnosis status $A$ and covariates $X$, among children who meet the inclusion criteria. We use $p_{M \mid \Delta = 1, A, X}(m \mid a, x)$ to denote a density with respect to Lebesgue measure. For simplicity, all subsequent densities are also defined with respect to the Lebesgue measure. %We denote by $q_{\delta}(m \mid a, x)$ the density function for the $A=a, X = x$ conditional motion distribution with respect to an appropriate measure $\nu_q$. 
We define  $p_{M \mid A, X, Z}(m \mid a, x, z)$ as the conditional density of $M$ given $A, X, Z$ and $p_{Z \mid A, X}(z \mid a, x)$ as the conditional density of $Z$ given $A, X$.

Let $\mu_{Y \mid A, M, X, Z}(a, m, x, z)$ denote the conditional mean functional connectivity given $A = a$, $M = m$, $X = x, Z = z$ and let $\pi_a(x)$ denote the probability that $A = a$ given $X = x$. We use an $n$-subscript to denote an estimate, e.g., $\pi_{n,a}$ is an estimate of $\pi_a$.
\vspace{-0.2em}

\subsection{Defining a model-agnostic target parameter for group comparisons in fMRI studies}\label{sec:MoCo_estimand}

We propose a framework motivated by limitations in previous studies of developmental and neurological disorders. The prevailing approach (e.g., \citealt{di2014autism}) is to remove participants that fail motion quality control ($\Delta_i=0$), then fit a linear regression with functional connectivity ($Y_i$, correlation between a seed region and another brain region) as the response variable and autism diagnosis ($A_i$), mean framewise displacement ($M_i$), and a limited set of demographic variables as predictors $
Y_i = \beta_0 + \beta_1 A_i + \beta_2 M_i + \sum_{j=1}^p \beta_{j+2} X_{ij} + \epsilon_i, i \in \{i : \Delta_i=1\}$. 
Our goal is to address two issues with this approach: first, that the linear model does not adequately capture the effects of motion \citep{power2014methods}; and second that removing participants that fail motion quality control introduces selection bias \citep{nebel2022accounting}. To that end, our approach considers nonparametric adjustment for $A$, $M$, $X$, and $Z$ as part of an outcome model $Y_i = \mu_{Y \mid A, M, X, Z}(A_i,M_i, X_i, Z_i)+\epsilon_i$.
Because our choice of model is nonparametric, $\mu_{Y \mid A, M, X,Z}$ is an infinite-dimensional function of $M, X,$ and $Z$. However, our primary interest lies in assessing differences in functional connectivity outcomes between diagnosis groups $A$. Thus, we choose to \emph{standardize} $(M, X, Z)$ over particular distributions to yield a motion- and covariate-controlled estimand. 

{\color{changed}In an ideal world, we would explicitly set motion equal to zero. The ideal estimand of group differences would be $\theta_1^I - \theta_0^I$, where
\begin{align}
\theta_a^I = \iint \mu_{Y\mid A,M,X,Z}(a,0,x,z) p_{Z\mid A,X}(z\mid a,x) p_X(x) \hs dz \hs dx \ .\label{eq:ideal}
\end{align}
Unfortunately, $m$ is never truly equal to zero in data applications, as all participants move at least some during a scanning session. Thus, this estimand is not identifiable. We propose to instead consider standardizing motion with respect to a given ``tolerable motion distribution'' and suggest that a natural choice is $p_{M \mid \Delta = 1, A, X}(m \mid 0, x)$, the $X$-conditional distribution of motion observed in non-ASD children that pass motion quality control. Then we define the Motion Controlled (MoCo) estimand of group differences, $\theta_1 - \theta_0$, where
\begin{equation}
\begin{aligned}
    \theta_a = &\iiint \mu_{Y\mid A,M,X,Z}(a,m,x,z) p_{Z\mid A,X}(z\mid a,x) p_{M\mid \Delta=1,A,X}(m\mid 0,x) p_X(x) \hs dz \hs dm \hs dx \ . \label{eq:moco}
\end{aligned}
\end{equation}
In this estimand, we average $Z$ values over their conditional distribution given $A$ and $X$, $p_{Z \mid A, X}$. We average motion values over their conditional distribution given $\Delta = 1$, $A = 0$, and $X$, $p_{M \mid \Delta = 1, A = 0, X}$. Finally, we average covariates $X$ over their marginal distribution $p_X$. Relative to existing approaches, our approach explicitly (i) averages over a tolerable distribution of motion and (ii) adjusts for $Z$. We provide motivation for these choices below.

\textit{\underline{Remark.}} The gap between the ideal estimand and target estimand is $|(\theta_1^{I} - \theta_0^{I}) - (\theta_1 - \theta_0)|$. In some semiparametric and parametric models, this gap is equal to zero. For example, if $\mu_{Y|A,M,X,Z} = \beta_0 + \beta_1 A + \beta_2 M + \beta_3 X + \beta_4 Z$, we have $\theta_1^{I} - \theta_0^{I} = \theta_1 - \theta_0 = \beta_1 + \beta_4 [E\{E(Z \mid A=1,X)-E(Z \mid A=0,X)\}]$. In contrast, if there is an interaction between motion and diagnosis, or between symptom severity and diagnosis, then the gap is non-zero. For example, if $\mu_{Y \mid A,M,X,Z} = \beta_0 + \beta_1 A + \beta_2 M + \beta_3 X + \beta_4 Z + \beta_5 A*M + \beta_6 X*M + \beta_7 Z*M$, then the gap is equal to $\beta_5 \int m \, p_{M \mid \Delta=1,A,X}(m \mid 0,x) p_X(x)dx + \beta_7 \iint z m \, \{p_{Z \mid A,X}(z \mid 1, x) - p_{Z \mid A,X}(z \mid 0, x) \} p_{M \mid \Delta=1,A,X}(m \mid 0,x) p_X(x) \hs dz \hs dm \hs dx$. If there is no interaction between diagnosis and motion and no interaction between symptom severity and motion, the ideal estimand and MoCo coincide. Specifically, when the outcome model admits the decomposition $\mu_{Y \mid A,M,X,Z} = g_1(a,x,z)+g_2(m,x)$ for functions $g_1$ and $g_2$, then the gap is zero. The gap is in general not equal to zero if the motion artifact is modified by diagnosis. See Supplement Section 1 for additional details. It seems biologically plausible to assume $\mu_{Y \mid A,M,X,Z} = g_1(a,x,z)+g_2(m,x)$, since if two children have identical motion in the scanner, we expect the motion artifacts to be equivalent regardless of their diagnosis. In this sense, the MoCo estimand is a reasonable approximation to the ideal estimand.}

\begin{figure}[!p]
	\centering
    \includegraphics[width = 0.95\textwidth]{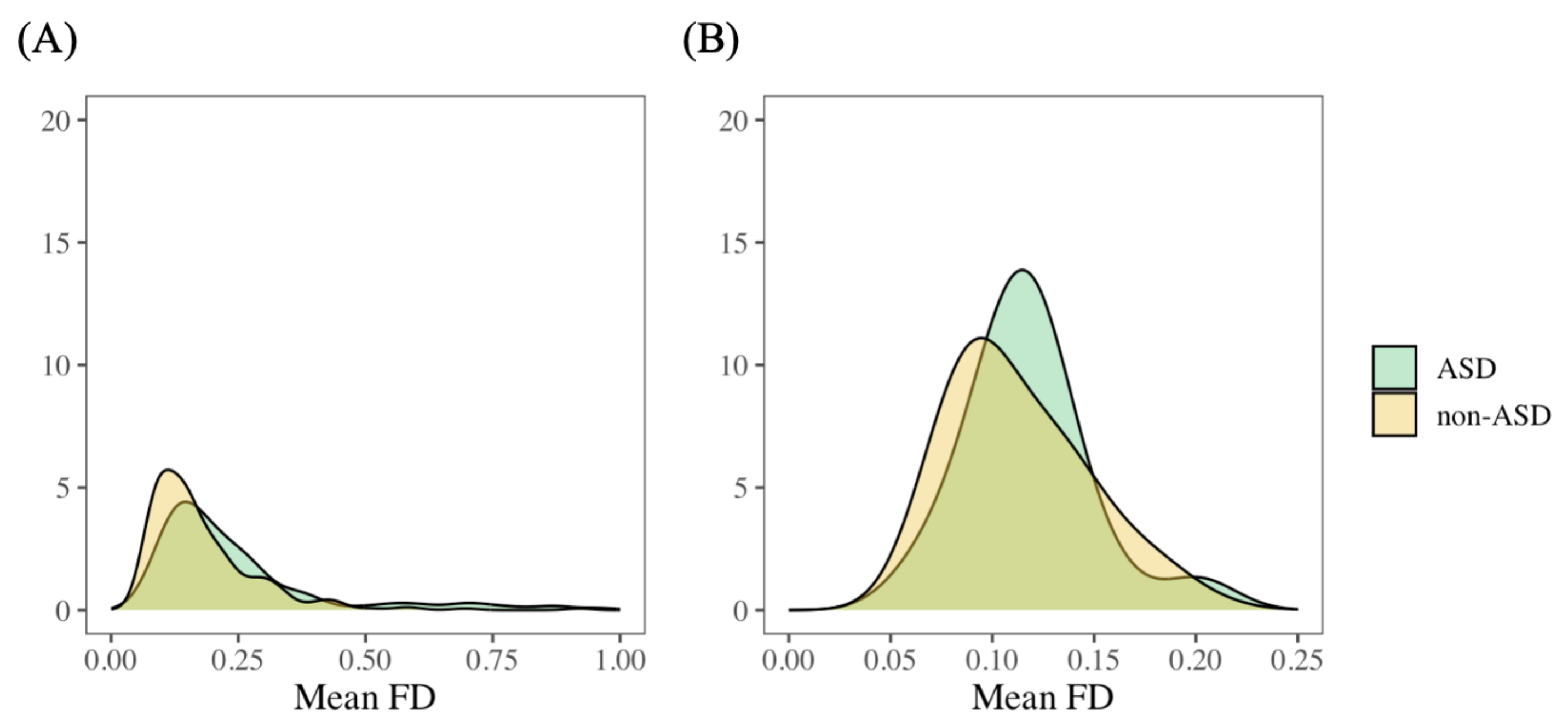}
    \caption{Distributions of mean framewise displacement (FD) in the school-age children dataset. Panel A shows the distribution of mean FD over all children. 
    Panel B shows the distribution of mean FD over children who meet the inclusion criteria. The distribution of motion in non-ASD children that pass motion quality control differs from the distribution of motion in children with ASD that pass motion quality control.} 
    %ASD children with usable data still have higher motion than non-ASD children, which motivates additional training in our hypothetical experiment such that the motion distribution in the ASD group (shown in green) would be identical to the motion distribution in the non-ASD group with $\Delta = 1$ (shown in yellow).}
	\label{fig:motion_distri} 
\end{figure}

{\color{changed} \emph{Justification for tolerable motion distribution}. In our application, the distribution of motion in children that pass motion quality control differs between diagnostic groups (\cref{fig:motion_distri}), with the non-ASD group demonstrating lower marginal motion than the ASD group. Allowing the tolerable motion distribution to depend on $X$ adheres to our principle that differences in $X$ across diagnosis groups are not of primary interest in our analysis. Thus, marginalizing both diagnosis groups with respect to the same tolerable motion distribution ensures that any residual artifacts attributable to motion within the tolerable range are appropriately balanced across diagnosis groups.} 

\emph{Justification for adjusting for $Z$}. Consider an outcome model that omits $Z$ from its formulation, while adjusting only for $A$, $M$, and $X$: $Y_i = \mu_{Y \mid A, M, X}(A_i,M_i,X_i)+\epsilon_i$.
We argue that the fact that this estimand ignores $Z$ would lead to undesirable consequences and potentially biased analyses. If we were to adopt the same standardizing approach based on $\mu_{Y \mid A, M, X}$, we might consider the parameter $\iint \mu_{Y\mid A,M,X}(a,m,x) p_{M\mid \Delta=1,A,X}(m\mid 0,x) p_X(x) dm\;dx$,
which by the law of total probability equals
\begin{equation}
\begin{aligned}
\iiint \mu_{Y \mid A,M,X,Z}(a, m, x, z) p_{Z \mid A,M,X}(z \mid a, m, x)p_{M \mid \Delta=1,A,X}(m \mid 0,x) p_X(x) dz\;dm\;dx \ . \label{eq:stochasticwrong}
\end{aligned}
\end{equation}
%However, the above presentation is an oversimplification because developmental and neurological disorders are heterogeneous. In most studies, we have variables that characterize the severity and/or other dimensions of a diagnosis (variables $Z$). We thus consider a non-parametric model for the conditional mean outcome including these variables:
%$$
%Y_i = \mu_{Y \mid A,M,X,Z}(A_i,M_i,X_i,Z_i)+\epsilon_i.
%$$
%Now consider marginalizing $\mu_{Y \mid A,M,X,Z}$ for a particular diagnosis group $a$ via
%\begin{equation}
%\begin{aligned}
%\iiint \mu_{Y \mid A,M,X,Z}(a, m, x, z) p_{Z \mid A,M,X}(z \mid a, m, x)p_{M \mid \Delta=1,A,X}(m \mid 0,x) p_X(x) dz\;dm\;dx \ , \label{eq:stochasticwrong}
%\end{aligned}
%\end{equation}
%where $p_{Z \mid A,M,X}$ is the conditional distribution of $Z$ given $(A, M, X)$. 
This view indicates that excluding $Z$ may implicitly yield residual motion artifacts within the tolerable range that are \emph{imbalanced} within a diagnosis group across levels of $Z$. To understand why this may be undesirable, note that in children with neurodevelopmental challenges ($A=1$), {\color{changed}more} severe symptomatology (as may be indicated by components of $Z$) {\color{changed}is associated with higher} motion \citep{cosgrove2022limits,nebel2022accounting}. Thus, considering \eqref{eq:stochasticwrong} with $a = 1$, we find that the {\color{changed}weights formed from the }product of densities $p_{Z\mid A,M,X}(z\mid 1,m,x) p_{M\mid \Delta=1,A,X}(m\mid 0,x)$ will tend to place more weight on children with severe symptoms and high motion than they place on children with severe symptoms and {\color{changed}low} motion. This is illustrated in Supplement Section {\color{changed}2}. On the other hand, our proposed estimand uses $p_{Z \mid A, X}$ to standardize over the distribution of $Z$. By removing the dependence of this distribution on motion, we are able to appropriately balance motion artifacts within each diagnosis group across levels of $Z$.

The motion-controlled estimand is well-defined and nonparametrically estimable if: \begin{enumerate}[align=left]
\item[(A1.1)] for every $x$ such that $p_X(x)>0$, we also have $\pi_a(x) > 0$ for $a = 0, 1$; for every $x$ such that $p_X(x)>0$, we also have $P(\Delta = 1 \mid A = 0, X = x) > 0$.
\item[(A1.2)] for every $(m, x, z)$ such that $p_{Z \mid A, X}(z \mid a, x)p_{M \mid \Delta = 1, A, X}(m \mid 0, x)p_X(x) > 0$, we also have that $p_{M \mid A, X, Z}(m \mid a, x, z) > 0$ for $a = 0,1$.
\end{enumerate}
(A1.1) states that, at a population level, there cannot be values of $X$ that are observed exclusively in the ASD group or exclusively in the non-ASD group and that there cannot be values of $X$ that exclusively lead to non-usable scans in the non-ASD group. In our application, $X$ consists of age, sex, and handedness, which do not perfectly predict ASD nor scan usability, and therefore assumption (A1.1) is plausible.
%To clarify the nature of Assumption (A1.2), consider the \emph{range} of motion values that are possible to observe for non-ASD children who meet the inclusion criteria, with a particular covariate value $X=x$.
(A1.2) stipulates that for both $a = 0, 1$, the conditional mean $\mu_{Y \mid A, M , X, Z}(a, m, x, z)$ must be well defined for every term in the integrand (\ref{eq:moco}) that is given non-zero weight by the product of the densities $p_{Z \mid A,X}(z \mid a,x)p_{M \mid \Delta=1,A,X}(m \mid 0,x) p_X(x)$. Thus, we require that within each diagnosis group $a$, for any demographic variable value $x$ in the marginal support of $X$, any behavioral variable value $z$ in the support of $p_{Z \mid A,X}$ for that diagnosis group, and for any motion $m$ on the support of the $p_{M \mid \Delta=1,A,X}(m \mid 0,x)$, it must be possible to observe the motion value $m$ for all values of $z$ that are observed in $A=a$ with $X = x$. This assumption would be violated for example if $Z$ included a measure of social disability and children in the ASD group with the highest levels of social disability never generate motion values comparable to the the motion values observed in the non-ASD group.
%(A1.2) stipulates that, for both ASD and non-ASD children with the same value $x$ of $X$, it is possible to observe the \emph{same range} of motion values \emph{irrespective} of the value of $Z$. Recall that $Z$ includes a measure of social disability, and we expect that children with higher support needs will move more in the scanner. This assumption requires that it is possible to obtain low-motion data even in these more challenging cases. This assumption can be scrutinized by studying the distribution of an estimate of the ratio of motion distributions, described in (\ref{eq:motion_dens_ratio}) in the next section. 
% In our data analysis, we will examine the plausibility of these assumptions. %(Supplementary Material Section 8.2).
 
In the Supplement Section {\color{changed}3}, we further describe connections between our estimand and those used in causal inference and include a causal graph illustrating relationships between the various variables used in our analysis.
%\section{Estimation and Inference}
%\vspace{-4em}
\subsection{Efficiency theory}\label{sec:efficiencytheory}

A key step in developing our estimator is deriving the \emph{efficient influence function} (EIF) of regular, asymptotically linear estimators of $\theta_a$. See Supplement Section {\color{changed}4}.1 for a short review of efficiency theory. To characterize this EIF, we define $\pi_{\Delta=1 \mid A, X}(0, x) = P(\Delta = 1 \mid A = 0, X = x)$ as the probability of a non-ASD child with covariate value $x$ having usable data. We introduce the shorthand $\bar{\pi}_{0}(x) = \pi_0(x) \pi_{\Delta = 1 \mid A, X}(0, x)$ as the probability that $A = 0$ and $\Delta = 1$ conditional on $X = x$. We denote the indicator function $\iv_{a}(A_i)$ equal to 1 if $A_i = a$ and zero otherwise; $\iv_{0,1}(A_i, \Delta_i)$ equal to 1 if $A_i = 0$ and $\Delta_i = 1$ and equals zero otherwise. We also define for $a = 0, 1$ 
    \begin{align}
    r_a(m, x, z) &= \frac{p_{M \mid \Delta = 1, A, X}(m \mid 0, x)}{p_{M \mid A, X, Z}(m \mid a, x, z)} \ , \label{eq:motion_dens_ratio} \\
    \eta_{\mu \mid A, Z, X}(a, z, x) &= \int \mu_{Y \mid A, M, X, Z}(a, m, x, z) \hs p_{M \mid \Delta = 1, A, X}(m \mid 0, x) \hs dm \ , \label{eq:eta1_def} \\
    \xi_{a, \eta \mid X}(x) &= \iint \mu_{Y \mid A, M, X, Z}(a, m, x, z) \hs p_{M \mid \Delta = 1, A, X}(m \mid 0, x) \hs p_{Z \mid A, X}(z \mid a, x) \hs dm \hs dz \ . \label{eq:xi_def}
\end{align}
In these definitions, we use a subscript notation for the functional parameters $\eta$ and $\xi$ that attempts to make explicit both the integrand in the parameter's definition, as well as the random variables that are arguments of the function. For example, the definition of $\eta_{\mu \mid A, Z, X}$ (\ref{eq:eta1_def}) involves integrating $\mu_{Y \mid A, M, X, Z}$, while $\eta_{\mu \mid A, Z, X}$ is a function of the random variables appearing in the subscript, $A$, $Z$ and $X$. 
%In real-world terms, $\eta_{\mu \mid A, Z, X}$ is the motion-standardized functional connectivity given the diagnosis and all covariates, and $\eta_{\mu \mid A, M, X}$ is the $Z$-standardized functional connectivity given diagnosis, motion, and demographic covariates $X$. The motion- and $Z$-standardized functional connectivity given diagnosis and demographic covariates is represented by $\xi_{a, \eta \mid X}$ (\ref{eq:xi_def}). 
\vspace{-1.4em}
\begin{theorem} \label{theorem:eif}
(Efficient Influence Function). In a nonparametric model, the efficient influence function for $\theta_a$ evaluated on a typical observation $O_i$ is $D_{P,a}(O_i)$ defined as
\begin{equation}
\begin{aligned}
    &\frac{\iv_a(A_i)}{\pi_a(X_i)} r_a(M_i, X_i, Z_i)\left\{Y_i - \mu_{Y \mid A, M, X, Z}(a, M_i, X_i, Z_i)\right\}\\
    &\hspace{2em} + \frac{\iv_a(A_i)}{\pi_a(X_i)} \left\{\eta_{\mu \mid A, Z, X}(a, X_i, Z_i) - \xi_{a, \eta \mid X}(X_i) \right\} \\
    &\hspace{4em}+ \frac{\iv_{0,1}(A_i, \Delta_i)}{\bar{\pi}_{0}(X_i)} \left\{ \eta_{\mu \mid A, M, X}(a, M_i, X_i) - \xi_{a, \eta \mid X}(X_i)\right\} + \xi_{a, \eta \mid X}(X_i) - \theta_a.
    \end{aligned} \label{eq:eif}
\end{equation}
\end{theorem}
A proof is included in the Supplement Section {\color{changed}4}.2. Fubini's theorem allows us to write $\xi_{a, \eta \mid X}$ in terms of either $\eta_{\mu \mid A, Z, X}$ or $\eta_{\mu \mid A, M, X}$, $ \xi_{a, \eta \mid X}(x) = \int \eta_{\mu \mid A, Z, X}(a, x, z) \hs p_{Z \mid A, X}(z \mid a, x) \hs dz = \int \eta_{\mu \mid A, M, X}(a, m, x) \hs p_{M \mid \Delta = 1, A, X}(m \mid 0, x) \hs dm$, where $\eta_{\mu \mid A, M, X}(a, m, x)=$ $\int \mu_{Y \mid A, M, X, Z}(a, m, x, z)$ $\hs p_{Z \mid A, X}(z \mid a, x) \hs dz$. 

We use the one-step estimation framework to define efficient estimators of $\theta_a$  \citep{bickel1993efficient}. Suppose we have an estimate of $\xi_{a, \eta \mid X}$ available, say $\xi_{n, a, \eta \mid X}$. An estimate of $\theta_a$ can be obtained by marginalizing $\xi_{n, a, \eta \mid X}$ over the empirical distribution of $X$, yielding the plug-in estimate 
%leading to an estimate of the form 
$\theta_{n, a} = n^{-1} \sum_{i=1}^n \xi_{n, a, \eta \mid X}(X_i)$. 
%We refer to $\theta_{n, a}$ as a \emph{plug-in estimate}. 
A \emph{one-step estimator} of $\theta_a$ can be constructed as $\theta_{n, a}^+ = \theta_{n, a} + n^{-1} \sum_{i=1}^n D_{n, a}(O_i)$, where $D_{n, a}$ is an estimate of $D_{P,a}$. Thus, to construct a one-step estimate of $\theta_a$, we require as an intermediate step estimates of the various parameters of $P$ that appear in $D_{P,a}$. We refer to these quantities as \emph{nuisance parameters}, parameters that need to be estimated as an intermediate step in the estimation of $\theta_a$.

Examining Theorem \ref{theorem:eif}, we find several nuisance parameters in $D_{P,a}$ for which we will require estimates.% to construct our estimate $D_{n,a}$ of $D_{P,a}$. 
Estimation of several of these parameters is straightforward. For example, $\mu_{Y \mid A, M, X, Z}$ could be estimated using mean regression of $Y$ on $A, M, X, Z$. On the other hand, the $\eta$ and $\xi$ parameters involve integration and conditional densities, which generally present challenges in implementation. Our approach emphasizes two key points: (i) wherever possible mean regression with pseudo-outcomes is used to avoid numeric integration and conditional density estimation and (ii) flexible estimation techniques are used. 

We choose to emphasize the use of mean regression because it is a technique familiar to many applied statisticians and there are widely available tools. In our application, we focus on a flexible framework for regression, known as regression stacking or super learning \citep{van2007super}. Super learning uses cross-validation to build a weighted combination of candidate regression estimators, with large sample theory indicating that the ensemble estimator is essentially as good or better than any of the individual candidate regressions considered. %We used super learning in our analysis to provide estimates of $\pi_a, \pi_{\Delta = 1 \mid 0, X}, \mu_{Y\mid a, M, X, Z},\eta_{\mu \mid a, Z, X}, \eta_{\mu \mid a, M, X}$, and $\xi_{\eta \mid a, X}$, described below. 
Unfortunately, mean regression cannot be used exclusively in the estimation of $\theta_a$ and we require estimates of conditional motion distributions described below. For this purpose, we utilize a version of the highly adaptive lasso (HAL) specifically tailored for conditional density estimation, as implemented in the \texttt{haldensify} R package \citep{hejazi2022haldensify}. To circumvent numerical integration in our estimation, we make use of a technique proposed by \citet{diaz2021nonparametric} that re-casts these estimation problems that involve integrals and densities as an estimation problem that can be solved using mean regression with pseudo-outcomes. A detailed description of the implementation of our estimator is included in Supplement Section {\color{changed}5}.2.

% For this, we utilize the highly adaptive lasso, a flexible semiparametric conditional density estimator \citep{hejazi2022haldensify}. % \citep{hejazi2022haldensify-rpkg}.

%In this problem, it is desirable to avoid the integration of estimated quantities for several reasons. First, numerical integration can introduce instability into the estimation process. Second, the nuisance parameter $\eta_{\mu \mid a, M, X}$ as defined in (\ref{eq:eta2_def}) involves an integral with respect to $p_{Z \mid A, X}$ and $Z$ may be high dimensional. Estimation of multivariate densities is a challenging problem and it may be difficult to propose estimates that satisfy the sufficient conditions outlined in Theorem \ref{theorem:one-step} below. 

%\vspace{-1em}

\subsection{Inference}

Below we present two theorems establishing the consistency and asymptotic linearity, respectively, of the one-step estimator $\theta^+_{n,a}$. We define $\lVert \cdot \rVert$ to be the $L^2(P)$ norm of a given function $f$ defined as $\lVert f \rVert = E[f(O)^2]^{1/2}$. We note that for the purposes of this definition, the function $f$ is treated as given, even if it involves estimated quantities. Theorem \ref{theorem2} assumes:
\begin{itemize}[align=left]
\item[(B1)] \emph{Boundedness:} $\pi_{n, a}$ is bounded below by some $\epsilon_1 > 0$, $\bar{\pi}_{n, 0}$ is bounded below by some $\epsilon_2 > 0$, and $p_{n, M \mid A, X, Z}(m \mid a, x, z)$ is bounded below by some $\epsilon_3 > 0$.
\item[(B2)] \emph{$o_p(1)$-convergence of certain combinations of nuisance parameters}: certain subsets of the nuisance parameters are consistently estimated, as described in \cref{tab:web_supp_multiplerobust}.

\begin{table}[h]
\caption{Assumption (B2) of Theorem 3.2 (multiple robustness). Each row indicates a setting for consistency, where check marks indicate the nuisance parameters which, when they converge to true functions combined with assumptions (B1), (B3) and (B4), result in the consistency of $\theta_{n,a}^+$.}
\label{tab:web_supp_multiplerobust}
\begin{tabular}{lccccccc}
 & $\mu_{n, Y \mid A, M, X, Z}$ & $\eta_{n, \mu \mid A, M, X}$ & $\xi_{n, a, \eta \mid X}$ & $\bar{\pi}_{n, 0}$ & $\pi_{n,a}$ & $p_{n,M \mid \Delta = 1, A, X}$ & $p_{n,M \mid A, X, Z}$ \\ \hline
(B2.1) &  &  &  & & \cmark & \cmark & \cmark \\  
(B2.2) &  &  & \cmark &  &  & \cmark & \cmark \\ 
(B2.3) & \cmark & \cmark &  & \cmark & \cmark &  &  \\ 
(B2.4) & \cmark &  &  &  & \cmark & \cmark &  \\ 
(B2.5) & \cmark &  & \cmark &  &  & \cmark & \\ 
\bottomrule
\end{tabular}
\end{table}

% (i) $\lVert p_{n, M \mid A, X, Z}(m \mid a, x, z) - p_{M \mid A, X, Z}(m \mid a, x, z) \rVert = o_p(1)$ and $\lVert p_{n, M \mid \Delta = 1, A, X}(m \mid 0, x) - p_{M \mid \Delta = 1, A, X}(m \mid 0, x) \rVert = o_p(1)$ and either $\lVert 
%  \pi_{n,a} - \pi_a \rVert = o_p(1)$ or $\lVert \xi_{n, a, \eta \mid X} - \xi_{a, \eta \mid X}\rVert = o_p(1)$; \\
% (ii) $\lVert \mu_{n, Y \mid A, M, X, Z}(a, m, x, z) - \mu_{Y \mid A, M, X, Z}(a, m, x, z) \rVert = o_p(1)$, $\lVert \pi_{n, a} = \pi_a \rVert = o_p(1)$, $\lVert \bar{\pi}_{n, 0}(x) - \bar{\pi}_{0}(x) \rVert = o_p(1)$, and $\lVert \eta_{n, \mu \mid A, M, X}(a, m, x) - \eta_{\mu \mid A, M, X}(a, m, x) \rVert = o_p(1)$; \\
% (iii) $\lVert \mu_{n, Y \mid A, M, X, Z}(a, m, x, z) - \mu_{Y \mid A, M, X, Z}(a, m, x, z)\rVert = o_p(1)$ and $\lVert p_{n, M \mid \Delta = 1, A, X}(0, x) - p_{M \mid \Delta = 1, A, X}(m \mid 0, x) \rVert = o_p(1)$ and either $\lVert \pi_{n,a } - \pi_a \rVert = o_p(1)$ or $\lVert \xi_{n, a, \eta \mid X} - \xi_{a, \eta \mid X} \rVert = o_p(1)$. 

\item[(B3)] \emph{$L^2(P)$-consistent influence function estimate}: $E[\{D_{P_{\ell}, a}(O) - D_{n, a}(O)\}^2] = o_P(1)$, where $D_{P_\ell,a}$ denotes the in-probability limit of $D_{n,a}$ as $n$ approaches infinity and $D_{n,a}$ is treated as a fixed function of $O$ in this expression.

\item[(B4)] \emph{Glivenko Cantelli influence function estimate}: the probability that $D_{n, a}$ falls in a $P$-Glivenko Cantelli class tends to one as $n \rightarrow \infty$. 
\end{itemize}
Assumption (B1) guarantees that estimated propensities and motion densities are appropriately bounded so that the one-step estimator is never ill-defined. Assumption (B2) stipulates consistent estimations of the nuisance parameters. Assumptions (B3) and (B4) are necessary to ensure the negligibility of an empirical process term.% \citep{van1996weak}. 

% consistency
\begin{theorem}
\emph{(Multiple robustness)}. Under (B1) - (B4), $\theta_{n, a}^+ - \theta_a = o_p(1)$.  \label{theorem2}
\end{theorem}
According to Theorem \ref{theorem2}, our one-step estimators will only require \emph{some} of the nuisance parameters to be consistently estimated to achieve consistency of $\theta_{n, a}^+$. For example, (B2.1) implies that obtaining consistent estimates of $p_{M \mid \Delta = 1, A, X}$ and $p_{M \mid A, X, Z}$, and $\pi_a$ is sufficient to ensure a consistency of $\theta_{n, a}^+$. For a proof, see Supplement Section {\color{changed}5}. 

\begin{theorem} \label{theorem:one-step}
\emph{(Asymptotic linearity)}. Under (B1), (B3), and 
\begin{enumerate}[align=left]
% \item[(C1)] \emph{Bounded estimates:} $\pi_{n, a}$ is bounded below by some $\epsilon_1 > 0$, $\bar{\pi}_{n, 0}$ is bounded below by some $\epsilon_2 > 0$, and $p_{n, M \mid A, X, Z}(m \mid a, x, z)$ is below by some $\epsilon_3 > 0$.

\item[(C1)] \emph{$n^{1/2}$-convergence of second order terms}: $\lVert \xi_{n, a, \eta \mid X} - \xi_{a, \eta \mid X} \rVert \lVert \pi_{n, a} - \pi_{a} \rVert = o_P(n^{-1/2})$,  $\lVert \mu_{n, Y \mid A, M, X, Z}- \mu_{Y \mid A, M, X, Z}\rVert \{\lVert p_{n, M \mid \Delta = 1, A, X}(\cdot \mid 0, \cdot) - p_{M \mid \Delta = 1, A, X}(\cdot \mid 0, \cdot) \rVert + \lVert p_{n, M \mid A, X, Z} - p_{M \mid A, X, Z} \rVert \}  = o_P(n^{-1/2})$ and $\lVert p_{n, M \mid \Delta = 1, A, X}(\cdot \mid 0, \cdot) - p_{M \mid \Delta = 1, A, X}(\cdot \mid 0, \cdot) \rVert \{ \lVert \eta_{n, \mu \mid A, M, X} - \eta_{\mu \mid A, M, X} \rVert +  \lVert \pi_{n, a} - \pi_{a} \rVert + \lVert \bar{\pi}_{n, 0} - \bar{\pi}_{0} \rVert\} = o_P(n^{-1/2})$.

% \item[(C3)] \emph{$L^2(P)$-consistent influence function estimate}: $E[\{D_{P_{\ell}, a}(o) - D_{n, a}(o)\}^2] = o_P(1)$, where $D_{P_\ell,a}$ denotes the in-probability limit of $D_{n,a}$ as $n$ approaches infinity and $D_{n,a}$ is treated as a fixed function of $O$ in this expression.

\item[(C2)] \emph{Donsker estimates}: $D_{n, a}$ falls in a $P$-Donsker class with probability $\rightarrow 1$ as $n \rightarrow \infty$. 
\end{enumerate} then $\theta_{n, a}^+ - \theta_a = \frac{1}{n}\sum_{i=1}^n D_{P,a}(O_i) + o_P(n^{-1/2}),$ and
 $n^{1/2} (\theta_{n, a}^+ - \theta_a) \Rightarrow N(0,E[D_{P,a}(O)^2]).$
\end{theorem}
%A direct corollary of Theorem \ref{theorem:one-step} is that $n^{1/2} (\theta_{n, a}^+ - \theta_a)$ converges to a mean-zero Normal random variable with variance $E[D_{P,a}(O)^2]$.

Assumption (C1) states that nuisance estimates converge to their true values at a sufficiently fast rate, while (C2) ensures large-sample negligibility of a certain second-order empirical process term (so-called \emph{Donsker conditions}.%, \citealt{van1996weak}). 
(C2) can be eliminated through the use of cross-fitting (Supplement Section {\color{changed}7}). We study the benefits of this approach in our simulation. For further discussion of assumptions and a proof of the theorem, see Supplement Section {\color{changed}8}.

When all nuisance regressions are consistently estimated, $\sigma_n^2 = (n-1)^{-1} \sum_{i=1}^n \{ D_{n, a}(O_i) - n^{-1} \sum_{j=1}^n D_{n,a}(O_j) \} ^2$ provides a consistent estimate of $E[D_{P,a}(O)^2]$. % the asymptotic variance of $\theta_{n,a}^+$. 
Thus, an asymptotically justified $1-\alpha$ confidence interval for $\theta_a$ is $\theta_{n, a}^+ \pm n^{-1/2} z_{1-\alpha/2} \sigma_n$, where $z_{1-\alpha/2}$ denotes the $(1-\alpha/2)$-quantile of a standard Normal distribution. Similarly, by Theorem \ref{theorem:one-step} implies the limiting distribution of $n^{1/2}\{(\theta^+_{n, 1}-\theta^+_{n, 0})-(\theta_1-\theta_0)\}$ is $N(0, \tau^2)$, with $\tau^2 = \text{Var}(D_{P,1}(O)-D_{P,0}(O))$. The estimate $\tau_n^2 = (n-1)^{-1} \sum_{i=1}^n \{D_{n,1}(O_i) - D_{n,0}(O_i) - n^{-1}\sum_{j=1}^n(D_{n,1}(O_j) - D_{n,0}(O_j))\}^2$ will be consistent for $\tau^2$ and can be used to generate a confidence interval for the association of ASD with brain connectivity in a single brain region, $(\theta_{n, 1}^+ - \theta_{n, 0}^+) \pm n^{-1/2} z_{1-\alpha/2} \tau_n$.

\subsection{Simultaneous inference for associations}
\label{sec:simultaneous_conf}

To control family-wise error rate across hundreds of regions, we conduct testing using simultaneous confidence bands. % \citep{ruppert2003semiparametric}. 
Let $j=1,\dots,J$ index the region. In our application, $J=399$ as we examine the association between a seed region and 399 other regions. Let $\theta_{a,j}$ denote the MoCo estimand in group $a$ and region $j$, and let $\theta_{n, a, j}^+$ denote its estimate. Let  $D_{P,a,j}$ denote the EIF for diagnosis group $a$ and region $j$, and let $\tau_{n,j}^2$ denote the region-specific estimate of the asymptotic variance. By Theorem \ref{theorem:one-step},
\begin{align}
    \label{eq:simultaneous_1} 
    n^{1/2} \left\{ 
    \begingroup
    \renewcommand*{\arraystretch}{0.7}
    \begin{pmatrix} \theta_{n, 1, 1}^+ - \theta_{n, 0, 1}^+ \\ \smash{\vdots} \\ \theta_{n, 1, J}^+ - \theta_{n, 0, J}^+ \end{pmatrix} - \begin{pmatrix} \theta_{1, 1}-\theta_{0, 1} \\  \smash{\vdots} \\ \theta_{1, J}-\theta_{0, J} \end{pmatrix} 
    \endgroup
    \right\}
    \Rightarrow N\left\{
    \begingroup
    \renewcommand*{\arraystretch}{0.7}
    \begin{pmatrix} 0  \\ \smash{\vdots}  \\  0 \end{pmatrix}, \text{Cov}\begin{pmatrix} D_{P, 1, 1}(O)-D_{P, 0, 1}(O)  \\ \smash{\vdots}  \\  D_{P, 1, J}(O)-D_{P, 0, J}(O) \end{pmatrix}
    \endgroup
    \right\}.
\end{align} 
An approximate $1-\alpha$ simultaneous confidence interval is $
(\theta_{n, 1, 1}^+ - \theta_{n, 0, 1}^+, \dots, \theta_{n, 1, J}^+ - \theta_{n, 0, J}^+)^\top \pm z_{\rm{max},1-\alpha} (\tau_{n,1}, \dots, \tau_{n,J})^\top$, where $z_{\rm{max},1-\alpha}$ is the $1-\alpha$ quantile of
$\text{max}_{1 \leq j \leq J} \hs \{ n^{1/2} \lvert (\theta_{n, 1, j}^+ - \theta_{n, 0, j}^+) - (\theta_{1, j} - \theta_{0, j})\rvert \hs / \hs \tau_{n,j} \},$ which depends on the covariance matrix in \eqref{eq:simultaneous_1}.

To approximate $z_{\rm{max},1-\alpha}$, Monte-Carlo integration is performed by taking $10^5$ independent draws of a mean-zero $J$-variate normal random variable with covariance equal to the sample correlation matrix of the vectors $(D_{n, 1, 1}(O_i)-D_{n, 0, 1}(O_i), \dots, D_{n, 1, J}(O_i)-D_{n, 0, J}(O_i))^\top$, $i=1,\dots,n$. For each of the draws, the maximal absolute value of the components of the vector is calculated. The critical value $z_{\rm{max},1-\alpha}$ is approximated by the $(1-\alpha)$-quantile. Wald hypothesis tests controlling family-wise error rate at level $\alpha$ are conducted by rejecting the null hypothesis of no association between diagnosis group and functional connectivity in the $j$-th region whenever $n^{1/2} \lvert \theta_{n, 1, j}^+ - \theta_{n, 0, j}^+ \rvert \hs / \hs \tau_{n,j}$ is larger than the estimated value of $z_{\text{max}, 1-\alpha}$. 

\section{Simulation study}
\label{sec:simulation}
To mirror our data analysis, we simulated {\color{changed}1000 datasets with 400 children. Details are in Supplement Section 9.2. Briefly, for $X=[X_1,X_2,X_3]^\top$, we simulated values for age, sex and handedness with marginal distributions similar to the ABIDE dataset. For $Z=[Z_1,Z_2,Z_3,Z_4]^\top$, we simulated values for autism diagnostic observation schedule (ADOS), full-scale IQ (FIQ), indicator of stimulant usage and indicator of other medication usage. For $M$ given $A$, $X$ and $Z$, we simulated a conditional log-normal distribution. We simulated functional connectivity between a seed region equal to the default mode network and six other resting-state parcels using linear models with no interactions between diagnosis and motion or between symptoms and motion. In this design, the ideal $M=0$ estimand ($\theta_1^I - \theta_0^I$) is equal to MoCo ($\theta_1 - \theta_0$) (see Section \ref{sec:MoCo_estimand}). For $Y_1,Y_2,Y_3,Y_4$, we set the coefficients for $A,Z_1,Z_2,Z_3,$ and $Z_4$ equal to 0. For $Y_5$ and $Y_6$, coefficients were selected to result in} large and small negative associations existed between the diagnosis group and $Y_5$ and $Y_6$, respectively. For $Y_5$ and $Y_6$, our data generating process also included quadratic associations between motion and observed functional connectivity to examine the ability of Super Learner to account for possible non-linear relationships. {\color{changed}For example, $\mu_{Y_5 \mid A, M, X, Z}(a, m, x, z) = -0.20 - 0.03a + 1.50m - 0.61m^2 + 0.02(x_1 - z_4) - 0.002x_2 + 0.03(x_3 - z_3) - 0.0005z_1 + 0.0003z_2$. Other formulas are in the Supplement Section 9.2.} 

{\color{changed} We compared MoCo to four approaches. 1) The na{\"i}ve approach that removes high-motion participants, which targets the estimand $E(Y \mid A=1, \Delta=1)-E(Y \mid A=0, \Delta=1)$. 2) The na{\"i}ve approach that does not remove any participants, which targets $E(Y \mid A=1)-E(Y \mid A=0)$. 3) Inverse Probability of Treatment Weighting (IPTW), which targets $E\{E( Y \mid \Delta=1, A=1, X)\} - E\{E( Y \mid \Delta=1, A=0, X) \}$. 4) The method proposed by \cite{nebel2022accounting}, which regresses $Y$ against $M$ and $X$ and subsequently uses the residuals as input to doubly robust targeted minimum loss based estimation in which the removed high-motion scans are treated as missing data. %Ignoring the preprocessing step, this would target the estimand $E \{ E(Y \mid \Delta=1, A=1, Z) \mid A=1 \} - E \{ E(Y \mid \Delta=1, A=0, Z) \mid A=0 )$. 
With the preprocessing step, it targets $E ( E[ \{Y- \mu_{Y \mid M, X}(M,X)\} \mid \Delta=1, A=1, Z] \mid A=1 )- E ( E[\{ Y - \mu_{Y \mid M, X}(M,X) \}\mid \Delta=1, A=0, Z ] \mid A= 0 )$. Further details can be found in Supplement Section {\color{changed}9}.1. Arguably, the target estimands for both IPTW and Nebel's method are approximations to the ideal ($M=0$) estimand. Our simulation results reflect the gaps between the target and ideal estimands for respective methods in addition to estimation error.} 

\begin{figure}[h!]
	\centering
	\includegraphics[width=\textwidth]{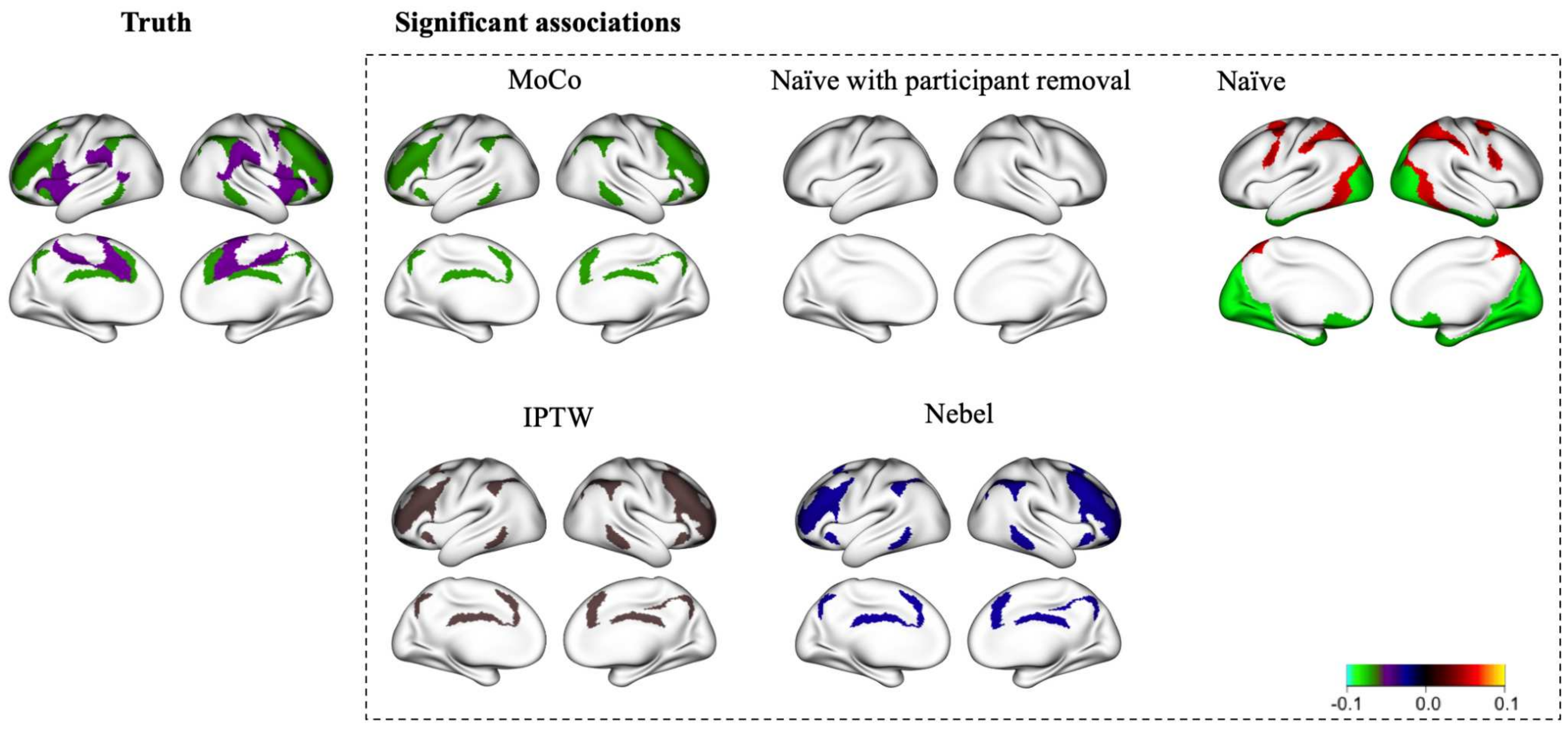}
	\caption{Example from a typical simulated dataset. The true association is marked in dark green and purple, while other regions have zero associations. MoCo identified one of the two true associations correctly. However, the na{\"i}ve method with participant removal failed to detect either of the two regions with true associations. The na{\"i}ve method with all data had two false positives and failed to recover either of the regions with true associations. IPTW detected one of the true regions, but it was a biased estimate. Nebel's method was able to detect one of the two true regions, but its estimate was highly biased because it could not capture the nonlinear relationship between motion and functional connectivity.}
	\label{fig:sim_plot}
\end{figure}

MoCo demonstrated advantages in terms of bias, MSE, type I error, and power relative to other approaches (Table \ref{sim:eval2}). It had the lowest type I error in regions 1 to 4, and lower bias than the na{\"i}ve methods and IPTW. MoCo achieved the lowest MSE in three of four zero-association regions and, in regions with true associations, demonstrated greater power and lower bias than all other methods. Nebel’s method was accurate in regions 1 to 4 in which motion impacts were linear, but exhibited substantial bias in regions 5 and 6 in which motion effects were nonlinear. In contrast, MoCo effectively captured both linear and nonlinear relationships, making it more robust across scenarios. Figure \ref{fig:sim_plot} illustrates the results of MoCo with cross-fitting on one simulated dataset and demonstrates its performance.

Additional simulations demonstrate that (i) MoCo has low bias but also lower power for small sample sizes (Supplement Section {\color{changed}9}.3); (ii) treating FIQ as $X$ instead of $Z$ tends to increase MoCo's MSE, but otherwise has minimal impact (Section {\color{changed}9}.4); (iii) that MoCo remains accurate when simulating time series motion that causes correlation between two regions (Section {\color{changed}9}.5); and (iv) that MoCo is multiply robust as shown in Theorem \ref{theorem2}.

\begin{table}[H]
\caption{Simulation results comparing MoCo, the na{\"i}ve approach with participant removal, the na{\"i}ve approach including all participants, IPTW, and \cite{nebel2022accounting}'s method. Bolded values indicate the lowest bias, standard deviation, MSE, and Type I error, and the highest power across methods.}
\label{sim:eval2}
\centering
\resizebox{0.88\textwidth}{!}{
\begin{tabular}{@{}llllllll@{}}
\toprule
 & Truth & Metric & MoCo & Na{\"i}ve removal & Na{\"i}ve &  {IPTW} &  {Nebel} \\ 
 \midrule
\multirow{4}{*}{Region 1} & \multirow{4}{*}{0.0000} 
& Bias & 0.0005 & -0.0190 & -0.0644 &  {-0.0107} &  {\textbf{0.0002}} \\
& & SD & 0.0372 & {\textbf{0.0191}} & 0.0204 &  0.0232 &  {0.0193} \\
& & MSE$\times 10^3$ & 1.3839 & 0.7283 & 4.5657 &  {0.6441} &  {\textbf{0.3720}} \\
& & Type I Error & \textbf{0.0110} & 0.1070 & 0.8670 &  {0.0610} &  {0.0740} \\ \midrule

\multirow{4}{*}{Region 2} & \multirow{4}{*}{0.0000} 
& Bias & 0.0048 & 0.0176 & 0.0611 &  {0.0097} &  {\textbf{0.0004}} \\
& & SD & 0.0238 & 0.0234 & {\textbf{0.0221}} &  0.0282 &  {0.0289} \\
& & MSE$\times 10^3$ & \textbf{0.5894} & 0.8576 & 4.2200 &  {0.6456} &  {0.8359} \\
& & Type I Error & \textbf{0.0100} & 0.0730 & 0.7220 &  {0.0610} &  {0.0910} \\ \midrule

\multirow{4}{*}{Region 3} & \multirow{4}{*}{0.0000} 
& Bias & 0.0044 & 0.0153 & 0.0553 &  {0.0088} &  {\textbf{-0.0002}} \\
& & SD & 0.0183 & {\textbf{0.0179}} & 0.0182 &  0.0223 &  {0.0211} \\
& & MSE$\times 10^3$ & \textbf{0.3554} & 0.5542 & 3.3941 &  {0.6464} &  {0.4441} \\
& & Type I Error & \textbf{0.0080} & 0.0680 & 0.8320 &  {0.0690} &  {0.0840} \\ \midrule

\multirow{4}{*}{Region 4} & \multirow{4}{*}{0.0000} 
& Bias & -0.0034 & -0.0179 & -0.0663 &  {-0.0105} &  {\textbf{-0.0001}} \\
& & SD & 0.0204 & {\textbf{0.0199}} & 0.0204 &  0.0234 &  {0.0208} \\
& & MSE$\times 10^3$ & {\textbf{0.4275}} & 0.7180 & 4.8182 &  0.6456 &  {0.4312} \\
& & Type I Error & \textbf{0.0100} & 0.1160 & 0.8870 &  {0.0870} &  {0.0880} \\ \midrule

\multirow{4}{*}{Region 5} & \multirow{4}{*}{-0.0484} 
& Bias & \textbf{0.0065} & 0.0213 & 0.0695 &  {0.0165} &  {0.0312} \\
& & SD & 0.0214 & {\textbf{0.0208}} & 0.0212 &  0.0250 &  {0.0264} \\
& & MSE$\times 10^3$ & \textbf{0.4990} & 0.8848 & 5.2748 &  {1.6316} &  {1.6748} \\
& & Power & \textbf{0.3790} & 0.1700 & 0.1260 &  {0.2410} &  {0.2272} \\ \midrule

\multirow{4}{*}{Region 6} & \multirow{4}{*}{-0.0682} 
& Bias & \textbf{0.0063} & 0.0241 & 0.0796 &  {0.0186} &  {0.0517} \\
& & SD & 0.0203 & {\textbf{0.0178}} & 0.0214 &  0.0220 &  {0.0261} \\
& & MSE$\times 10^3$ & \textbf{0.4523} & 0.8979 & 6.7937 &  {3.3802} &  {3.3521} \\
& & Power & \textbf{0.8690} & 0.5280 & 0.0670 &  {0.6000} &  {0.2490} \\ 
\bottomrule
\end{tabular}
}
\end{table}

Additional simulation results are available in the Supplement. In Supplement Section {\color{changed}9}.3, we show MoCo has low bias but also lower power for $n=50$ and $n=100$. In Supplement Section {\color{changed}9}.4, we show that under the same design described above, treating FIQ as $X$ instead of $Z$ tends to increase MoCo's MSE, but overall does not have a substantial impact. In Supplement Section {\color{changed}9}.5, we simulate a time series of motion that causes correlation between two regions, and we demonstrate that MoCo is accurate in this setting. Supplement Section {\color{changed}9}.6 presents additional simulations  (n = 50 to 4000) demonstrating multiple robustness.

\section{Data analysis of functional connectivity in ASD}
\label{sec:application}

%\subsection{Data and methods}

We conducted a functional connectivity analysis using a seed region in the default mode network. We applied our method to 377 resting-state fMRI data from children ages 8-13 in the ABIDE dataset \citep{di2014autism, di2017enhancing} (Supplement Table 8). Details are described in Supplement Section {\color{changed}10}. We compared na{\"i}ve estimates without participant removal, na{\"i}ve estimates with removal ($\Delta = 1$), IPTW, Nebel’s method, and MoCo with cross-fitting. FWER-critical values were determined using simultaneous confidence intervals (\cref{sec:simultaneous_conf}) derived from residual correlations for naïve methods, bootstrap replicates for IPTW, and EIFs for Nebel’s method {\color{changed}and MoCo}. {\color{changed}Nebel's method and} MoCo used the same super learner library as the simulations. To handle variability from cross-validation, we generated estimates 50 times and averaged z-statistics across runs. Positivity assumptions were assessed via histograms of inverse probability weights and density ratios (Supplement Section {\color{changed}10}.3). All estimated density ratios were below 4, suggesting the assumptions were reasonably satisfied. Results were visualized using the R package \texttt{ciftiTools} \citep{pham2022ciftitools}.

\begin{figure}[!p]
	\centering
	\includegraphics[width=0.92\textwidth]{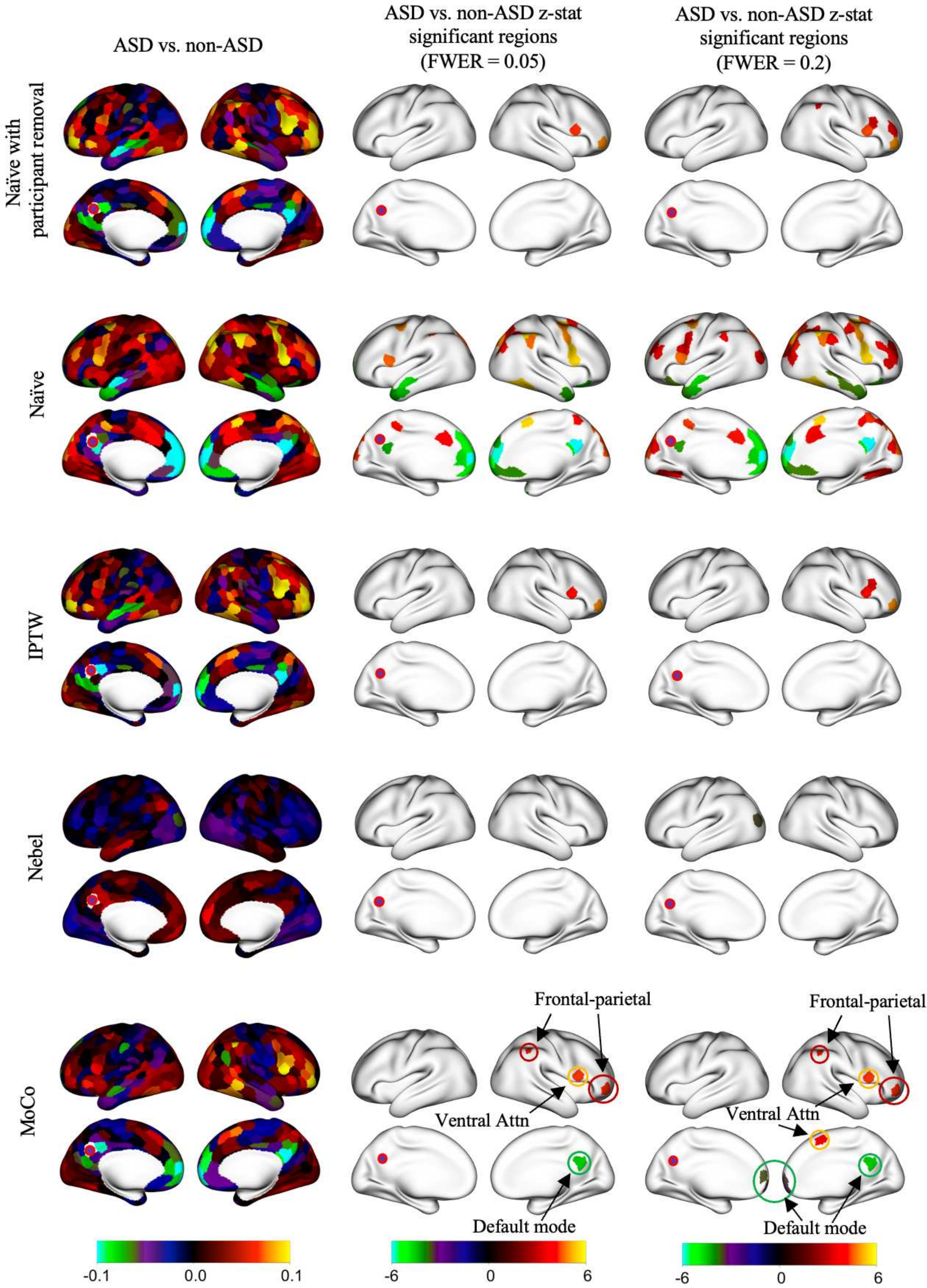}
	\caption{Z-statistics for the group difference (ASD$\;-\;$non-ASD) for a seed in the posterior default mode network (fuchsia point) in the ABIDE dataset. %Extensive hypoconnectivity between the seed region and anterior parts of the default mode network in the na{\"i}ve approach are possibly due to motion artifacts. While IPTW adjusts for motion and covariates via reweighting, it tends to identify a large number of regions, potentially due to biased estimates; some of these resemble the naïve results and may still reflect residual confounding. MoCo appears to effectively control motion artifacts, as these differences are reduced, and the results look more similar to the na{\"i}ve approach with participant removal. Nebel’s method, while theoretically rigorous, yields the fewest discoveries, possibly reflecting limited sensitivity under complex nonlinear confounding. Taken together, MoCo achieves a better balance, reducing bias while preserving sensitivity. At FWER=0.05, MoCo also identifies frontal-parietal hyperconnectivity not detected in the na{\"ive} approaches. %The na{\"i}ve approach with high-motion children generates spurious findings, while excluding them erases group differences.    MoCo highlights more significant differences than the na{\"i}ve approach with participant removal. 
    %Network labels from \cite{schaefer2018local}.  
 }
	\label{fig:zstats}
\end{figure}

MoCo {\color{changed} and} the na{\"i}ve approach use {\color{changed} the imaging} data from 377 participants, including 132 with ASD, while the na{\"i}ve approach with participant removal{\color{changed}, IPTW,} and Nebel's method use {\color{changed} the imaging data from} 160 participants, and only 34 with ASD. MoCo reveals four regions that differ in connectivity with the posterior default mode seed region in ASD versus non-ASD at FWER=0.05, including three regions of hyperconnectivity with distant frontal-parietal regions (\cref{fig:zstats}). The na{\"i}ve approach indicates more extensive differences than MoCo, including prominent default mode hypoconnectivity in ASD in long-distance correlations. These are possibly spurious differences due to motion, as long-distance correlations tend to be attenuated in high-motion participants \citep{ciric2017benchmarking}. These possible biases are also prominent in the mean connectivity estimates (Supplement Figure 4). The na{\"i}ve approach also selects some regions of hyperconnectivity in ASD with lateral regions of the frontal lobe. The na{\"i}ve approach with participant removal produces more conservative results, and it suffers from substantial loss of power due to reduced sample size. IPTW {\color{changed}behaves similarly to the na{\"i}ve approach with participant removal.} Nebel’s method yields the fewest discoveries, possibly reflecting low sensitivity from the small number of children that pass motion quality control. Overall, MoCo detects more regions than the na{\"i}ve with participant removal, {\color{changed}IPTW,} and Nebel's method, while remaining more selective than the na{\"i}ve approach. This may be due to improved sensitivity relative to {\color{changed} methods that do not use all imaging data (na{\"i}ve with participant removal, IPTW, and Nebel's method)} and improved specificity relative to the na{\"i}ve approach.

\vspace{-1em}

\section{Discussion}
\label{sec:dis}

{\color{changed}We introduce MoCo, a method for controlling motion in fMRI studies to estimate the difference in functional connectivity between two groups that addresses the selection bias caused by motion quality control exclusion criteria. We use flexible machine-learning techniques for parameter estimation with simultaneous confidence intervals for controlling FWER across hundreds of brain connections. MoCo improves statistical power and lowers type I error rate.}

In our data application, our findings differ greatly from the na{\"i}ve approach including all participants, which suggested hypoconnectivity across many DMN regions. The na{\"i}ve approach with participant removal suggests these differences were due to motion artifacts, but it is difficult to disentangle this from power loss and selection biases, as only 34 ASD children passed motion quality control. MoCo contributes to the ASD literature by flexibly modeling motion artifacts while including all the phenotypic variability in the study sample, providing stronger evidence that the hypoconnectivity differences were due to motion artifacts. MoCo recovered more regions than the na{\"i}ve approach with motion removal (four versus two at FWER=0.05), although the overall picture suggests minor differences between autistic and non-ASD children in correlations with the default mode network seed region. %Previous studies have shown that individuals with ASD exhibit hypoconnectivity between anterior and posterior DMN regions \citep{assaf2010abnormal}. 

An important decision in the modeling process is to designate variables that could or should have been balanced through careful recruitment ($X$) versus variables biologically related to diagnosis group ($Z$). In the data analysis, we consider age, sex, and handedness as $X$. We treat FIQ as a diagnosis-specific variable, which on average was lower in the ASD group. However, FIQ is highly variable in autism, and whether or not it should be considered as a part of $X$ or as a part of $Z$ is debatable. In our dataset, the child with the highest FIQ was also diagnosed with autism (Supplement Table 8). Neural diversity in autism is associated with strengths like unique perspectives, problem-solving skills, intense focus, attention to detail, and other traits that extend beyond a single measure of intelligence. %A limitation in our data analysis is that we only considered a limited set of behavioral and diagnosis-specific covariates, which was driven by the covariates available in the two ABIDE study sites. Additional research into the associations between functional connectivity and neural diversity may help elucidate the neurological underpinnings in ASD. %Researchers may choose to include specific variables from $Z$ to $X$ as long as they adhere to the positivity conditions and align with existing scientific knowledge. 

There are a number of limitations and directions for future research. %Although MoCo does not exclude any children due to motion during the rs-fMRI scan, there were a few scans that failed the fMRIPrep cortical segmentation pipeline due to issues with the structural scan, and effectively have missing fMRI data (Supplement Section 7, Web Supplement Figure 1). MoCo could be extended to incorporate the behavioral and demographic information from these participants. 
First, we use machine learning to predict functional connectivity from an overall measure of motion, mean FD, which does not use the time series structure of motion. Future research can investigate the use of machine learning to predict the BOLD time series from the motion alignment parameters, followed by the calculation of the functional connectivity, although this would be computationally demanding.
Second, our study fits functional connectivity for a seed-based analysis, rather than simultaneously analyzing the functional connectivity matrix. A matrix-variate approach could be designed to exploit low-rank or sparse structure, which may improve efficiency. Efficiency may also be improved in small samples by considering more stable nuisance parameter estimates, such as those based on low-dimensional, working parametric models. This approach may lead to improvements in the smallest sample sizes, where our methods showed considerable under coverage. % \citep{ran2024unveiling}. %Additionally, we analyze correlations, which have been shown to have much higher reproducibility than partial correlations \citep{mahadevan2021evaluating}. However, other authors prefer partial correlations \citep{smith2011network}, which may reveal different insights in autism. 
Finally, while we focus on rs-fMRI, MoCo could may be effective in other neuroimaging modalities, where motion can similarly induce spurious effects in morphometry and diffusion MRI.% \citep{yendiki2014spurious}.

\noindent {\color{changed} \textbf{Acknowledgments.} We thank Xiyan Tan, Liangkang Wang, and Zihang Wang for assistance with processing and quality control of the ABIDE data.}

\noindent \textbf{Funding.} This work was supported by R01 MH129855 (BR, DB). %The content is solely the responsibility of the authors and does not necessarily represent the official views of the National Institutes of Health.

\noindent {\color{changed} \textbf{Supplementary Materials.} Web Appendices, Tables, and Figures referenced in Sections 2-5 and simulations code are available with this paper at the Biometrics website on Oxford Academic. The R package MoCo is available on \url{https://github.com/thebrisklab/MoCo}.} 

\noindent \textbf{Data {\color{changed} Availability Statement.}} The Autism Brain Imaging Data Exchange (ABIDE) data are from \url{https://fcon_1000.projects.nitrc.org/indi/abide/}.

\vspace{-2em}

\processdelayedfloats

\newpage

%% ============================================================
%% SUPPLEMENTARY MATERIALS
%% ============================================================
\setcounter{section}{0}
\setcounter{figure}{0}
\setcounter{table}{0}
\setcounter{equation}{0}
\setcounter{theorem}{0}
\setcounter{lemma}{0}
\renewcommand{\thesection}{S\arabic{section}}
\renewcommand{\thesubsection}{S\arabic{section}.\arabic{subsection}}
\renewcommand{\thefigure}{S\arabic{figure}}
\renewcommand{\thetable}{S\arabic{table}}
\renewcommand{\theequation}{S\arabic{equation}}
\renewcommand{\theHsection}{supp.\arabic{section}}
\renewcommand{\theHsubsection}{supp.\arabic{section}.\arabic{subsection}}
\renewcommand{\theHfigure}{supp.\arabic{figure}}
\renewcommand{\theHtable}{supp.\arabic{table}}
\renewcommand{\theHequation}{supp.\arabic{equation}}
\renewcommand{\theHtheorem}{supp.\arabic{theorem}}

\makeatletter
\renewenvironment{figure}[1][htbp]{\@float{figure}[#1]}{\end@float}
\renewenvironment{table}[1][htbp]{\@float{table}[#1]}{\end@float}
\makeatother

\begin{center}
{\Large\bf Supplementary Materials for}\\[6pt]
{\large\bf Nonparametric Motion Control in Functional Connectivity Studies in Children with Autism Spectrum Disorder}\\[6pt]
{\large Jialu Ran, Sarah Shultz, Benjamin B. Risk, and David Benkeser}
\end{center}
\vspace{1em}

{\color{changed}
\section{Examining the gap between the ideal estimand and the MoCo estimand}

\subsection{Linear model with no interaction between diagnosis and motion and no interaction between symptom severity and motion}
    
    Consider the linear outcome model for the conditional mean $\mu_{Y|A,M,X,Z}(a,m,x,z)$:
    \begin{equation*}
        \mu_{Y\mid A,M,X,Z}(a,m,x,z) = \beta_0 + \beta_1 a + \beta_2 m + \beta_3 x + \beta_4 z.
    \end{equation*}
    In the ideal estimand, we fix motion $m = 0$. Substituting the linear model into the definition:
    \begin{align*}
        \theta_a^I &= \iint (\beta_0 + \beta_1 a + \beta_2*0 + \beta_3 x + \beta_4 z) p_{Z\mid A,X}(z\mid a,x) p_X(x) \, dz \, dx \\
        &= \beta_0 + \beta_1 a + \beta_3 E[X] + \beta_4 \iint z p_{Z\mid A,X}(z\mid a,x) p_X(x) \, dz \, dx \\
        &= \beta_0 + \beta_1 a + \beta_3 E[X] + \beta_4 E[E(Z \mid A=a, X)].
    \end{align*}
    
    The group difference is:
    \begin{align*}
        \theta_1^I - \theta_0^I &= \beta_1 + \beta_4 [E\{E(Z \mid A=1, X) - E(Z \mid A=0, X)\}].
    \end{align*}
    
    In MoCo, we average over the tolerable motion distribution $p_{M\mid \Delta=1, A=0, X}(m \mid 0, x)$:
    \begin{align*}
        \theta_a &= \iiint (\beta_0 + \beta_1 a + \beta_2 m + \beta_3 x + \beta_4 z) p_{Z\mid A,X} (z \mid a, x) p_{M\mid \Delta=1, A=0, X} ( m \mid  0, x) p_X (x) \, dz \, dm \, dx \\
        &= \beta_0 + \beta_1 a + \beta_2 \iiint m p_{M\mid \Delta=1, A=0, X} (m \mid 0, x) p_X (x) \, dm \, dx + \beta_3 E[X] + \beta_4 E[E(Z \mid A=a, X)].\\
    \end{align*}
    
    When calculating the difference $\theta_1 - \theta_0$, the terms for the intercept ($\beta_0$), the motion effect ($\beta_2$), and the covariates ($\beta_3$) are identical for both $a=1$ and $a=0$.
    \begin{align*}
        \theta_1 - \theta_0 &= \beta_1 + \beta_4 [E\{E(Z \mid A=1, X) - E(Z \mid A=0, X)\}].
    \end{align*}

\subsection{Semiparametric additive model with no interactions between motion and autism and no interactions between motion and symptoms}
Consider the semiparametric model $\mu_{Y \mid A,M,X,Z}(a,m,x,z) = g_1(a,x,z) + g_2(m,x)$. 

In the ideal estimand ($m=0$), the group difference is:
\begin{align*}
    \theta_1^I - \theta_0^I &= \iint \{g_1(1,x,z) + g_2(0,x)\} p_{Z\mid A,X}(z\mid 1,x) p_X(x) \, dz \, dx \\
    &\quad - \iint \{g_1(0,x,z) + g_2(0,x)\} p_{Z\mid A,X}(z\mid 0,x) p_X(x) \, dz \, dx \\
    &= \iint g_1(1,x,z) p_{Z\mid A,X}(z\mid 1,x) p_X(x) \, dz \, dx - \iint g_1(0,x,z) p_{Z\mid A,X}(z\mid 0,x) p_X(x) \, dz \, dx,
\end{align*}
where the $g_2(0,x)$ terms cancel because $\int p_{Z\mid A,X}(z\mid a,x) \, dz = 1$ for both $a \in \{0, 1\}$.

In the MoCo estimand, we average over the tolerable motion distribution $p_{M\mid \Delta=1, A=0, X}(m \mid 0, x)$:
\begin{align*}
    \theta_a &= \iiint \{g_1(a,x,z) + g_2(m,x)\} p_{Z\mid A,X}(z\mid a,x) p_{M\mid \Delta=1, A=0, X}(m \mid 0, x) p_X(x) \, dz \, dm \, dx \\
    &= \iint g_1(a,x,z) p_{Z\mid A,X}(z\mid a,x) p_X(x) \, dz \, dx + \iint g_2(m,x) p_{M\mid \Delta=1, A=0, X}(m \mid 0, x) p_X(x) \, dm \, dx.
\end{align*}

When calculating the difference $\theta_1 - \theta_0$, the second integral is identical for both $a=1$ and $a=0$:
\begin{align*}
    \theta_1 - \theta_0 &= \iint g_1(1,x,z) p_{Z\mid A,X}(z\mid 1,x) p_X(x) \, dz \, dx - \iint g_1(0,x,z) p_{Z\mid A,X}(z\mid 0,x) p_X(x) \, dz \, dx.
\end{align*}
Thus, $\theta_1^I - \theta_0^I = \theta_1 - \theta_0$, and the gap is zero.
}

\section{Illustration of imbalance of motion artifacts}

For simplicity, suppose $X = \emptyset$ and that $Z$ and $M$ are both binary with $Z = 0$ denoting less severe ASD symptoms, $Z = 1$ denoting more severe ASD symptoms, $M = 0$ denoting ``lower motion'' and $M = 1$ denoting ``higher motion'' within a tolerable motion distribution, which for simplicity we denote as $\tilde{p}_m$ for $m=0,1$. Suppose the joint distribution of $Z, M$ in the ASD group ($A = 1$) is given by the 2x2 table shown in Table \ref{tab:joint_distr_Z_M}, where we see that $Z$ and $M$ are correlated with more severe symptomology ($Z = 1$) associated with greater chance of higher motion ($M = 1$). 

\begin{table}[ht]
\centering
\begin{tabular}{c|c|c}
            &  $Z = 0$ & $Z = 1$ \\
\hline
 $M = 0$    &  0.4     & 0.1     \\
  $M = 1$    &  0.1     & 0.4     \\
\end{tabular}
\caption{Joint distribution of $Z$ and $M$ {\color{changed}given $A=1$.}}
\label{tab:joint_distr_Z_M}
\end{table}

We can easily compute that $p_{Z \mid M, A}(1 \mid 1, 1) = P(Z = 1 \mid M = 1, A = 1) = 0.8$ and $p_{Z \mid M, A}(1 \mid 0, 1) = P(Z = 1 \mid M = 0, A = 1) = 0.2$, while $p_{Z \mid A}(1 \mid 1) = P(Z = 1 \mid A = 1) = 0.5$ and $p_{Z \mid A}(0 \mid 1) = 0.5$. Consider the weights $p_{Z \mid M, A}(z \mid m, 1) \times \tilde{p}_m$ versus the weights $p_{Z \mid A}(z \mid 1) \times \tilde{p}_m$ shown in Table \ref{tab:weight_comparison}. Note that when we use $p_{Z \mid M, A}$ (second column), the children with more severe symptoms and low motion $(z=1, m=0)$ receive less weight than the children with less severe symptoms and low motion $(z=0, m=0)$. On the other hand, when we use $p_{Z \mid A}$ (third column), we are able to appropriately balance these residual motion artifacts across levels of disease severity. 

\begin{table}[ht]
\centering
\begin{tabular}{c|c|c}
            &  $p_{Z \mid M, A}(z \mid m, 1) \times \tilde{p}_m$ & $p_{Z \mid A}(z \mid 1) \times \tilde{p}_m$ \\
\hline
 $z = 1, m = 0$    &   $0.2 \times \tilde{p}_0$     & $0.5 \times \tilde{p}_0$  \\
 $z = 0, m = 0$    &   $0.8 \times \tilde{p}_0$     & $0.5 \times \tilde{p}_0$  \\
 $z = 1, m = 1$    &   $0.8 \times \tilde{p}_1$     & $0.5 \times \tilde{p}_1$  \\
 $z = 0, m = 1$    &   $0.2 \times \tilde{p}_1$     & $0.5 \times \tilde{p}_1$  \\
\end{tabular}
\caption{Comparison of weights used when conditioning on $M$ versus not.}
\label{tab:weight_comparison}
\end{table}

\newpage

\section{Connections to causal mediation analysis}%\label{sec:identification}

Our approach can be motivated from a purely statistical point of view, in terms of balancing certain covariates across groups to appropriately account for differential motion between diagnosis groups. It is also possible to motivate our approach from an explicitly causal viewpoint. In this section, we describe connections to existing causal literature and highlight several places where connections to causal inference may be more tenuous for our problem.

\subsection{Directed acyclic graph}

\begin{figure}[h]
\centering
\includegraphics[width=0.35\textwidth]{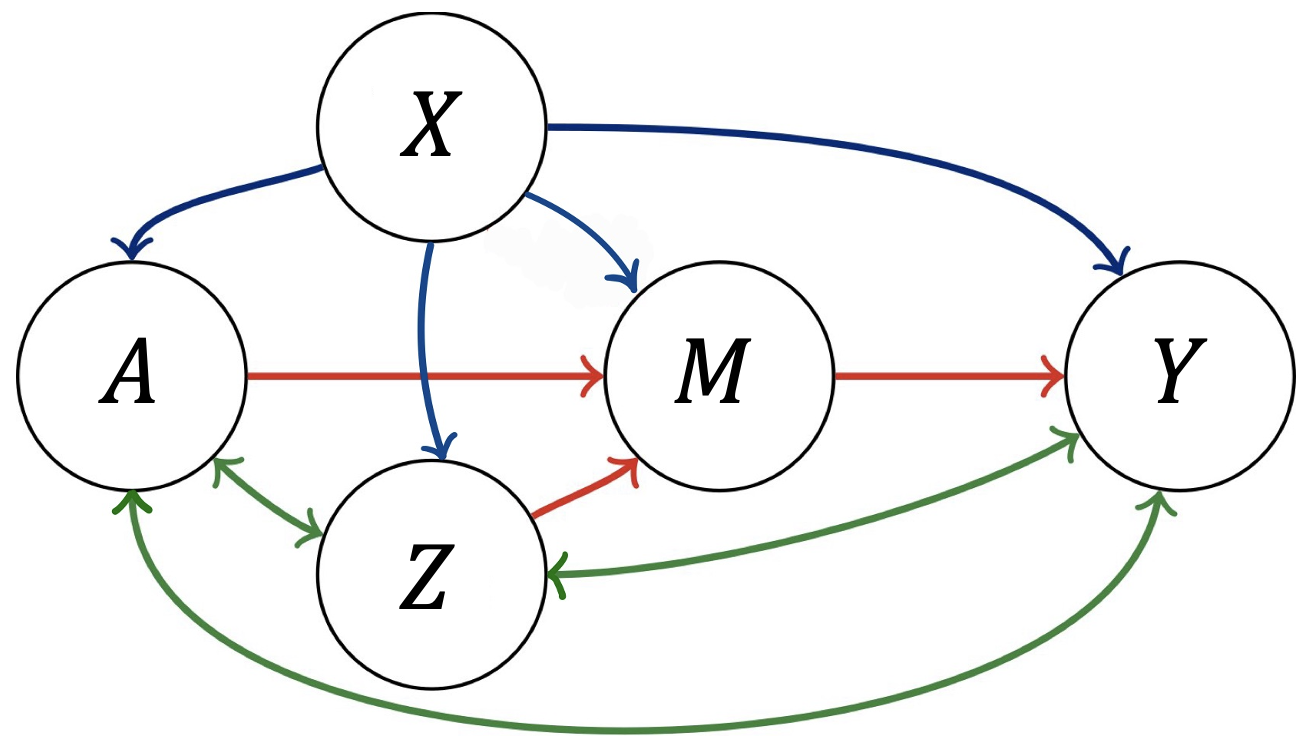}
\caption{Graphical representation of relationships between variables in analysis}
\label{fig:DAG}
\end{figure}

A fundamental aspect of causal inference is the usage of directed acyclic graphs to represent data structures and causal relationships between variables. A similar graph can be formulated for our problem (Figure \ref{fig:DAG}). In this graph, $A$ represents the diagnosis of Autism Spectrum Disorder (ASD): a value of 0 indicates a non-ASD child without autism, while 1 indicates a child with ASD. $M$ is a continuous motion variable such as mean FD, and $Y$ represents the functional connectivity between two brain locations, measured as a scalar element of the functional connectivity matrix. $X$ are covariates that may be considered ``confounders'' of the treatment, mediator, and outcome, such as age, sex, and handedness. $Z$ includes variables intrinsically related to the diagnostic group, such as IQ score, disease severity, and medication status. For this reason, we have elected to use double-edged arrows to denote that these variables do not represent temporally causal relationships and that $A$, $Z$, and $Y$ are simply measuring different aspects of a single condition that leads to a diagnosis of ASD. 

Our graph also departs from DAGs used in traditional causal inference in that $A$ represents a diagnosis status. Therefore, direct intervention on $A$ is not possible,  we cannot assign someone to have or not have ASD under any hypothetical experiment. Thus, even below when we describe hypothetical experiments that manipulate motion values, it is still not appropriate to describe differences between groups as causal effects and we instead out for an associational interpretation. 

%causal mediation analysis, whereby motion is considered a mediating variable. We aim to disentangle the impact of motion in the pathway between ASD and functional connectivity. 

\subsection{Formalizing causal estimands}

To formalize our approach using machinery from causal inference, we imagine a hypothetical experiment in which we are able to manipulate the motion of children during the scanning session. The hypothetical intervention is motivated by empirical studies in which children complete training in a mock scanner. In the data set used in our analysis, children received at least one mock training, but oftentimes this training was insufficient to adequately control motion in the scanner. Therefore, we consider defining counterfactuals based on the hypothetical intervention that could be given to children prior to scanning that would successfully reduce motion during the scan. The hypothetical training program is applied to all children irrespective of diagnosis category. Our methods mathematically formalize the nature of the training program in terms of its impact on scanner motion, define the counterfactual data that would be observed under the hypothetical training program, describe assumptions under which inference can be made pertaining to the counterfactual data, and provide estimators of associations between diagnosis and functional connectivity that appropriately account for motion artifacts.

We first define the \emph{total association} between ASD diagnosis and functional connectivity as $\theta_{\text{O}, 1} - \theta_{\text{O}, 0}$, where $\theta_{\text{O}, a} = E(E(Y \mid A = a, X))$, where the outer expectation is with respect to $P_X$, the marginal distribution of $X$. The total association can be interpreted as the mean difference functional connectivity between diagnosis groups after balancing measured confounders $X$ and making \emph{no effort} to control for motion. If $X$ were balanced across diagnosis groups, i.e., if $X \ind A$, then $\theta_{\text{O}, 1} - \theta_{\text{O}, 0} = E(Y \mid A = 1) - E(Y \mid A = 0)$. Therefore, the total association is adjusting for differences in the distribution of $X$ across diagnosis groups, but nothing else. The subscript O denotes an association in the {\it observed} data distribution that may include potentially inappropriate and imbalanced levels of motion.

Using a mediation-inspired approach, we then aim to isolate the associations of biological interest between ASD and functional connectivity from the total association by mitigating the influences of motion differences between diagnosis groups. This association of biological interest, which we term a \emph{motion-controlled association} (MoCo), is an analogue to the classic direct effect in the causal mediation literature \citep{pearl2014interpretation}. However, in this case, we do not view ASD as something that is inherently manipulable, so we restrict the interpretation of our analysis to describing direct \emph{associations} of ASD with functional connectivity. On the other hand, we can view motion as something that \emph{is} inherently manipulable. 

Since all children have at least \emph{some} motion during a scanning session, as a practical alternative, we suggest using an estimand inspired by a \emph{stochastic intervention} in which motion is a mediating variable. In causal inference, \emph{static interventions} imagine counterfactual scenarios wherein a variable is set to the same fixed value for everyone. On the other hand, a stochastic intervention is one in which the value of a variable is drawn at random from a user-specified distribution \citep{diaz2021nonparametric}. We define an \emph{acceptable motion distribution} and imagine a counterfactual scenario where all children, irrespective of diagnosis category, receive sufficient training such that the distribution of motion that would be observed in the counterfactual scenario corresponds to this acceptable motion distribution. The acceptable motion distribution may be allowed to depend on covariates and should be selected to represent a distribution of motion that, if present in the experimental data, would still yield biologically meaningful readouts of functional connectivity. 

Let $M_0 \sim P_{M \mid \Delta = 1, A, X}(m \mid 0, x)$ be the counterfactual motion that would be observed for a child in the counterfactual scenario under our stochastic intervention. Let $Y(M_0)$ denote the counterfactual functional connectivity that would be measured under the training provided in our hypothetical experiment. The counterfactual data unit generated in this scenario is $O_{\text{C}} = (A,M_0,X, Z, Y(M_0)) \sim P_{\text{C}}$, %Note that the joint distribution of $X, Z, A$ is the same under $P_{\text{C}}$ as under $P$
where the probability distribution of evaluated at an observation $o_c = (a, m_0, x, z, y)$ can be written $P_{\text{C}}(o_c) = P_{C, Y(M_0) \mid A, M, X, Z}(y \mid a, m_0, x, z) P_{M \mid \Delta = 1, A, X}(m_0 \mid 0, x) P_{Z \mid A, X}(z \mid a, x) P_{A \mid X}(a \mid x) P_X(x)$. 
The scanner training program in our hypothetical experiment leads to a motion distribution under $P_{\text{C}}$ that is described by $P_{M \mid \Delta = 1, A, X}(\cdot \mid 0, x)$ as opposed to the motion distribution $P_{M \mid A, X, Z}(\cdot \mid a, x, z)$ observed in the real experiment. The counterfactual motion is conditionally independent of diagnostic group ($A$) and diagnosis-specific variables ($Z$) given demographic covariates ($X$), which is
consistent with our goal to control for the impact of motion in the two groups. Similarly, the conditional distribution of measured functional connectivity $Y(M_0)$ may be altered under $P_{\text{C}}$ relative to the measured functional connectivity $Y$ under $P$.

Using this counterfactual construction, we then define a motion-controlled association $\theta_{\text{C}, 1} - \theta_{\text{C}, 0}$, where $\theta_{\text{C}, a} = E\{E_{\text{C}}[Y(M_{0}) \mid A = a,  X]\}$, and we use $E_{\text{C}}$ to denote expectation under $P_{\text{C}}$ with the subscript C denoting an association defined with respect to a counterfactual distribution. The motion-controlled association compares the ASD and non-ASD group while controlling for (i) $X$ differences between diagnosis groups; \emph{and additionally} (ii) differences in motion between the two diagnosis groups by ensuring that the $X$-conditional motion distribution is the same across diagnosis groups. 

\subsection{Identifying the motion-controlled association of ASD and brain connectivity} \label{sec:identification}

Identifiability of $\theta_{\text{C}, a}$ for $a=0,1$ can be established under the following assumptions:
\begin{itemize}
\item[(A1)] {\it Positivity}: (A1.1) for every $x$ such that $p_X(x)>0$, we also have $\pi_a(x) > 0$ for $a = 0, 1$; for every $x$ such that $p_X(x)>0$, we also have $P(\Delta = 1 \mid A = 0, X = x) > 0$. (A1.2) for every $(x, z, m)$ such that $p_X(x) p_{Z \mid A, X}(z \mid a, x)p_{M \mid \Delta = 1, A, X}(m \mid 0, x) > 0$, we also have that $p_{M \mid A, X, Z}(m \mid a, x, z) > 0$ for $a = 0,1$.
\item[(A2)] {\it Mean exchangeability}: for all $m$ such that $P\{ p_{M \mid \Delta = 1, A, X} (m \mid 0, X) > 0 \} > 0$, $E_{\text{C}}\{Y(m) \mid A = a, X, Z\} = E_{\text{C}}\{Y(m) \mid A = a, M = m, X, Z\}$ a.e.-$P$.
\item[(A3)] {\it Causal Consistency}: for any child with observed motion value $M = m$, the observed functional connectivity measurement $Y$ is equal to the counterfactual functional connectivity measurement $Y(m)$.
\end{itemize}
At the end of this section, we describe how our proposal yields biologically relevant inference even when (A2) and (A3) do not hold.

Assumption (A1) is described in the main manuscript.

Assumption (A2) implies that conditioned on $X$, $Z$, and diagnosis status $A$, there is no unmeasured confounding between $Y(m)$ and $M$. %The plausibility of this assumption could be scrutinized, for example, by establishing conditional \emph{d}-separation of $Y$ and $M$ in a graph \citep{pearl1988probabilistic}. 
%Because this assumption is fundamentally an assumption on the counterfactual distribution $P_{\text{C}}$, it cannot be fully verified empirically.
Assumption (A3) stipulates that the observed functional connectivity from children who naturally have motion level $m$ is the same as the functional connectivity that would have been observed under our hypothetical experiment where children receive training in the scanner. This assumption would be violated, for example, if the hypothetical scanner training received by children had an impact on the underlying functional connectivity of the child's brain. This assumption seems plausible but is not verifiable empirically.% BEN TO ADD A SENTENCE ON PLAUSIBILITY?

\begin{theorem}
Under (A1)-(A3), the counterfactual $\theta_{\text{C}, a}$ is identified by $\theta_a$, where \\$\theta_a = \iiint \mu_{Y|A, Z, M, X}(a, z, m, x) \hs p_{Z \mid A, X}(z \mid a, x) \hs p_{M \mid \Delta = 1, A, X}(m \mid 0, x) \hs p_X(x) dz \hs dm \hs dx$.
\end{theorem}

\textit{Proof} Let $\theta_{\text{C}, a} = E\{E_{\text{C}}[Y(M_0) \mid A = a, X]\}$, where $M_0\sim P_{M \mid \Delta = 1, A, X}(m \mid 0, x)$. We have:

{\footnotesize
    \begin{align*}
    &\theta_{\text{C}, a} = E\{E_{\text{C}}[Y(M_0) \mid A = a, X]\} \\
    &\text{=} \int E_{\text{C}}(Y(M_0) \mid A=a, X = x) \hs p_{X}(x) \hs dx\\
    &\stackrel{\text{tower rule}}{=} \iint E_{\text{C}}(Y(m) \mid M_0 = m, A = a, X = x) \hs p_{M \mid \Delta = 1, A, X}(m \mid 0, x) \hs p_{X}(x) \hs dm \hs dx  \\
    &\stackrel{\text{defn}}{=} \iint E_{\text{C}}(Y(m) \mid A = a, X = x) \hs p_{M \mid \Delta = 1, A, X}(m \mid 0, x) \hs p_{X}(x) \hs dm \hs dx  \\
    &\stackrel{\text{tower rule}}{=} \iiint E_{\text{C}}(Y(m) \mid A = a, X = x, Z = z) \hs p_{Z \mid A, X }(z \mid a, x) \hs p_{M \mid \Delta = 1, A, X}(m \mid 0, x) \hs p_{X}(x) dz \hs dm \hs dx  \\
    &\stackrel{\text{assumption (A2)}}{=} \iiint E_{\text{C}}(Y(m) \mid M = m, A = a, X = x, Z = z) \hs p_{Z \mid A, X }(z \mid a, x) \hs p_{M \mid \Delta = 1, A, X}(m \mid 0, x) \hs p_{X}(x) dz \hs dm \hs dx  \\
    &\stackrel{\text{assumption (A3)}}{=} \iiint E(Y \mid M = m, A = a, X = x, Z = z) \hs p_{Z \mid A, X }(z \mid a, x) \hs p_{M \mid \Delta = 1, A, X}(m \mid 0, x) \hs p_{X}(x) dz \hs dm \hs dx  \ .
    \end{align*}
}
The fourth equality results from the fact that by the construction of $M_0$, we have that $Y(m) \ind M_0 \mid A, X$ for all $m$. The sixth equality results from the assumption that $Y(m) \ind M \mid A, X, Z$.

The parameter $\theta_a$ involves integrating the conditional mean functional connectivity $\mu_{Y \mid A, M, X, Z}$ over the distributions of $Z$, $M$, and $X$ in a sequential manner. We note that the integration over $Z$ is specific to diagnosis group $a$, while the integration over $M$ and $X$ is the same irrespective of diagnosis group. Thus, a comparison of $\theta_1$ and $\theta_0$ provides a marginal associative measure that describes the joint impact of diagnosis category $A$ and diagnosis-specific variables $Z$ on functional connectivity while controlling for both motion $M$ and covariates $X$.

We conclude this section by noting that inference on $\theta_1 - \theta_0$ may be biologically relevant \emph{even in settings} where the fundamentally untestable assumptions (A2) and (A3) do not hold. To make this argument, we define \begin{equation}
        \eta_{\mu \mid A, M, X}(a, m, x) = \int \mu_{Y \mid A, M, X, Z}(a, m, x, z) \hs p_{Z \mid A, X}(z \mid a, x) \hs dz \ ,
\end{equation}
and note that $\eta_{\mu \mid A, M, X}(1, m, x) - \eta_{\mu \mid A, M, X}(0, m, x)$ describes an $m$- and $x$-specific difference in functional connectivity between diagnosis groups. Thus,
$\theta_1 - \theta_0 = \iint \{ \eta_{\mu \mid A, M, X}(1, m, x) - \eta_{\mu \mid A, M, X}(0, m, x) \}$ \\ $ p_{M \mid \Delta = 1, A, X}(m \mid 0, x) \hs p_X(x) \hs dm \hs dx$, 
simply standardizes these $m$- and $x$-specific associations over the selected acceptable motion distribution and distribution of covariates, thereby controlling for motion and covariate differences between diagnostic groups. We argue that this is likely still a biologically relevant parameter for describing differences in functional connectivity between diagnosis groups even when (A2)-(A3) do not hold.

\section{Efficient influence function Theorem 1}
\label{eif}

\subsection{Overview of efficiency theory}

The efficient influence function can be used to characterize the nonparametric efficiency bound, i.e., the smallest asymptotic variance of any regular, \emph{asymptotically linear} estimator of $\theta_a$ \citep{bickel1993efficient}. An estimator $\theta_{n, a}$ of $\theta_a$ based on $O_1, \dots, O_n \stackrel{\text{i.i.d}}{\sim}P$ is said to be asymptotically linear if there exists a function $o \mapsto \Tilde{D}_{P,a}(o)$ such that $E[\Tilde{D}_{P,a}(O)] = 0$, $E[ \Tilde{D}_{P,a}^2(O)] < \infty$, and $\theta_{n, a}=\theta_a + n^{-1} \sum_{i=1}^n \Tilde{D}_{P,a}(O_i) + o_P(n^{-1/2})$. We refer to $\Tilde{D}_{P,a}$ as the \emph{influence function} of $\theta_{n, a}$. Under this representation, the asymptotic study of $\theta_{n, a}$ reduces to the study of the sample mean $n^{-1} \sum_{i=1}^n \Tilde{D}_{P,a}(O_i)$ whose large sample behavior can be described using standard results such as the weak law of large numbers and the central limit theorem. The latter implies $n^{1/2}(\theta_{n, a}-\theta_a)$ converges to a mean-zero normal random variable with variance equal to $E[\Tilde{D}_{P,a}^2(O)]$. Due to the fact that the asymptotic variance of an asymptotically linear estimator is characterized by the variance of the influence function, the influence function that has the smallest variance amongst all influence functions of regular estimators is called the \emph{efficient influence function}. An estimator with an influence function equal to the efficient influence function is, by definition, asymptotically efficient.

\subsection{Proof of Theorem 1}

\textit{Proof} Let $\mathcal{P}$ be the model for the true distribution. Let $o=(a, m, \delta, x, z, y)$ denote values of the observed vector of variable $O=(A, M, \Delta, X, Z, Y)$, and $O \sim P \in \mathcal{P}$. We use $\Psi_a: \mathcal{P} \rightarrow \mathbb{R}$ to denote a parameter as a functional that maps the distribution $P$ in the model $\mathcal{P}$ to the real number $\theta_a$. 

\begin{align*}
    \Psi_a(P) & =\iiint \mu_{Y\mid A,M,X,Z}(a,m,x,z) p_{Z \mid A,X}(z \mid a,x) p_{M \mid \Delta=1,A,X}(m \mid 0, x) p_X(x) d z d m d x \\
    & =\int y d P_{Y \mid A,M,X,Z}(y \mid a,m,x,z) d P_{Z \mid A,X}(z \mid a,x) d P_{M \mid \Delta=1,A,X}(m \mid 0, x) d P_X(x) \ .
\end{align*}

We consider a collection of submodels through $P$ at $\epsilon = 0$ in the direction $S$, $\{ P_{\epsilon} \in \mathcal{P}$, $dP_{\epsilon} = (1 + \epsilon S)dP\}$ where $S$ is an element of the Hilbert space $L^2_0(P)$, the space of all functions of $O$ such that $\int S(o)dP(o) = 0$, $\int S(o)^2 dP(o) < \infty$ equipped with inner product $<f, g> = \int f(o) g(o) dP(o)$. We consider the derivative of the parameter mapping along the path $P_{\epsilon}$. We can view this derivative as a bounded functional on $L^2_0(P)$, which, by the Reisz representation theorem, will have an inner-product form $<s, D_P>$ for a unique element $D_P \in L^2_0(P)$. $D_P$ is referred to as the \emph{canonical gradient} of $\Psi_a$. This gradient will also be the efficient influence function of regular asymptotically linear estimators of $\Psi_a$. Thus, to derive the efficient influence function, we may study the following derivative of $\Psi_a$ and write that derivative in an inner product form \citep{levy2019tutorial}. Below, we use the notation $\int f(o) dP_0(o)$ interchangably with $\int f(o) P_0(do)$ to denote the Lebesgue integral of a $P_0$-measurable function $f$ with respect to probability measure $P_0$.
{\footnotesize
  \begin{align*}
    \left.\frac{\partial}{\partial \epsilon} \Psi_a\left(P_{\epsilon}\right)\right|_{\epsilon=0} 
    & =\left.\frac{\partial}{\partial \epsilon} \int y d P_{\epsilon, Y \mid A,M,X,Z}(y \mid a,m,x,z) d P_{\epsilon, Z \mid A,X}(z \mid a,x) d P_{\epsilon, M \mid \Delta=1,A,X}(m \mid 0,x) d P_{\epsilon, X}(x)\right|_{\epsilon=0} \\
    & =\int y S_{Y \mid A, M, X, Z}(y \mid a,m,x,z) d P_{Y \mid A,M,X,Z}(y \mid a,m,x,z) d P_{Z \mid A,X}(z \mid a,x) d P_{M \mid \Delta=1,A,X}(m \mid 0,x) d P_X(x) \tag{2.1} \\
    & +\int y S_{Z \mid A,X}(z \mid a,x) d P_{Y \mid A,M,X,Z}(y \mid a,m,x,z) d P_{Z \mid A,X}(z \mid a,x) d P_{M \mid \Delta=1,A,X}(m \mid 0,x) d P_X(x) \tag{2.2} \\
    & +\int y S_{M \mid \Delta=1,A,X}(m \mid 0,x) d P_{Y \mid A,M,X,Z}(y \mid a,m,x,z) d P_{Z \mid A,X}(z \mid a,x) d P_{M \mid \Delta=1,A,X}(m \mid 0,x) d P_X(x) \tag{2.3} \\
    & +\int y S_X(x) d P_{Y \mid A,M,X,Z}(y \mid a,m,x,z) d P_{Z \mid A,X}(z \mid a,x) d P_{M \mid \Delta=1,A,X}(m \mid 0,x) d P_X(x) \tag{2.4}
    \end{align*} 
where
  \begin{align*}
& d P_{\epsilon, Y \mid A,M,X,Z}(y \mid a,m,x,z) = \frac{\int_{\delta} (1 + \epsilon S(o)) P(a, m, d\delta, x, z, y) }{\int_{\delta, y} (1 + \epsilon S(o)) P(a, m, d\delta, x, z, dy) }, \\
& S_{Y \mid A, M, X, Z}(y \mid a,m,x,z) = \frac{\partial \log dP_{\epsilon, Y \mid A,M,X,Z}(y \mid a,m,x,z)}{\partial \epsilon}\Big|_{\epsilon=0} = E(S(O) \mid y, a, m, x, z) - E(S(O) \mid a, m, x, z), \\
& d P_{\epsilon, Z \mid A,X}(y \mid a,x) = \frac{\int_{\delta,m,y} (1 + \epsilon S(o)) P(a, dm, d\delta, x, z, dy)}{\int_{\delta,m,y,z} (1 + \epsilon S(o)) P(a, dm, d\delta, x, dz, dy)}, \\
& S_{Z \mid A,X}(z \mid a,x) = \frac{\partial \log dP_{\epsilon, Z \mid A,X}(y \mid a,x)}{\partial \epsilon}\Big|_{\epsilon=0} = E(S(O) \mid z, a, x) - E(S(O) \mid a, x), \\
& d P_{\epsilon, M \mid \Delta=1,A,X}(m \mid 0,x) = \frac{\int_{z,y} \iv_{0, 1}(a, \delta)(1 + \epsilon S(o)) P(a, m, \delta, x, dz, dy)}{\int_{m, z,y} \iv_{0, 1}(a, \delta)(1 + \epsilon S(o)) P(a, dm, \delta, x, dz, dy)}, \\
& S_{ M \mid \Delta=1,A,X}(m \mid 0,x) = \frac{\partial \log P_{\epsilon, M \mid \Delta=1,A,X}(m \mid 0,x)}{\partial \epsilon}\Big|_{\epsilon=0} = E(S(O) \iv_{0, 1}(A, \Delta) \mid m, a, \delta, x) - E(S(O) \iv_{0, 1}(A, \Delta)\mid a, \delta, x), \\
& d P_{\epsilon, X}(x) = \int_{\delta,a,m,y,z} (1 + \epsilon S(o)) P(da, dm, d\delta, x, dz, dy), \\
& S_{X}(x) = E(S(O) \mid x) \ .
\end{align*} 
\par} 

\newpage 
{\footnotesize
Evaluating the derivative for the term (2.1), we have:

\begin{align*} 
& (2.1) \int y S_{Y \mid A, M, X, Z}(y \mid a,m,x,z) P_{Y \mid A,M,X,Z}(y \mid a,m,x,z) d P_{Z \mid A,X}(z \mid a,x) d P_{M \mid \Delta=1,A,X}(m \mid 0,x) d P_X(x) \\ 
& = \int \frac{\iv_a(a^{\prime})}{\pi_{a^{\prime}}(x)} y S_{Y \mid A, M, X, Z}\left(y \mid a^{\prime}, m, x, z\right) d P_{Y \mid A,M,X,Z}\left(y \mid a^{\prime}, m, x,z\right) d P_{Z \mid A,X}\left(z \mid a^{\prime}, x\right) d P_{M \mid \Delta=1,A,X}(m \mid 0, x) d P_{A, X}\left(a^{\prime}, x\right) \\
& = \int \frac{\iv_a(a^{\prime})}{\pi_{a^{\prime}}(x)} \frac{p_{M \mid \Delta=1, A,X}(m \mid 0,x)}{p_{M \mid A, X, Z}\left(m \mid a^{\prime},x, z\right)} y S_{Y \mid A, M, X, Z}\left(y \mid a^{\prime}, m, x, z\right) d P_{Y \mid A,M,X,Z}\left(y \mid a^{\prime}, m, x,z\right) \\ 
& \hspace{2em} dP_{M \mid A, X, Z}\left(m \mid a^{\prime},x, z\right) d P_{Z \mid A,X}\left(z \mid a^{\prime}, x\right) d P_{A, X}\left(a^{\prime}, x\right) \\
& =\int \frac{\iv_a(a^{\prime})}{\pi_{a^{\prime}}(x)}\frac{p_{M \mid \Delta=1, A,X}(m \mid 0,x)}{p_{M \mid A, X, Z}\left(m \mid a^{\prime},x, z\right)} y S_{Y \mid A, M, X, Z}\left(y \mid a^{\prime}, m, x,z\right) d P(o) \\ 
& \stackrel{\text{*}}{=}\int \frac{\iv_a(a^{\prime})}{\pi_{a^{\prime}}(x)}\frac{p_{M \mid \Delta=1, A,X}(m \mid 0,x)}{p_{M \mid A, X, Z}\left(m \mid a^{\prime},x, z\right)} \left(y - \mu_{Y \mid A, M, X, Z}(a, m, x, z)\right) S_{Y \mid A, M, X, Z}\left(y \mid a^{\prime}, m, x,z\right) d P(o) \\
& \stackrel{\text{**}}{=}\int \frac{\iv_a(a^{\prime})}{\pi_a(x)} \frac{p_{M \mid \Delta=1, A,X}(m \mid 0,x)}{p_{M \mid A, X, Z}\left(m \mid a,x, z\right)} \left(y - \mu_{Y \mid A, M, X, Z}(a, m, x, z)\right) S(o) d P(o) \ .
\end{align*}
\par} 

{\footnotesize
The reason for (*) is:
\begin{align*} 
\int \frac{\iv_a(a^{\prime})}{\pi_{a^{\prime}}(x)}\frac{p_{M \mid \Delta=1, A,X}(m \mid 0,x)}{p_{M \mid A, X, Z}\left(m \mid a^{\prime},x, z\right)} \mu_{Y \mid A, M, X, Z}(a, m, x, z) S_{Y \mid A, M, X, Z}\left(y \mid a^{\prime}, m, x,z\right) d P(o) = 0 \ .
\end{align*}

For (**), since 
\begin{align*} 
& S_{Y \mid A, M, X, Z}\left(y \mid a^{\prime}, m, x,z\right) = E(S(O) \mid y, a^{\prime}, m, x, z) - E(S(O) \mid a^{\prime}, m, x, z) \\
& S_{Y, A, M, X, Z}\left(y, a^{\prime}, m, x,z\right) = E(S(O) \mid y, a^{\prime}, m, x, z)
\\ 
& S_{A, M, X, Z}\left(a^{\prime}, m, x,z\right) \ =  E(S(O) \mid a^{\prime}, m, x, z), 
\end{align*} 
we have:
\begin{align*} 
& S_{Y, A, M, X, Z}\left(y, a^{\prime}, m, x,z\right) = S_{Y \mid A, M, X, Z}\left(y \mid a^{\prime}, m, x,z\right) + S_{A, M, X, Z}\left(a^{\prime}, m, x,z\right) \ . 
\end{align*}
and,
\begin{align*}
& \int \frac{\iv_a(a^{\prime})}{\pi_a(x)} \frac{p_{M \mid \Delta=1, A, X}(m \mid 0, x)}{p_{M \mid A, X, Z}(m \mid a, x, z)} \left( y - \mu_{Y \mid A, M, X, Z}(a, m, x, z) \right) S_{A, M, X, Z}(a^{\prime}, m, x, z) \, dP(o) = 0.
\end{align*}
then, 
\begin{align*}
& \int \frac{\iv_a(a^{\prime})}{\pi_a(x)} \frac{p_{M \mid \Delta=1, A, X}(m \mid 0, x)}{p_{M \mid A, X, Z}(m \mid a, x, z)} \left( y - \mu_{Y \mid A, M, X, Z}(a, m, x, z) \right) S_{Y, A, M, X, Z}\left(y, a^{\prime}, m, x,z\right) \, dP(o) = 0.
\end{align*}
Since 
\begin{align*} 
&S(o) = S_{Y, A, M, X, Z}\left(y, a^{\prime}, m, x,z\right) + S_{\Delta \mid Y, A, M, X, Z}\left(\delta \mid y, a^{\prime}, m, x,z\right) \,
\end{align*}
and, 
\begin{align*} 
& \int \frac{\iv_a(a^{\prime})}{\pi_a(x)} \frac{p_{M \mid \Delta=1, A, X}(m \mid 0, x)}{p_{M \mid A, X, Z}(m \mid a, x, z)} \left( y - \mu_{Y \mid A, M, X, Z}(a, m, x, z) \right) S_{\Delta \mid Y, A, M, X, Z}\left(\delta \mid y, a^{\prime}, m, x,z\right) \, dP(o) = 0,
\end{align*}
then, 
\begin{align*} 
& \int \frac{\iv_a(a^{\prime})}{\pi_a(x)} \frac{p_{M \mid \Delta=1, A,X}(m \mid 0,x)}{p_{M \mid A, X, Z}\left(m \mid a,x, z\right)} \left(y - \mu_{Y \mid A, M, X, Z}(a, m, x, z)\right) S(o) d P(o) \ .
\end{align*}
\par} 

{\footnotesize
The same logic can be applied to evaluate terms (2.2)-(2.4). We have, 

\begin{align*} 
& (2.2) \int yS_{Z \mid A, X}(z \mid a,x) dP_{Y \mid A, M, X, Z}(y \mid a,m,x,z) d P_{Z \mid A, X}(z \mid a,x) d P_{M \mid \Delta=1, A,X}(m \mid 0,x) d P_X(x) \\ 
= & \int \eta_{\mu \mid A,X,Z}(a,x,z)S_{Z \mid A, X}(z \mid a,x) d P_{Z \mid A, X}(z \mid a,x) d P_X(x) \\ 
= & \int \frac{\iv_a(a^{\prime})}{\pi_{a^{\prime}}(x)} \eta_{\mu \mid A,X,Z}\left(a^{\prime}, x,z\right) S_{Z \mid A, X}\left(z \mid a^{\prime}, x\right) d P_{Z \mid A, X}\left(z \mid a^{\prime}, x\right) d P\left(a^{\prime}, x\right) \\ 
= & \int \frac{\iv_a(a^{\prime})}{\pi_{a^{\prime}}(x)}\left(\eta_{\mu \mid A,X,Z}\left(a^{\prime}, x,z\right)-\xi_{a,\eta \mid X}(x)\right) S_{Z \mid A, X}\left(z \mid a^{\prime}, x\right) d P(o) \\ = & \int \frac{\iv_a(a^{\prime})}{\pi_a(x)}\left(\eta_{\mu \mid A,X,Z}(a,x, z)-\xi_{a,\eta \mid X}(x)\right) S(o) d P(o) \ .
\end{align*}
\par} 

{\footnotesize
\begin{align*} 
& (2.3) \int y S_{M \mid \Delta=1,A,X}(m \mid 0,x) d P_{Y \mid A,M,X,Z}(y \mid a,m,x,z) d P_{Z \mid A,X}(z \mid a,x) d P_{M \mid \Delta=1,A,X}(m \mid 0,x) d P_X(x) \\ 
= & \int \eta_{\mu \mid A,M,X}(a,m,x) S_{M \mid \Delta=1,A,X}(m \mid 0,x) d P_{M \mid \Delta=1,A,X}(m \mid 0,x) d P_X(x) \\ 
= & \int \frac{\iv_{0, 1}(a^{\prime}, \Delta)}{\bar{\pi}_0(x)} \eta_{\mu \mid A,M,X}(a,m,x) S_{M \mid \Delta=1,A,X}(m \mid 0,x) d P_{M, \Delta=1, A, X}\left(m, \delta = 1, a^{\prime}=0,x\right) \\ = & \int \frac{\iv_{0, 1}(a^{\prime}, \Delta)}{\bar{\pi}_0(x)}\left(\eta_{\mu \mid A,M,X}(a,m,x)-\xi_{a,\eta \mid X}(x)\right) S_{M \mid \Delta=1,A,X}(m \mid 0,x) d P(o) \\ 
= & \int \frac{\iv_{0, 1}(a^{\prime}, \Delta)}{\bar{\pi}_0(x)}\left(\eta_{\mu \mid A,M,X}(a,m,x)-\xi_{a,\eta \mid X}(x)\right) S(o) d P(o) \ .
\end{align*}
\par} 

{\footnotesize
\begin{align*} 
& (2.4) \int y S_X(x) d P_{Y \mid A,M,X,Z}(y \mid a,m,x,z) d P_{Z \mid A,X}(z \mid a,x) d P_{M \mid \Delta=1,A,X}(m \mid 0,x) d P_X(x) \\ = & \int \xi_{a,\eta \mid X}(x) S_X(x) d P_X(x) \\ 
= & \int \left( \xi_{a,\eta \mid X}(x) - \theta_a \right) S(o) d P(o) \ .
\end{align*}
\par} 

{\footnotesize
Putting the results together, we have:
\begin{align*}
    &\left.\frac{\partial}{\partial \epsilon} \Psi_a\left(P_{\epsilon}\right)\right|_{\epsilon=0} = \int \frac{\iv_a(a^{\prime})}{\pi_a(x)} \frac{p_{M \mid \Delta=1, A,X}(m \mid 0,x)}{p_{M \mid A, X, Z}\left(m \mid a,x, z\right)} \left(y - \mu_{Y \mid A, M, X, Z}(a, m, x, z)\right) S(o) d P(o) \\
    &\hspace{8em}+ \int \frac{\iv_a(a^{\prime})}{\pi_a(x)}\left(\eta_{\mu \mid A,X,Z}(a,x, z)-\xi_{a,\eta \mid X}(x)\right) S(o) d P(o) \\ 
    &\hspace{8em}+ \int \frac{\iv_{0, 1}(a^{\prime}, \Delta)}{\bar{\pi}_0(x)}\left(\eta_{\mu \mid A,M,X}(a,m,x)-\xi_{a,\eta \mid X}(x)\right) S(o) d P(o) \\
    &\hspace{8em}+ \int \left( \xi_{a,\eta \mid X}(x) - \theta_a \right) S(o) d P(o) \ .
\end{align*}
\par} 

Thus, we have expressed the derivative of $\Psi_a$ along a path $P_{\epsilon}$ as an inner product between $S$ and the gradient:
\begin{align*}
    D_{P,a}(O_i) &= \frac{\iv_a(A_i)}{\pi_a(X_i)} \frac{p_{M \mid \Delta = 1, A, X}(M_i \mid 0, X_i)}{p_{M \mid A, X, Z}(M_i \mid A_i, X_i, Z_i)}\left\{Y_i - \mu_{Y \mid A, M, X, Z}(a, M_i, X_i, Z_i)\right\}\\
    &\hspace{2em} + \frac{\iv_a(A_i)}{\pi_a(X_i)} \left\{\eta_{\mu \mid A, Z, X}(a, X_i, Z_i) - \xi_{a, \eta \mid X}(X_i) \right\} \\
    &\hspace{4em}+ \frac{\iv_{0,1}(A_i, \Delta_i)}{\bar{\pi}_{0}(X_i)} \left\{ \eta_{\mu \mid A, M, X}(a, M_i, X_i) - \xi_{a, \eta \mid X}(X_i)\right\} \\
    &\hspace{6em} + \xi_{a, \eta \mid X}(X_i) - \theta_a \ .
\end{align*}
As the tangent space of our model is $L^2_0(P_0)$, there is only a single gradient for $\Psi_a$. Thus, this gradient is by definition the efficient gradient and the efficient influence function for regular asymptotically Normal estimators of $\Psi_a(P_0)$.

\section{Additional details on estimator}

\subsection{Avoiding numerical integration}

To circumvent numerical integration, we make use of a technique proposed by \citet{diaz2021nonparametric} that re-casts estimation problems that involve integrals and densities as an estimation problem that can be solved using mean regression with pseudo-outcomes. This technique can be motivated for our problem by the fact that $\eta_{\mu \mid A, Z, X}(a, z, x)$ in (6) of the main manuscript is equivalent to 
\begin{align*}
&\int \mu_{Y \mid A, M, X, Z}(a, m, x, z) \frac{p_{M \mid \Delta = 1, A, X}(m \mid 0, x)}{p_{M \mid \Delta = 1, A, X, Z}(m \mid a, x, z)}\hs p_{M \mid \Delta = 1, A, X, Z}(m \mid a, x, z) \hs dm \\
&\hspace{1em} = E \left[ \mu_{Y \mid A, M, X, Z}(A, M, X, Z) \frac{p_{M \mid \Delta = 1, A, X}(M \mid 0, X)}{p_{M \mid \Delta = 1, A, X, Z}(M \mid A, X, Z)} \;  \bigm| \; \Delta = 1, A = a, X = x, Z = z\right] \ .
\end{align*}
This equivalence suggests that $\eta_{\mu \mid A, Z, X}$ could be estimated using mean regression, where a pseudo-outcome $\mu_{n,Y \mid A, M, X, Z}(A, M, X, Z) \frac{p_{n, M \mid \Delta = 1, A, X}(M \mid 0, X)}{p_{n, M \mid \Delta = 1, A, X, Z}(M \mid A, X, Z)}$ is regressed onto $A, X,$ and $Z$ using only the observations with $\Delta = 1$. In this way, we avoid  numeric integration and instead utilize super learning-based mean regression. Similar techniques involving pseudo-outcomes can be applied for the estimation of $\eta_{\mu \mid A, M, X}$ and $\xi_{a, \eta \mid X}$, as described below.

\subsection{Detailed description of estimator implementation}
Our estimator can be implemented in the following steps.

\noindent 1. \emph{Estimate mean functional connectivity $\mu_{Y \mid A, M, X, Z}$}. Fit a super learner regression using $Y$ as the outcome and including $A$, $M$, $X$, and $Z$ as predictors. Evaluate the fitted value, $\mu_{n, Y \mid A, M, X, Z}(a, M_i, X_i, Z_i)$ for $i = 1, \dots, n$ and for $a = 0, 1$. For a particular value of $a$, this can be achieved by predicting from the fitted super learner using the observed values of $M, X,$ and $Z$, but replacing the observed value of $A$ with the constant value $a$. Below we refer to this process as \emph{evaluating the fitted value from the regression, setting $A$ to $a$}.

\noindent 2. \emph{Estimate motion distributions $p_{M \mid A, X}$, $p_{M \mid \Delta = 1, A, X}$, $p_{M \mid A, X, Z}$}, and $p_{M \mid \Delta = 1, A, X, Z}$. Estimate densities using the highly adaptive LASSO and evaluate $p_{n, M \mid A, X}(M_i \mid a, X_i)$, $p_{n, M \mid \Delta = 1, A, X}(M_i \mid a, X_i)$, $p_{n, M \mid A, X, Z}(M_i \mid a, X_i, Z_i)$, $p_{n,M \mid \Delta = 1, A, X, Z}(M_i \mid a, X_i, Z_i)$ for $a = 0, 1$ and $i = 1, \dots, n$.

\noindent 3. \emph{Estimate motion-standardized functional connectivity $\eta_{\mu \mid A, Z, X}$}. Using estimates obtained in steps 1 and 2, for $i = 1, \dots, n$ create the pseudo-outcome $\hat{Y}_{M,i} = \mu_{n, Y \mid A, M, X, Z}(A_i, M_i, X_i, Z_i) \times \frac{p_{n, M \mid \Delta = 1, A, X}(M_i \mid 0, X_i)}{p_{n, M \mid \Delta = 1, A, X, Z}(M_i \mid A_i, X_i, Z_i)}.$
Using only observations with $\Delta_i = 1$, fit a super learner regression using $\hat{Y}_{M}$ as the outcome and including $A$, $Z$, and $X$ as predictors. Evaluate the fitted value from this regression setting $A$ to $a$ to obtain $\eta_{n,\mu \mid A, Z, X}(a, Z_i, X_i)$ for $i = 1, \dots, n$.

\noindent 4. \emph{Estimate $Z$-standardized functional connectivity $\eta_{\mu \mid A, M, X}$}. Use estimates obtained in steps 1 and 2, and for $i = 1, \dots, n$ to create the pseudo-outcome $\hat{Y}_{Z,i} = \mu_{n, Y \mid A, M, X, Z}(A_i, M_i, X_i, Z_i) \times \frac{p_{n, M \mid A, X}(M_i \mid A_i, X_i)}{p_{n, M \mid A, X, Z}(M_i \mid A_i, X_i, Z_i)}.$ Fit a super learner regression using $\hat{Y}_{Z}$ as the outcome and including $M$, $X$, and $A$ as predictors. Evaluate the fitted value from this regression setting $A$ to $a$ to obtain $\eta_{n, \mu \mid A, M, X}(a, M_i, X_i)$ for $i = 1,\dots,n$.

\noindent 5. \emph{Estimate motion- and $Z$-standardized functional connectivity $\xi_{a, \eta \mid X}$}. Fit a super learner regression using $\eta_{n, \mu \mid A, Z, X}$ as the outcome and including $A$ and $X$ as predictors in the super learner. For $a = 0, 1$, evaluate the fitted value from this regression setting $A$ to $a$ to obtain $\xi_{n, a, \eta \mid X}(X_i)$ for $i = 1, \dots, n$.

%Alternatively, considering only observations with $\Delta_i = 1$, fit a super learner regression using $\eta_{n, \mu \mid a, X, M}$ as the outcome and include diagnosis category $A$ and demographic covariates $X$ as predictors in the super learner.

\noindent 6. \emph{Calculate plug-in estimate}. Compute the plug-in estimate $\theta_{n, a} = n^{-1} \sum_{i=1}^n \xi_{n, a, \eta \mid X}(X_i)$.

\noindent 7. \emph{Estimate diagnosis distribution $\pi_a$ and inclusion probability $\pi_{\Delta = 1 \mid A, X}$}. Fit a super learner regression using $A$ as the outcome and including $X$ as predictors. Evaluate the fitted value $\pi_{n,1}(X_i)$ for $i = 1,\dots,n$ and set $\pi_{n,0}(X_i) = 1 - \pi_{n,1}(X_i)$. Then fit a super learner using $\Delta$ as the outcome and including $A$ and $X$ as predictors. Evaluate the fitted value from this regression setting $A$ to $0$ to obtain $\pi_{n,\Delta = 1 \mid A, X}(0, X_i)$ for $i=1,\dots, n$. Compute $\bar{\pi}_{n,0}(X_i) = \pi_{n,0}(X_i)\pi_{n,\Delta = 1 \mid A, X}(0, X_i)$ for $i = 1,\dots,n$.

\noindent 8. \emph{Evaluate estimated efficient influence function $D_{n, a}(O_i)$}. For $a = 0, 1$ and each $i = 1, \dots, n$, evaluate $D_{n, a}(O_i)$ by substituting the fitted values based on the estimated nuisance parameters obtained in steps 1-7 into equation (6) of the main manuscript. 

\noindent 9. \emph{Compute the one-step estimator}. For $a = 0, 1$, compute $\theta_{n,a}^+ = \theta_{n,a} + n^{-1} \sum_{i=1}^nD_{n, a}(O_i)$.

\section{Proof of Theorem 2}
\label{multirobust}

\subsection{Part I}
For fixed $a$, we define $D^{*}_{P,a} = D_{P,a} + \theta_a$. $D^{*}_{P,a}$ corresponds to the first three lines and the first term of the fourth line in equation (3.5) of the main manuscript. We rewrite $D^*_{P,a}$ as 

{\footnotesize
\begin{equation*}
\begin{aligned}
    D^{*}_{P,a}(O) &= \frac{\iv_a(A)}{\pi_a(X)} \frac{p_{M \mid \Delta = 1, A, X}(M \mid 0, X)}{p_{M \mid A, X, Z}(M \mid A, X, Z)}\left\{Y - \mu_{Y \mid A, M, X, Z}(a, M, X, Z)\right\}\\
    &\hspace{2em} + \frac{\iv_a(A)}{\pi_a(X)} 
    \eta_{\mu \mid A, Z, X}(a, X, Z) \\
    &\hspace{4em}+ \frac{\iv_{0,1}(A, \Delta)}{\bar{\pi}_{0}(X)} \left\{ \eta_{\mu \mid A, M, X}(a, M, X) - \xi_{a, \eta \mid X}(X)\right\} \\
    &\hspace{6em} + \xi_{a, \eta \mid X}(X) - \frac{\iv_a(A)}{\pi_a(X)} \xi_{a, \eta \mid X}(X) \ . 
    \end{aligned} 
\end{equation*}
\par} 

Consider a probability distribution $P' \in \mathcal{P}$. Next we show that, if any one of the conditions in assumption (B2) holds, then 

{\footnotesize
    \begin{align*}
    E\left[D^{*}_{P', a}(O)\right] - \theta_a &= E\left[\frac{\iv_a(A)}{\pi'_a(X)} \frac{p'_{M \mid \Delta = 1, A, X}(M \mid 0, X)}{p'_{M \mid A, X, Z}(M \mid A, X, Z)}\left\{Y - \mu'_{Y \mid A, M, X, Z}(a, M, X, Z)\right\}\right] \\
    &\hspace{1em} + E \left[ \frac{\iv_a(A)}{\pi'_a(X)} 
    \eta'_{\mu \mid A, Z, X}(a, X, Z) \right]\\
    &\hspace{2em} + E \left[ \frac{\iv_{0,1}(A, \Delta)}{\bar{\pi}'_{0}(X)} \left\{ \eta'_{\mu \mid A, M, X}(a, M, X) - \xi'_{a, \eta \mid X}(X)\right\} \right] \\    
    &\hspace{3em} + E \left[ \xi'_{a, \eta \mid X}(X) - \frac{\iv_a(A)}{\pi'_a(X)} \xi'_{a, \eta \mid X}(X) \right] \\
    &\hspace{4em} - \theta_a \ .
    \end{align*}
\par} We derive the precise expression for each term:
{\footnotesize
    \begin{align*}
    &E\left[\frac{\iv_a(A)}{\pi'_a(X)} \frac{p'_{M \mid \Delta = 1, A, X}(M \mid 0, X)}{p'_{M \mid A, X, Z}(M \mid A, X, Z)}\left\{Y - \mu'_{Y \mid A, M, X, Z}(a, M, X, Z)\right\}\right] \\ 
    &=\int \frac{\iv_a(a^{*})}{\pi'_{a}(X)} \frac{p'_{M \mid \Delta = 1, A, X}(m \mid 0, x)}{p'_{M \mid A, X, Z}(m \mid a^{*}, x, z)}\left\{y - \mu'_{Y \mid A, M, X, Z}(a, m, x, z)\right\}p_{A, M, X, Z, Y}(a^{*}, m, x, z, y) da^{*} dm dz dx dy\\
    &=\int \frac{\iv_a(a^{*})}{\pi'_{a}(X)} \frac{p'_{M \mid \Delta = 1, A, X}(m \mid 0, x)}{p'_{M \mid A, X, Z}(m \mid a, x, z)}\left\{ \int y p_{Y \mid A, M, X, Z}(y \mid a, m, x, z) dy - \mu'_{Y \mid A, M, X, Z}(a, m, x, z)\right\} \\
    & \hspace{2em} p_{A, M, X, Z}(a, m, x, z) da^{*} dm dz dx\\
    &=\int \frac{\int \iv_a(a^{*}) \pi_{a}(x) da^{*}}{\pi'_{a}(X)} \frac{p'_{M \mid \Delta = 1, A, X}(m \mid 0, x)}{p'_{M \mid A, X, Z}(m \mid a, x, z)}\left\{\mu_{Y \mid A, M, X, Z}(a, m, x, z)- \mu'_{Y \mid A, M, X, Z}(a, m, x, z)\right\} \\
    & \hspace{2em} p_{M \mid A, X, Z}(m \mid a, x, z) p_{Z \mid A, X}(z \mid a, x) p_X(x) dm dz dx\\  
    &= \int \frac{\pi_a(x)}{\pi'_a(x)}\frac{p'_{M \mid \Delta = 1, A, X}(m \mid 0, x)}{p'_{M \mid A, X, Z}(m \mid a, x, z)}\left\{\mu_{Y \mid A, M, X, Z}(a, m, x, z) - \mu'_{Y \mid A, M, X, Z}(a, m, x, z)\right\} \\
    & \hspace{2em} p_{M \mid A, X, Z}(m \mid a, x, z)p_{Z \mid A, X }(z \mid a, x) p_X(x) dm dz dx \ . \tag{3.1}
    \end{align*}
\par}
 
{\footnotesize
    \begin{align*}
    &E \left[ \frac{\iv_a(A)}{\pi'_a(X)} 
    \eta'_{\mu \mid A, Z, X}(a, X, Z) \right] \\ 
    &=\int \frac{\iv_a(a^{*})}{\pi'_{a}(X)} \eta'_{\mu \mid A, Z, X}(a, x, z) p_{A, X, Z}(a^{*}, x, z) da^{*} dz dx \\
    &=\int \frac{\iv_a(a^{*})}{\pi'_{a}(X)} \eta'_{\mu \mid A, Z, X}(a, x, z) p_{Z \mid A, X}(a, x) \pi_a(x) p_X(x) da^{*} dz dx \\
    &=\int \frac{\int \iv_a(a^{*}) \pi_{a}(x) da^{*}}{\pi'_{a}(X)} \eta'_{\mu \mid A, Z, X}(a, x, z) p_{Z \mid A, X}(a, x)  p_X(x) dz dx\\
    &=\int \frac{\pi_a(x)}{\pi'_a(x)} \eta'_{\mu \mid A, Z, X}(a, x, z) p_{Z \mid A, X}(a, x)  p_X(x) dz dx \ .\tag{3.2}
    \end{align*}
\par}

{\footnotesize
    \begin{align*}
    &E \left[ \frac{\iv_{0,1}(A, \Delta)}{\bar{\pi}'_{0}(X)} \left\{ \eta'_{\mu \mid A, M, X}(a, M, X) - \xi'_{a, \eta \mid X}(X)\right\} \right] \\ 
    &=\int  \frac{\iv_{0,1}(a^{*}, \Delta)}{\bar{\pi}'_{0}(X)} \eta'_{\mu \mid A, M, X}(a, m, x) p_{A, \Delta, M, X}(a^{*}, \Delta, m, x) da^{*} d\Delta dm dx \\
    & \hspace{2em} -\int  \frac{\iv_{0,1}(a^{*}, \Delta)}{\bar{\pi}'_{0}(X)} \xi'_{a, \eta \mid X}(x) p_{A, \Delta, X}(a^{*}, \Delta, x) da^{*} d\Delta dx \\
    &= \int  \frac{\iv_{0,1}(a^{*}, \Delta)}{\bar{\pi}'_{0}(X)} \eta'_{\mu \mid A, M, X}(a, m, x) p_{M \mid \Delta=1, A, X}(m \mid 0, x) \bar{\pi}_{0}(x) p_X(x) da^{*} d\Delta dm dx \\
    & \hspace{2em} -\int  \frac{\iv_{0,1}(a^{*}, \Delta)}{\bar{\pi}'_{0}(X)} \xi'_{a, \eta \mid X}(x) \bar{\pi}_{0}(x) p_X(x) da^{*} d\Delta dx \\
    &=\int \frac{\int \iv_{0,1}(a^{*}, \Delta) \bar{\pi}_{0}(x) da^{*}d\Delta}{\bar{\pi}'_{0}(X)} \eta'_{\mu \mid A, M, X}(a, m, x) p_{M \mid \Delta=1, A, X}(m \mid 0, x) p_X(x) dm dx\\
    & \hspace{2em} -\int  \frac{\int \iv_{0,1}(a^{*}, \Delta) \bar{\pi}_{0}(x) da^{*}d\Delta}{\bar{\pi}'_{0}(X)} \xi'_{a, \eta \mid X}(x) p_X(x)  dx \\
    &=\int \frac{\bar{\pi}_{0}(x)}{\bar{\pi}'_{0}(x)} \eta'_{\mu \mid A, M, X}(a, m, x)p_{M \mid \Delta = 1, A, X}(m \mid 0, x) p_X(x) dm dx \\
    & \hspace{2em} -\int \frac{\bar{\pi}_{0}(x)}{\bar{\pi}'_{0}(x)} \eta'_{\mu \mid A, M, X}(a, m, x) p'_{M \mid \Delta = 1, A, X}(m \mid 0, x) p_X(x)  dmdx \\
    &=\int \frac{\bar{\pi}_{0}(x)}{\bar{\pi}'_{0}(x)} \eta'_{\mu \mid A, M, X}(a, m, x)\left\{ p_{M \mid \Delta = 1, A, X}(m \mid 0, x) - p'_{M \mid \Delta = 1, A, X}(m \mid 0, x)\right\}p_X(x) dm dx  \ . \tag{3.3}
    \end{align*}
\par}

{\footnotesize
    \begin{align*}
    &E \left[ \xi'_{a, \eta \mid X}(X) - \frac{\iv_a(A)}{\pi'_a(X)} \xi'_{a, \eta \mid X}(X) \right] \\ 
    &=\int \xi'_{a, \eta \mid X}(x) p_X(x) dx - \int \frac{\iv_a(a^{*})}{\pi'_{a}(X)} \xi'_{a, \eta \mid X}(x) p_{A, X}(a^{*}, x) da^{*} dx  \\
    &=\int \xi'_{a, \eta \mid X}(x) p_X(x) dx - \int \frac{\pi_a(x)}{\pi'_a(x)} \xi'_{a, \eta \mid X}(x) p_X(x) dx \\
    &=\int \xi'_{a, \eta \mid X}(x)\left\{ 1 - \frac{\pi_a(x)}{\pi'_a(x)} \right\} p_X(x) dx \ . \tag{3.4} \\
    \end{align*}
\par}

{\footnotesize
    \begin{align*}
    -\theta_a = -\int \xi_{a, \eta \mid X}(x) p_X(x) dx  \ . \tag{3.5} \\
    \end{align*}
\par}

Therefore, 
{\footnotesize
    \begin{align*}
    E[D^{*}_{P', a}(O)] - \theta_a  = (3.1) + (3.2) + (3.3) + (3.4) + (3.5)  \ .
    \end{align*}
\par}

% {\footnotesize
%     \begin{align*}
%     &E[D^{*}_{P', a}(O)] - \theta_a  =\\ &\int \frac{\pi_a(x)}{\pi'_a(x)}\frac{p'_{M \mid \Delta = 1, A, X}(m \mid 0, x)}{p'_{M \mid A, X, Z}(m \mid a, x, z)}\left\{\mu_{Y \mid A, M, X, Z}(a, m, x, z) - \mu'_{Y \mid A, M, X, Z}(a, m, x, z)\right\}p_{M \mid A, X, Z}(m \mid a, x, z)p_{Z \mid A, X }(z \mid a, x) p_X(x) dm dz dx \tag{1}\\
%     &+ \int \frac{\pi_a(x)}{\pi'_a(x)} \eta'_{\mu \mid A, X, Z}(a, x, z)p_{Z \mid A, X }(z \mid a, x) p_X(x) dz dx \tag{2}\\
%     &+ \int \frac{\bar{\pi}_{0}(x)}{\bar{\pi}'_{0}(x)} \eta'_{\mu \mid A, M, X}(a, m, x)\{p_{M \mid \Delta = 1, A, X}(m \mid 0, x) - p'_{M \mid \Delta = 1, A, X}(m \mid 0, x)\}p_X(x) dm dx \tag{3}\\    
%     &+ \int \xi'_{a, \eta \mid X}(x)\left\{ 1 - \frac{\pi_a(x)}{\pi'_a(x)} \right\} p_X(x) dx \tag{4} \\
%     &-\int \xi_{a, \eta \mid X}(x) p_X(x) dx . \tag{5}
%     \end{align*}
% \par}

First, note that 

{\footnotesize
    \begin{align*}
    (3.4) + (3.5) &= 
    \int \xi'_{a, \eta \mid X}(x)\left\{1-\frac{\pi_a(x)}{\pi'_a(x)}\right\} p_X(x) dx -\int \xi_{a, \eta \mid X}(x)p_X(x) dx \\
    &= \int \left\{\xi'_{a, \eta \mid X}(x) - \xi_{a, \eta \mid X}(x)\right\} \left\{1-\frac{\pi_a(x)}{\pi'_a(x)}\right\} p_X(x) dx \tag{3.6}\\
    &\text{ }-\int \xi_{a, \eta \mid X}(x)\frac{\pi_a(x)}{\pi'_a(x)}p_X(x) dx  \ .
    \end{align*}
\par}

Using the definition of $\xi_{a, \eta \mid X}(x)$ in the main manuscript, we obtain

{\footnotesize
    \begin{align*}
    \int &\xi_{a, \eta \mid X}(x)\frac{\pi_a(x)}{\pi'_a(x)}p_X(x) dx =\\ & 
    \int \frac{\pi_a(x)}{\pi'_a(x)} \frac{p_{M \mid \Delta = 1, A, X}(m \mid 0, x)}{p_{M \mid A, X, Z}(m \mid a, x, z)}\left\{\mu_{Y \mid A, M, X, Z}(a, m, x, z) - \mu'_{Y \mid A, M, X, Z}(a, m, x, z)\right\} p_{M \mid A, X, Z}(m \mid a, x, z) p_{Z \mid A, X }(z \mid a, x) p_X(x) dmdzdx \\
    &+ \int \frac{\pi_a(x)}{\pi'_a(x)}p_{M \mid \Delta = 1, A, X}(m \mid 0, x)\mu'_{Y \mid A, M, X, Z}(a, m, x, z)p_{Z \mid A, X }(z \mid a, x) p_X(x) dmdzdx \ .\tag{3.7}
    \end{align*}
\par}

Thus 

{\footnotesize
	\begin{align*}
    (3.1) + (3.4) + (3.5) = &(3.6) - (3.7) \\
    &+ \int \frac{\pi_a(x)}{\pi'_a(x)} \left\{\frac{p'_{M \mid \Delta = 1, A, X}(m \mid 0, x)}{p'_{M \mid A, X, Z}(m \mid a, x, z)} - \frac{p_{M \mid \Delta = 1, A, X}(m \mid 0, x)}{p_{M \mid A, X, Z}(m \mid a, x, z)}\right\} \times \\ &\left\{\mu_{Y \mid A, M, X, Z}(a, m, x, z) - \mu'_{Y \mid A, M, X, Z}(a, m, x, z)\right\} p_{M \mid A, X, Z}(m \mid a, x, z) p_{Z \mid A, X }(z \mid a, x) p_X(x) dmdzdx \ . \tag{3.8}
    \end{align*}
\par}

We also have 

{\footnotesize
	\begin{align*}
    (3.2) - (3.7) = \int \frac{\pi_a(x)}{\pi'_a(x)}\left\{p'_{M \mid \Delta = 1, A, X}(m \mid 0, x) - p_{M \mid \Delta = 1, A, X}(m \mid 0, x)\right\}\mu'_{Y \mid A, M, X, Z}(a, m, x, z)p_{Z \mid A, X }(z \mid a, x) p_X(x) dmdzdx  \ ,
    \end{align*}
\par}
which yields

{\footnotesize
    \begin{align*}
    (3.2) &+ (3.3) - (3.7) \\ 
    = &\int \frac{\bar{\pi}_{0}(x)}{\bar{\pi}'_{0}(x)} \eta'_{\mu \mid A, M, X}(a, m, x)\left (p_{M \mid \Delta = 1, A, X}(m \mid 0, x) - p'_{M \mid \Delta = 1, A, X}(m \mid 0, x)\right )p_X(x) dmdx  \\ 
    &+  \int \frac{\pi_a(x)}{\pi'_a(x)}\left\{p'_{M \mid \Delta = 1, A, X}(m \mid 0, x) - p_{M \mid \Delta = 1, A, X}(m \mid 0, x)\right\}\mu_{Y \mid A, M, X, Z}(a, m, x, z)p_{Z \mid A, X }(z \mid a, x) p_X(x) dmdzdx \\
    &+  \int \frac{\pi_a(x)}{\pi'_a(x)}\left\{p'_{M \mid \Delta = 1, A, X}(m \mid 0, x) - p_{M \mid \Delta = 1, A, X}(m \mid 0, x)\right\}\left\{\mu'_{Y \mid A, M, X, Z}(a, m, x, z)-\mu_{Y \mid A, M, X, Z}(a, m, x, z)\right\}\\&p_{Z \mid A, X }(z \mid a, x) p_X(x) dmdzdx \\
    = &-\int \frac{\bar{\pi}_{0}(x)}{\bar{\pi}'_{0}(x)} \eta'_{\mu \mid A, M, X}(a, m, x)\left \{p'_{M \mid \Delta = 1, A, X}(m \mid 0, x)-p_{M \mid \Delta = 1, A, X}(m \mid 0, x)\right \}p_X(x) dmdx \\
    &+ \int \frac{\pi_a(x)}{\pi'_a(x)}\eta_{\mu \mid A, M, X}(a, m, x)\left\{p'_{M \mid \Delta = 1, A, X}(m \mid 0, x) - p_{M \mid \Delta = 1, A, X}(m \mid 0, x)\right\}p_X(x) dmdx\\
    &+  \int \frac{\pi_a(x)}{\pi'_a(x)}\left\{p'_{M \mid \Delta = 1, A, X}(m \mid 0, x) - p_{M \mid \Delta = 1, A, X}(m \mid 0, x)\right\}\{\mu'_{Y \mid A, M, X, Z}(a, m, x, z)-\mu_{Y \mid A, M, X, Z}(a, m, x, z)\}\\&p_{Z \mid A, X }(z \mid a, x) p_X(x) dmdzdx \\
    = &\int \left\{\frac{\pi_a(x)}{\pi'_a(x)}\eta_{\mu \mid A, M, X}(a, m, x)-\frac{\bar{\pi}_{0}(x)}{\bar{\pi}'_{0}(x)} \eta'_{\mu \mid A, M, X}(a, m, x)\right\}\left\{p'_{M \mid \Delta = 1, A, X}(m \mid 0, x)-p_{M \mid \Delta = 1, A, X}(m \mid 0, x)\right\}p_X(x) dmdx  \tag{3.9}\\
    &+ \int \frac{\pi_a(x)}{\pi'_a(x)}\left\{p'_{M \mid \Delta = 1, A, X}(m \mid 0, x) - p_{M \mid \Delta = 1, A, X}(m \mid 0, x)\right\}\left\{\mu'_{Y \mid A, M, X, Z}(a, m, x, z)-\mu_{Y \mid A, M, X, Z}(a, m, x, z)\right\}\\&p_{Z \mid A, X }(z \mid a, x) p_X(x) dmdzdx  \ . \tag{3.10}
    \end{align*}
\par}

Putting everything together yields

{\footnotesize
	\begin{align*}
    (3.1) + (3.2) + (3.3) + (3.4) + (3.5) = (3.6) + (3.8) + (3.9) + (3.10) \ .
    \end{align*}
\par}

Thus,

{\footnotesize
\begin{align*}
    &E\left[D^{*}_{P', a}(O)\right] - \theta_a \\ 
    =&\int \left\{\xi'_{a, \eta \mid X}(x) - \xi_{a, \eta \mid X}(x)\right\}\left(1-\frac{\pi_a(x)}{\pi'_a(x)}\right)p_X(x) dx\\
    &+ \int \frac{\pi_a(x)}{\pi'_a(x)} \left\{\frac{p'_{M \mid \Delta = 1, A, X}(m \mid 0, x)}{p'_{M \mid A, X, Z}(m \mid a, x, z)} - \frac{p_{M \mid \Delta = 1, A, X}(m \mid 0, x)}{p_{M \mid A, X, Z}(m \mid a, x, z)}\right\} \times \\ &\left\{\mu_{Y \mid A, M, X, Z}(a, m, x, z) - \mu'_{Y \mid A, M, X, Z}(a, m, x, z)\right\} p_{M \mid A, X, Z}(m \mid a, x, z) p_{Z \mid A, X }(z \mid a, x) p_X(x) dmdzdx \\
    &+\int \left\{\frac{\pi_a(x)}{\pi'_a(x)}\eta_{\mu \mid A, M, X}(a, m, x)-\frac{\bar{\pi}_{0}(x)}{\bar{\pi}'_{0}(x)} \eta'_{\mu \mid A, M, X}(a, m, x)\right\}\left\{p'_{M \mid \Delta = 1, A, X}(m \mid 0, x)-p_{M \mid \Delta = 1, A, X}(m \mid 0, x)\right\}p_X(x) dmdx \\
    &+ \int \frac{\pi_a(x)}{\pi'_a(x)}\left\{p'_{M \mid \Delta = 1, A, X}(m \mid 0, x) - p_{M \mid \Delta = 1, A, X}(m \mid 0, x)\right\}\left\{\mu'_{Y \mid A, M, X, Z}(a, m, x, z)-\mu_{Y \mid A, M, X, Z}(a, m, x, z)\right\}\\&p_{Z \mid A, X }(z \mid a, x) p_X(x) dmdzdx \ .
    \end{align*}
\par}

\subsection{Part II}

Let \(P_n\) denote the empirical measure of $O_1, \dots, O_n$. Let $P_n^{'}$ be any estimator of $P_0$ compatible with the nuisance models used in the estimation of $D_{P,a}$, we have:
\begin{equation*}
    \begin{aligned}
        \Psi_a(P_n^{'}) - \Psi_a(P) &= (P_n^{'} - P)D_{P_n^{'}, a} + R_2(P, P_n^{'}) \\
        &= -PD_{P_n^{'}, a} + R_2(P, P_n^{'}) \\
        &= -P_n D_{P_n^{'}, a} + (P_n - P)D_{P,a} + (P_n - P)(D_{P_n^{'}, a} - D_{P,a}) + R_2(P, P_n^{'}) \ ,
    \end{aligned}
\end{equation*}
where $R_2(P, P_n^{'})$ is the second-order remainder term. As the one-step estimator is defined as $\Psi_a(P_n^{'}) + P_n D_{P_n^{'}, a}$, we have:
\begin{equation*}
    \theta_{n,a}^{+} - \theta_a = (P_n - P)D_{P,a} + (P_n - P)(D_{P_n^{'}, a} - D_{P,a}) + R_2(P, P_n^{'}) \ .
\end{equation*}
Next, we need to show $(P_n - P)(D_{P_n^{'}, a} - D_{P,a}) = o_P(1)$ and $R_2(P, P_n^{'}) = o_P(1)$. The former will hold under conditions (B3) and (B4) of the theorem. 

As for the second order reminder term, we have:

\begin{equation*}
    R_2(P, P_n^{'}) = \Psi_a(P_n^{'}) + PD_{P_n^{'}, a} - \Psi_a(P) = E\left[D^{*}_{P_n^{'}, a}(O)\right] - \theta_a \ .
\end{equation*} Referring back to Part I, we have
{\footnotesize
\begin{align*}
    &R_2(P, P_n^{'})  \\ 
    =&\int \left\{\xi_{n, a, \eta \mid X}(x) - \xi_{a, \eta \mid X}(x)\right\}\left(1-\frac{\pi_a(x)}{\pi_{n,a}(x)}\right)p_X(x) dx\\
    &+ \int \frac{\pi_a(x)}{\pi_{n,a}(x)} \left\{\frac{p_{n, M \mid \Delta = 1, A, X}(m \mid 0, x)}{p_{n, M \mid A, X, Z}(m \mid a, x, z)} - \frac{p_{M \mid \Delta = 1, A, X}(m \mid 0, x)}{p_{M \mid A, X, Z}(m \mid a, x, z)}\right\} \times \\ &\left\{\mu_{Y \mid A, M, X, Z}(a, m, x, z) - \mu_{n, Y \mid A, M, X, Z}(a, m, x, z)\right\} p_{M \mid A, X, Z}(m \mid a, x, z) p_{Z \mid A, X }(z \mid a, x) p_X(x) dmdzdx \\
    &+\int \left\{\frac{\pi_a(x)}{\pi_{n,a}(x)}\eta_{\mu \mid A, M, X}(a, m, x)-\frac{\bar{\pi}_{0}(x)}{\bar{\pi}_{n,0}(x)} \eta_{n, \mu \mid A, M, X}(a, m, x)\right\}\left\{p_{n, M \mid \Delta = 1, A, X}(m \mid 0, x)-p_{M \mid \Delta = 1, A, X}(m \mid 0, x)\right\}p_X(x) dmdx \\
    &+ \int \frac{\pi_a(x)}{\pi_{n, a}(x)}\left\{p_{n, M \mid \Delta = 1, A, X}(m \mid 0, x) - p_{M \mid \Delta = 1, A, X}(m \mid 0, x)\right\}\left\{\mu_{n, Y \mid A, M, X, Z}(a, m, x, z)-\mu_{Y \mid A, M, X, Z}(a, m, x, z)\right\}\\&p_{Z \mid A, X }(z \mid a, x) p_X(x) dmdzdx \ .
    \end{align*}
\par}

Using Cauchy-Schwartz inequality and assumptions (B1) and (B2) of the theorem, we can show $R_2(P, P_n^{'})=o_P(1)$. To be more specific, for the first line of the term,
{\footnotesize
\begin{equation*}
    \begin{aligned}
        &\int \left\{\xi_{n, a, \eta \mid X}(x) - \xi_{a, \eta \mid X}(x)\right\}\left(1-\frac{\pi_a(x)}{\pi_{n,a}(x)}\right)p_X(x) \, dx \\
        & \leq  \int \left| \left\{\xi_{n, a, \eta \mid X}(x) - \xi_{a, \eta \mid X}(x)\right\}\left(\frac{\pi_{n,a}(x) - \pi_a(x)}{\pi_{n,a}(x)}\right)p_X(x) \right| \, dx \\
        & \leq \left\{ \sup_{x}\frac{1}{\pi_{n,a}(x)} \right\} \int \left| \{\xi_{n, a, \eta \mid X}(x) - \xi_{a, \eta \mid X}(x)\} \left\{\pi_{n,a}(x) - \pi_a(x)\right\} \right|p_X(x) \, dx \ .
    \end{aligned}
\end{equation*}
\par}
Then applying assumption (B1) to the supremum and Cauchy-Schwarz to the integration, we have 
\begin{align*}
    & \leq \frac{1}{\epsilon_1}  \lVert \xi_{n, \eta \mid a, X} - \xi_{a, \eta \mid X} \rVert \lVert \pi_{n, a} - \pi_{a} \rVert \\
    &= o_P(1) \ ,
\end{align*}
where the last line follows from assumption (B2). The same reasoning can be extended to the other terms in $R_2(P, P_n^{'})$.

For the second line of the term,

{\footnotesize
\begin{equation*}
    \begin{aligned}
        &\int \frac{\pi_a(x)}{\pi_{n,a}(x)} \left\{\frac{p_{n, M \mid \Delta = 1, A, X}(m \mid 0, x)}{p_{n, M \mid A, X, Z}(m \mid a, x, z)} - \frac{p_{M \mid \Delta = 1, A, X}(m \mid 0, x)}{p_{M \mid A, X, Z}(m \mid a, x, z)}\right\} \times \\ &\left\{\mu_{Y \mid A, M, X, Z}(a, m, x, z) - \mu_{n, Y \mid A, M, X, Z}(a, m, x, z)\right\} p_{M \mid A, X, Z}(m \mid a, x, z) p_{Z \mid A, X }(z \mid a, x) p_X(x) dmdzdx \\
        & = \int \frac{\pi_a(x)}{\pi_{n,a}(x)} \left\{\frac{p_{n, M \mid \Delta = 1, A, X}(m \mid 0, x) - p_{M \mid \Delta = 1, A, X}(m \mid 0, x)}{p_{n, M \mid A, X, Z}(m \mid a, x, z)} \right\} \times \\ &\left\{\mu_{Y \mid A, M, X, Z}(a, m, x, z) - \mu_{n, Y \mid A, M, X, Z}(a, m, x, z)\right\} p_{M \mid A, X, Z}(m \mid a, x, z) p_{Z \mid A, X }(z \mid a, x) p_X(x) dmdzdx \\
        & + \int \frac{\pi_a(x)}{\pi_{n,a}(x)} \left\{\frac{p_{M \mid A, X, Z}(m \mid a, x, z) - p_{n, M \mid A, X, Z}(m \mid a, x, z)}{p_{n, M \mid A, X, Z}(m \mid a, x, z)} \right\} \times \\ &\left\{\mu_{Y \mid A, M, X, Z}(a, m, x, z) - \mu_{n, Y \mid A, M, X, Z}(a, m, x, z)\right\} p_{M \mid \Delta = 1, A, X}(m \mid 0, x) p_{Z \mid A, X }(z \mid a, x) p_X(x) dmdzdx \\
        & \leq \Bigl\{ \sup_{m, a, x, z}\frac{1}{\pi_{n,a}(x)p_{n, M \mid A, X, Z}(m \mid a, x, z)} \Bigr\} \int \Bigl| \Bigl \{p_{n, M \mid \Delta = 1, A, X}(m \mid 0, x) - p_{M \mid \Delta = 1, A, X}(m \mid 0, x)\Bigr \} \times \\ 
        &\hspace{2em} \Bigl\{\mu_{Y \mid A, M, X, Z}(a, m, x, z) - \mu_{n, Y \mid A, M, X, Z}(a, m, x, z)\Bigr\} \Bigr| p_{M \mid A, X, Z}(m \mid a, x, z) p_{Z \mid A, X }(z \mid a, x) p_X(x) \, dx
         \\
        &+ \Bigl\{ \sup_{m, a, x, z}\frac{1}{\pi_{n,a}(x)p_{n, M \mid A, X, Z}(m \mid a, x, z)} \Bigr\} \int \Bigl| \Bigl \{p_{M \mid A, X, Z}(m \mid a, x, z) - p_{n, M \mid A, X, Z}(m \mid a, x, z)\Bigr \} \times \\
        &\hspace{2em} \Bigl\{\mu_{Y \mid A, M, X, Z}(a, m, x, z) - \mu_{n, Y \mid A, M, X, Z}(a, m, x, z)\Bigr\} \Bigr| p_{M \mid \Delta = 1, A, X}(m \mid 0, x) p_{Z \mid A, X }(z \mid a, x) p_X(x) \, dx \\ 
        & \leq  \frac{1}{\epsilon_1 \epsilon_3}  \lVert p_{n, M \mid \Delta = 1, A, X}(m \mid 0, x) - p_{M \mid \Delta = 1, A, X}(m \mid 0, x) \rVert \lVert \mu_{Y \mid A, M, X, Z}(a, m, x, z) - \mu_{n, Y \mid A, M, X, Z}(a, m, x, z) \rVert \\
        & + \frac{1}{\epsilon_1 \epsilon_3}  \lVert p_{M \mid A, X, Z}(m \mid a, x, z) - p_{n, M \mid A, X, Z}(m \mid a, x, z) \rVert \lVert \mu_{Y \mid A, M, X, Z}(a, m, x, z) - \mu_{n, Y \mid A, M, X, Z}(a, m, x, z) \rVert \\
        & = o_p(1) \ .
    \end{aligned}
\end{equation*}
\par}

For the third line of the term,
{\footnotesize
\begin{equation*}
    \begin{aligned}
    &\int \left\{\frac{\pi_a(x)}{\pi_{n,a}(x)}\eta_{\mu \mid A, M, X}(a, m, x)-\frac{\bar{\pi}_{0}(x)}{\bar{\pi}_{n,0}(x)} \eta_{n, \mu \mid A, M, X}(a, m, x)\right\}\left\{p_{n, M \mid \Delta = 1, A, X}(m \mid 0, x)-p_{M \mid \Delta = 1, A, X}(m \mid 0, x)\right\}p_X(x) dmdx  \\
    &= \int \left\{\frac{\pi_a(x) - \pi_{n,a}(x)}{\pi_{n,a}(x)}\eta_{\mu \mid A, M, X}(a, m, x) + \frac{\bar{\pi}_{n,0}(x) - \bar{\pi}_{0}(x)}{\bar{\pi}_{n,0}(x)} \eta_{\mu \mid A, M, X}(a, m, x) +  \frac{\bar{\pi}_{0}(x)}{\bar{\pi}_{n,0}(x)} \left\{ \eta_{\mu \mid A, M, X}(a, m, x) - \eta_{n, \mu \mid A, M, X}(a, m, x) \right\} \right\} \times \\ 
    &\hspace{2em} \left\{p_{n, M \mid \Delta = 1, A, X}(m \mid 0, x)-p_{M \mid \Delta = 1, A, X}(m \mid 0, x)\right\}p_X(x) dmdx  \\
    & \leq \left\{ \sup_{x}\frac{1}{\pi_{n,a}(x)} \right\} \int \left| \left\{ \pi_a(x) - \pi_{n,a}(x) \right\} \left\{p_{n, M \mid \Delta = 1, A, X}(m \mid 0, x)-p_{M \mid \Delta = 1, A, X}(m \mid 0, x)  \right\} \right| \eta_{\mu \mid A, M, X}(a, m, x) p_X(x) dmdx \\
    &+ \left\{ \sup_{x}\frac{1}{\bar{\pi}_{n,0}(x)} \right\} \int \left| \left\{ \bar{\pi}_{n,0}(x) - \bar{\pi}_{0}(x) \right\} \left\{p_{n, M \mid \Delta = 1, A, X}(m \mid 0, x)-p_{M \mid \Delta = 1, A, X}(m \mid 0, x)\right\} \right| \eta_{\mu \mid A, M, X}(a, m, x) p_X(x) dmdx \\
    &+ \left\{ \sup_{x}\frac{1}{\bar{\pi}_{n,0}(x)} \right\} \int \left| \left\{ \eta_{\mu \mid A, M, X}(a, m, x) - \eta_{n, \mu \mid A, M, X}(a, m, x) \right\} \left\{p_{n, M \mid \Delta = 1, A, X}(m \mid 0, x)-p_{M \mid \Delta = 1, A, X}(m \mid 0, x)\right\} \right| \bar{\pi}_{0}(x) p_X(x) dmdx \\
    & \leq \frac{1}{\epsilon_1} \lVert \pi_a(x) - \pi_{n,a}(x) \rVert \lVert p_{n, M \mid \Delta = 1, A, X}(m \mid 0, x)-p_{M \mid \Delta = 1, A, X}(m \mid 0, x) \rVert \\
    &+ \frac{1}{\epsilon_2} \lVert \bar{\pi}_{n, 0}(x) - \bar{\pi}_{0}(x) \rVert \lVert p_{n, M \mid \Delta = 1, A, X}(m \mid 0, x)-p_{M \mid \Delta = 1, A, X}(m \mid 0, x) \rVert \\
    &+ \frac{1}{\epsilon_2} \lVert \eta_{\mu \mid A, M, X}(a, m, x) - \eta_{n, \mu \mid A, M, X}(a, m, x) \rVert \lVert p_{n, M \mid \Delta = 1, A, X}(m \mid 0, x)-p_{M \mid \Delta = 1, A, X}(m \mid 0, x) \rVert \\
    & = o_p(1) \ .
    \end{aligned}
\end{equation*}
\par}

For the last line of the term,
{\footnotesize
\begin{equation*}
    \begin{aligned}
    &\int \frac{\pi_a(x)}{\pi_{n, a}(x)}\left\{p_{n, M \mid \Delta = 1, A, X}(m \mid 0, x) - p_{M \mid \Delta = 1, A, X}(m \mid 0, x)\right\}\left\{\mu_{n, Y \mid A, M, X, Z}(a, m, x, z)-\mu_{Y \mid A, M, X, Z}(a, m, x, z)\right\} \\
    & \hspace{2em} p_{Z \mid A, X }(z \mid a, x) p_X(x) \, dmdzdx   \\
    & \leq \left\{ \sup_{x}\frac{1}{\pi_{n,a}(x)} \right\} \int \Bigl| \Bigl \{ p_{n, M \mid \Delta = 1, A, X}(m \mid 0, x) - p_{M \mid \Delta = 1, A, X}(m \mid 0, x) \Bigr \} \times \\
    & \hspace{4em} \Bigl \{ \mu_{n, Y \mid A, M, X, Z}(a, m, x, z)-\mu_{Y \mid A, M, X, Z}(a, m, x, z)\Bigr \} \Bigr| p_{Z \mid A, X }(z \mid a, x) \pi_a(x) p_X(x) \, dmdx \\
    & \leq \frac{1}{\epsilon_1} \lVert p_{n, M \mid \Delta = 1, A, X}(m \mid 0, x) - p_{M \mid \Delta = 1, A, X}(m \mid 0, x) \rVert \lVert \mu_{n, Y \mid A, M, X, Z}(a, m, x, z)-\mu_{Y \mid A, M, X, Z}(a, m, x, z) \rVert\\
    &= o_p(1) \ .
    \end{aligned}
\end{equation*}
\par}

Thus $R_2(P, P_n^{'}) = o_p(1)$, thereby concluding the proof.

\section{Cross-fit one-step estimation} \label{sec:cross_fitting}

Cross-fitting avoids the necessity of (C2) in Theorem 3, which may afford the ability to utilize more aggressive machine learning techniques as part of the super learner, while maintaining well-calibrated inference in finite samples \citep{zivich2021machine}. The cross-fitting process involves randomly dividing the data set into $K$ parts, followed by separate cross-validation routines on each part. $K-1$ parts of the data are used to estimate the nuisance parameters appearing in the efficient influence function using super learner with $K'$-fold cross-validation. In practice, we use $K=5$ and $K'=10$ for all nuisance regressions and conditional density estimates. Consider the example of $\xi_{a, \eta \mid X}$. We denote by $\xi_{n, k, a, \eta \mid X}$ the estimate of $\xi_{a, \eta \mid X}$ obtained when the $k$-th part of the data is withheld from the nuisance estimation stage. Similarly, we denote by $D_{a, n, k}$ the efficient influence function evaluated at the nuisance parameters estimated without using the $k$-th part of the data. Denote by $\mathcal{I}_k$ the indices of observations in the $k$-th part of the data and denote the number of observations in this set by $n_k$. The cross-fit estimate of $\theta_a$ is $\theta_{n, a}^{\text{cf}} = \frac{1}{K} \sum_{k=1}^K \left[ \frac{1}{n_k} \sum_{i \in \mathcal{I}_k} \xi_{n, k, a, \eta \mid X}(X_{i}) + \frac{1}{n_k} \sum_{i \in \mathcal{I}_k} D_{a, n, k}(O_i) \right]$. 
The asymptotic linearity of the cross-fit one-step estimator follows using the same arguments as in Theorem 3, where nuisance estimates are replaced by their $k$-specific counterparts and assumption (C2) is removed \citep{van2011cross, chernozhukov2018double}.

\section{Theorem 3}

\subsection{Assumptions of theorem}

Assumption (B1) guarantees that estimated propensities and motion densities are appropriately bounded so that the one-step estimator is never ill-defined. Assumption (C1) stipulates convergence rate conditions on nuisance estimates in terms of $L^2(P)$ norms. This assumption would be satisfied, for example, if each nuisance estimate achieved a rate of at least $n^{-1/4}$ with respect to $L^2(P)$ norm. However, it is also possible for slower convergence rates attained by some nuisance estimators to be compensated for by faster convergence rates attained by others. We note that the $n^{-1/4}$ rate is slower than the standard parametric rate, which potentially allows for the use of flexible regression techniques. On the other hand, to achieve this rate if $X$ and/or $Z$ are high-dimensional may require additional smoothness assumptions on the underlying nuisance parameters. For example, the highly adaptive LASSO estimator achieves a sufficiently fast rate of convergence \emph{if} the underlying nuisance parameters have a \emph{bounded variation norm} \citep{benkeser2016highly}, an assumption that becomes more restrictive in higher dimensions. Assumptions (B3) and (C2) are necessary to ensure the negligibility of an empirical process term \citep{van1996weak}. Assumption (C2) can be eliminated by utilizing cross-fitting techniques \citep{van2011cross, chernozhukov2018double}.

\subsection{Proof of Theorem 3}

The proof of Theorem 3 closely parallels the proof of Theorem 2.

\begin{equation*}
\theta_{n,a}^{+} - \theta_a = (P_n - P)D_{P,a} + (P_n - P)(D_{P_n^{'}, a} - D_{P,a}) + R_2(P, P_n^{'}) \ .
\end{equation*}

For asymptotic linearity, we need to show $(P_n - P)(D_{P_n^{'}, a} - D_{P,a}) = o_P(1/\sqrt{n})$ and $R_2(P, P_n^{'}) = o_P(1/\sqrt{n})$. The former holds under conditions (B3) and (C2) of the theorem. The structure of the second-order term is presented in the proof of Theorem 2, Part I, and the convergence of the second-order term is established using logic similar to the proof of Theorem 2, Part II.

\section{Simulation details}

\subsection{Methods comparisons}
{\color{changed} In the na{\"i}ve approach that removes high-motion participants, we compared groups using a Welch's t-test applied to participants with $M \le 0.2$. In the na{\"i}ve approach that does not remove any participants, we compared groups using a Welch's t-test applied to all participants. }

For the IPTW estimand, we consider the form 
$$\psi_a = E(E(Y \mid \Delta = 1, A = a, X)) = E \left[ \frac{\mathbb{I}(A = a) \Delta Y}{\Pr(\Delta = 1 \mid A = a, X){\color{changed}\Pr(A = a | X)}} \right]$$ 
% not dealing with Z
and then the estimand is $\psi_1 - \psi_0.$
This parameter balances possible  confounders $X$ (sex, age, handedness) when comparing the difference in functional connectivity between ASD and non-ASD children. {\color{changed}Note it does not account for the possible difference in the motion distributions between autistic children that pass motion quality control and non-ASD children.} 

The IPTW estimator is expressed as:
$$
\hat{\psi}_a = \frac{1}{n} \sum_{i=1}^n \frac{\mathbb{I}(A_i = a) \Delta_i Y_i}{\hat{P}(\Delta_i = 1 \mid A_i = a,  X_i){\color{changed}\hat{P}(\mid A_i = a | X_i)}}.
$$
This estimator reweights the observed outcomes by the inverse of the missingness probability to account for the data removed during quality control. {\color{changed}We estimate the propensity models using logistic regression.} 

In \cite{nebel2022accounting}, the parameter of interest is the difference in functional connectivity between autistic and non-ASD children: $E^{*}[ Y(1) \mid A = 1] - E^{*}[ Y(1) \mid A = 0]$, where $E^{*}$ denotes
an expectation with respect to the probability measure of $\{Y(1), A, Z\}$. After identification, the estimand for $E^{*}[ Y(1) \mid A = a]$ is 
$$\psi_a = E(E(Y \mid \Delta = 1, A = a, Z) \mid A = a).$$ This estimand conditions on diagnosis-specific variables (measures of autism severity) that differ between groups. Variables that should be balanced between groups (confounders $X$ and motion $M$) are removed during a linear regression preprocessing step. After the linear regression preprocessing, \cite{nebel2022accounting} used the DRTMLE estimator with superlearner \citep{benkeser2017improved} for its analysis. This approach 1) removes all data that fail motion quality control, 2) linearly models $X$ and $M$, then 3) non-linearly models $Z$. {\color{changed}We estimate the nuisance models using using the same learners used in MoCo (mean of the outcome, multivariate adaptive regression splines, LASSO, generalized additive models, generalized linear models (with and without interactions, and with and without forward stepwise covariate selection), random forest, and xgboost).} 

\subsection{Simulation study details}
We set the sample size to $n = 400$ and simulated covariates that are similar in distribution to the covariates in the data in Section 4. The simulated demographic covariates $X$ had three dimensions similar to sex, age, and handedness. Sex $X_1$ was generated from a Bernoulli distribution with a probability of $0.75$ of female sex, age $X_2$ was generated by truncating a Gamma distribution with a shape of 25 and a rate of 2.5 to values between 8 and 13, and right-handedness $X_3$ was simulated from a Bernoulli distribution with success probability of 0.92. Given covariates $X = x$, diagnosis $A$ was drawn from a Bernoulli distribution with a success probability of expit($-0.11+0.71x_1-0.08x_2-0.19x_3$), where these coefficients were derived from a logistic regression fitted to the real data.

Diagnosis-specific covariates $Z$ included four components: ADOS $Z_1$ was assigned a value of 0 for simulated non-ASD participants, while for simulated ASD participants, a value was drawn from a Poisson distribution with mean 11.86; FIQ $Z_2$ was sampled from a Normal$(114.6, 11.6^2)$ distribution for simulated non-ASD participants and a Normal$(104.2, 17.4^2)$ distribution for simulated ASD participants, where these means and standard deviations were calculated based on the real data; stimulant and non-stimulant medication, $Z_3$ and $Z_4$, respectively, were assigned a value of 0 for $A=0$ participants and had a value drawn from a Bernoulli distribution with success probabilities of 0.2 and 0.17, respectively for $A=1$ participants. Given $A = a, X = x, Z = z$, the natural logarithm of mean framewise displacement $M$ was generated from a Normal$(-1.26 + 0.095a + 0.104x_1 - 0.0535x_2 - 0.12x_3 + 0.00675z_1 - 0.000255z_2 + 0.324z_3 + 0.064z_4, 0.56^2)$ distribution. We defined a tolerable motion level $\Delta = 1$ as the indicator that $M \le 0.2$.

We simulated the functional connectivity between a seed region equal to the default mode network and six other resting-state parcels. We denote by $Y_1, \dots, Y_6$ the simulated functional connectivity for these six parcels. 
%In the first scenario, the conditional means of the six functional connectivity outcomes were 
% \begin{align*}
%     \mu_{Y_1 \mid A, M, X, Z}(a, m, x, z) &= -0.22 + 0\times a + 0.49m + 0.06x_1 - 0.009x_2 - 0.03x_3, \\ 
%     \mu_{Y_2 \mid A, M, X, Z}(a, m, x, z) &= -0.20 + 0\times a + 0.55m + 0.03x_1 + 0.012x_2 - 0.02x_3, \\
%     \mu_{Y_3 \mid A, M, X, Z}(a, m, x, z) &= -0.37 + 0\times a + 0.93m + 0.07x_1 - 0.009x_2 - 0.034x_3, \\ 
%     \mu_{Y_4 \mid A, M, X, Z}(a, m, x, z) &= 0.17 + 0\times a - 0.03m - 0.02x_1 + 0.002x_2 + 0.04x_3,  \\ 
%     \mu_{Y_5 \mid A, M, X, Z}(a, m, x, z) & \\
%     &\hspace{-8em} = -0.40 + 0.16a + 0.97m + 0.05x_1 - 0.004x_2 + 0.02x_3 - 0.004z_1 + 0.0003z_2 - 0.048z_3 - 0.12z_4, \\ 
%     \mu_{Y_6 \mid A, M, X, Z}(a, m, x, z) &\\
%     &\hspace{-8em} = -0.12 + 0.12a + 1.02m + 0.03x_1 + 0.0007x_2 + 0.01x_3 - 0.0007z_1 + 0.0006z_2 - 0.042z_3 - 0.07z_4 \ .
% \end{align*}

We used processed ABIDE data (see below) to motivate realistic values for the coefficients. For each subject, we averaged the vertices that fell within each of the seven Yeo networks \citep{yeo2011organization}, and then calculated the correlation between the default mode network and the other six networks. We regressed functional connectivity on $A$, $M$, $X$, and $Z$, then calculated the sample covariance matrix of the residuals. This sample covariance matrix was used to parameterize the errors of $Y_1,\dots,Y_6$. The conditional means of the six functional connectivity outcomes were 
\begin{align*}
    \mu_{Y_1 \mid A, M, X, Z}(a, m, x, z) &= -0.22 + 0\times a - 0.98m - 0.06x_1 + 0.012x_2 + 0.03x_3, \\ 
    \mu_{Y_2 \mid A, M, X, Z}(a, m, x, z) &= -0.20 + 0\times a + 0.92m + 0.06x_1 - 0.009x_2 - 0.03x_3, \\ 
    \mu_{Y_3 \mid A, M, X, Z}(a, m, x, z) &= -0.37 + 0\times a + 0.86m + 0.04x_1 + 0.002x_2 + 0.04x_3, \\ 
    \mu_{Y_4 \mid A, M, X, Z}(a, m, x, z) &= 0.17 + 0\times a - 1.02m - 0.06x_1 + 0.002x_3 + 0.04x_3,  \\ 
    \mu_{Y_5 \mid A, M, X, Z}(a, m, x, z) & \\
    &\hspace{-8em} = -0.20 - 0.03a + 1.50m - 0.61m^2 + 0.02(x_1 - z_4) - 0.002x_2 + 0.03(x_3 - z_3) - 0.0005z_1 + 0.0003z_2, \\ 
    \mu_{Y_6 \mid A, M, X, Z}(a, m, x, z) &\\
    &\hspace{-8em} = -0.16 - 0.05a + 1.67m - 0.64m^2 + 0.03(x_1 - z_4) - 0.001x_2 + 0.02(x_3 - z_3) - 0.0005z_1 + 0.0003z_2.
\end{align*}

In MoCo, we included the mean of the outcome, multivariate adaptive regression splines, LASSO, generalized additive models, generalized linear models (with and without interactions, and with and without forward stepwise covariate selection), random forest, and xgboost as candidate regressions in super learner for nuisance regressions.

\subsection{Additional simulation with smaller sample size}

For the realistic data simulations, we also examined smaller samples with $n=50$ and $n=100$. Our method provides unbiased estimates of the true association (see Table \ref{sim:real_data_smaller_sample_size}). At $n=50$, the type 1 errors are near $\alpha=0.05$, although the power is low.  As the sample size increases, MoCo is better at estimating the standard deviation, yielding lower MSE and greater power in detecting the true effect.

\begin{table}[!h]
\caption{Simulation results for MoCo with varying sample sizes. Metrics include bias, standard deviation (sd), mean squared error (MSE), Type I error, and power.}
\label{sim:real_data_smaller_sample_size}
\begin{center}
\begin{tabular}{@{}lll lll@{}}
\toprule
Region & True association & Metric &   {MoCo (n = 50)} &   {MoCo (n = 100)} & MoCo (n = 400) \\ \midrule

\multirow{4}{*}{Region 1} & \multirow{4}{*}{0} 
& Bias         &   {0.0125} &   {-0.0032} & 0.0005 \\
& & sd           &   {0.0913} &   {0.0494} & 0.0372 \\
& & MSE$\times 10^3$   &   {8.4787} &   {2.4434} & 1.3839 \\
& & Type I error &   {0.0510} &   {0.0400} & 0.0110 \\ \midrule

\multirow{4}{*}{Region 2} & \multirow{4}{*}{0} 
& Bias         &   {0.0044} &   {0.0012} & 0.0048 \\
& & sd           &   {0.0996} &   {0.0619} & 0.0238 \\
& & MSE$\times 10^3$   &   {9.9344} &   {3.8285} & 0.5894 \\
& & Type I error &   {0.0400} &   {0.0370} & 0.0100 \\ \midrule

\multirow{4}{*}{Region 3} & \multirow{4}{*}{0} 
& Bias         &   {0.0068} &   {0.0010} & 0.0044 \\
& & sd           &   {0.0809} &   {0.0482} & 0.0183 \\
& & MSE$\times 10^3$   &   {6.5849} &   {2.3186} & 0.3554 \\
& & Type I error &   {0.0470} &   {0.0320} & 0.0080 \\ \midrule

\multirow{4}{*}{Region 4} & \multirow{4}{*}{0} 
& Bias         &   {-0.0042} &   {-0.0019} & -0.0034 \\
& & sd           &   {0.0830} &   {0.0483} & 0.0204 \\
& & MSE$\times 10^3$   &   {6.9011} &   {2.3307} & 0.4275 \\
& & Type I error &   {0.0580} &   {0.0350} & 0.0100 \\ \midrule

\multirow{4}{*}{Region 5} & \multirow{4}{*}{-0.0484} 
& Bias         &   {0.0021} &   {-0.0002} & 0.0065 \\
& & sd           &   {0.0874} &   {0.0529} & 0.0214 \\
& & MSE$\times 10^3$   &   {7.6322} &   {2.7922} & 0.4990 \\
& & Power        &   {0.0680} &   {0.1340} & 0.3790 \\ \midrule

\multirow{4}{*}{Region 6} & \multirow{4}{*}{-0.0682} 
& Bias         &   {-0.0046} &   {0.0008} & 0.0063 \\
& & sd           &   {0.0806} &   {0.0482} & 0.0203 \\
& & MSE$\times 10^3$   &   {6.5160} &   {2.3171} & 0.4523 \\
& & Power        &   {0.1610} &   {0.2890} & 0.8690 \\ 

\bottomrule
\end{tabular}
\end{center}
\end{table}

\subsection{Additional simulation with IQ in $X$}

To assess the impact of potential misclassification between $Z$ and $X$, we conducted an additional simulation in which IQ was instead included in $X$. We used the same simulation design such that IQ was generated to mimic the real data distribution with moderate imbalance between groups. Results (Table \ref{sim:iq_x}) indicate that the method remains robust to this  change. Note here, the positivity assumption holds. As expected, MSE and Type I error rates were modestly affected in some regions. More generally, for a variable designated as $X$ instead of $Z$, we speculate that mild to moderate imbalance in $Z$ alongside a mild to moderate relationship between $Z$ and $Y$ would have a limited impact, but severe imbalance and a stronger relationship may lead to larger sensitivity.

\begin{table}[!h]
\caption{Simulation results for MoCo with mispecifying IQ in $X$ versus correctly specifying it in $Z$. Metrics include bias, standard deviation (sd), mean squared error (MSE), Type I error, and power.}
\label{sim:iq_x}
\begin{center}
\begin{tabular}{@{}lll ll@{}}
\toprule
Region & True association & Metric & MoCo (IQ in $X$) & MoCo (IQ in $Z$) \\ \midrule

\multirow{4}{*}{Region 1} & \multirow{4}{*}{0} 
& Bias         & 0.0006              & 0.0005 \\
& & sd           & 0.0225              & 0.0372 \\
& & MSE$\times 10^3$   & 0.5069              & 1.3839 \\
& & Type I error & 0.0767             & 0.0110 \\ \midrule

\multirow{4}{*}{Region 2} & \multirow{4}{*}{0} 
& Bias         & 0.0034              & 0.0048 \\
& & sd           & 0.0283              & 0.0238 \\
& & MSE$\times 10^3$   & 0.8104              & 0.5894 \\
& & Type I error &  0.0788             & 0.0100 \\ \midrule

\multirow{4}{*}{Region 3} & \multirow{4}{*}{0} 
& Bias         & 0.0024              & 0.0044 \\
& & sd           & 0.0217              & 0.0183 \\
& & MSE$\times 10^3$   & 0.4750             & 0.3554 \\
& & Type I error & 0.0715             & 0.0080 \\ \midrule

\multirow{4}{*}{Region 4} & \multirow{4}{*}{0} 
& Bias         &  -0.0020              & -0.0034 \\
& & sd           & 0.0229             & 0.0204 \\
& & MSE$\times 10^3$   &  0.5293             & 0.4275 \\
& & Type I error & 0.0819             & 0.0100 \\ \midrule

\multirow{4}{*}{Region 5} & \multirow{4}{*}{-0.0484} 
& Bias         & 0.0105             & 0.0065 \\
& & sd           & 0.0251              & 0.0214 \\
& & MSE$\times 10^3$   & 0.7403             & 0.4990 \\
& & Power        & 0.4383             & 0.3790 \\ \midrule

\multirow{4}{*}{Region 6} & \multirow{4}{*}{-0.0682} 
& Bias         & 0.0098              & 0.0063 \\
& & sd           & 0.0216              & 0.0203 \\
& & MSE$\times 10^3$   & 0.5613             & 0.4523 \\
& & Power        & 0.8497             & 0.8690 \\ 

\bottomrule
\end{tabular}
\end{center}
\end{table}

\subsection{Additional simulation with time-series–level motion effects on functional connectivity}

We consider a simulation setting that simulates framewise displacement time series that directly impact the correlations between two brain regions. 
We consider $n = 400$ participants, each with $T = 120$ frames. The demographics $X$ and variables $Z$ related to the diagnosis group are generated in the manner described in Supplement Section 6.2.

For motion, let $f_r(m_{it})$ denote the effect of motion on the $r$th region. For clarity, we assume two regions. To induce correlation between regions due to temporal patterns of motion, we let $f_1(m_{it})=f_2(m_{it})=0.2m_{it}+0.05m_{it}^2$. Let $e_{itr}=\xi_{itr}+f(m_{it})$ denote the subject-specific BOLD time course for regions $r=1,2$, where $\xi_{itr}$ denotes the neural component of the BOLD signal. Let $\bm{\xi}_{it}=[\xi_{it1},\xi_{it2}]^\top$. We simulate signal temporal dependence (functional connnectivity) by 1)  $\bm{\xi}^*_{it}\overset{ind}{\sim} N(0,\bm{\Psi}_i)$ with $[\bm{\Psi}_i]_{11}=[\bm{\Psi}_i]_{22}=1$ and $[\bm{\Psi}_i]_{12}=[\bm{\Psi}_i]_{21}=-0.15 A_i+0.3$; 2) $\bm{\xi}_{it}=0.3\bm{\xi}_{i,t-1}+\sqrt{1-0.3^2}\bm{\xi}^*_{it}$. This VAR model is based on the design in \cite{qiu2023unveiling}, but our design is more realistic by incorporating the motion impacts. Then define $\rho_i = cor(e_{it1}, e_{it2})$. Finally, define the functional connectivity as 
\begin{align*}
Y_i &= -0.16 -0.05A_i + 1.67\times \rho_i - 0.64 \times \rho_i^2 + 0.03X_{1i} - 0.001X_{2i} + 0.02X_{3i} \\
&\;\;\;\;-0.0005Z_{1i} + 0.0003Z_{2i} - 0.02Z_{3i} - 0.03Z_{4i}.
\end{align*}

\begin{table}[!h]
\label{sim:motiontimeseries}
\centering

\begin{tabular}{@{}lllll@{}}
\toprule
\multirow{2}{*}{Truth} & \multicolumn{4}{l}{MoCo}                       \\ \cmidrule(l){2-5} 
                       & Estimation & Bias  & SD    & MSE $\times 10^3$ \\ \midrule
-0.2761                & -0.2741    & 0.002 & 0.023 & 0.5348            \\ \bottomrule
\end{tabular}
\caption{Simulation results for MoCo in the scenario of motion time-series}
\end{table}

The results show that the MoCo estimate ($-0.2741$) is very close to the true value ($-0.2761$), with a negligible bias, small standard deviation, and low mean squared error. This indicates that MoCo performs well when functional connectivity is correlated with temporal patterns of head motion.

\subsection{Confirm theoretical properties of estimators}
\label{additional_sim}

We evaluate the statistical properties of our estimators established by our theorems through Monte Carlo simulation. These simulations were conducted purely to confirm theoretical properties of the estimators and are not tied to the real data analysis context in any way. In this simulation, we generated covariate $X$ from $\text{Bernoulli}(1/2)$. Given $X = x$, we generated a binary variable $A$ according to a Bernoulli distribution with $\pi_1(x) = \text{expit}(x-1/4)$. Given $A = a$, we drew $Z$ from Bernoulli$(\text{expit}(5a/4-1/2))$. Given $A = a, X = x, Z = z$, we drew $M$ from a normal distribution $N(1+a+x/2-z/4, 1)$. Finally, given $A = a, X = x, Z = z, M = m$, we drew $Y$ is drawn from a normal distribution $N(-1+x/2-z/3-a/4+m/5, 1)$. We defined a tolerable $M$ level $\Delta = 1$ as the indicator that $M \le 2$. The true values of these parameters were -0.717 and -1.068, respectively, with variance bound $Var\{D_{a,P}(O)\}$ equal to 3.453 and 7.151, respectively.

For each sample size $n \in \{50,\, 200, 500, 1000, 2000, 4000\}$, we generated 1000 datasets based on the above data generating process, and used the resultant data to compute our proposed estimators along with their corresponding confidence interval, either with cross-fitting or without the use of cross-fitting. We evaluated the estimators based on their bias (scaled by $n^{1/2}$), their standard error (scaled by $n^{1/2}$), the ratio of the scaled standard error to the square root of the efficient variance, and the coverage of 95\% Wald-style confidence interval. 

We first evaluated estimators under the conditions of the theorem where all nuisance parameters are consistently estimated at appropriate rates. To achieve this, we used the logistic regression model for $\pi_a(x)$ and main term linear model $ \mu_{Y \mid A, M, X, Z}(a, m, x, z)$ and fully saturated (all possible interactions) regression models for the remaining nuisance parameters. In this scenario, the one-step estimators are expected to be consistent and asymptotically linear. 

When all the nuisance parameters are estimated consistently at appropriate rates, the bias of the estimators decreases to 0 as the sample size increases, and the coverage of the 95\% confidence interval increases to 0.95 (Table \ref{tab: sim1 cross}). At n=50, the estimator demonstrates coverage of the true value around 0.8 for both parameters. At a sample size of 200, the bias is approximately 2\% of the true value, and the worst-case coverage of 95\% confidence intervals is 86\%. % Similar results are achieved without the use of cross-fitting (Table \ref{tab: sim1 no_cross}).

We also studied the impact of inconsistent estimation of different combinations of nuisance parameters. We examined five situations in which only specific combinations of the nuisance parameters were correctly specified, as defined in column 1 of \cref{tab: sim2 multiple robust_cross}. Based on its multiple robustness properties, our one-step estimators are expected to maintain consistency across these five scenarios. Nuisance regressions that were incorrectly specified were modeled using only the intercept. Conditional densities that were incorrectly specified were modeled using a Gaussian distribution in which the mean was equal to the sample mean and the standard deviation equal to the sample standard deviation. 

Our results indicate that as sample size increases, the bias and standard error decrease across all settings considered, which supports the multiple robustness theory (Table \ref{tab: sim2 multiple robust_cross}).  %  and Table \ref{tab: sim2 multiple robust_no_cross}

\begin{table}[pt]
\centering
\begin{tabular}{@{}l|llll|llll@{}}
\toprule
\multicolumn{1}{c|}{\multirow{2}{*}{n}} & \multicolumn{4}{c|}{$\theta_{n, 0}^{\text{cf}}$}         & \multicolumn{4}{c}{$\theta_{n, 1}^{\text{cf}}$}         \\ 
                   & $n^{1/2}$ bias   & $n^{1/2}$ sd    & sd ratio & cover & $n^{1/2}$ bias   & $n^{1/2}$ sd    & sd ratio & cover \\ \midrule
   {50}                &    {-0.125} &    {2.551} &    {2.063}    &    {0.785} &    {0.209} &    {4.400} &    {2.343}    &    {0.830} \\ 
200                & -0.208 & 2.148 & 1.122    & 0.910 & -0.143 & 2.391 & 1.382    & 0.860 \\
500                & -0.169 & 1.926 & 0.996    & 0.942 & -0.165 & 2.319 & 1.172    & 0.912 \\
1000               & -0.149 & 1.918 & 0.987    & 0.945 & -0.119 & 2.325 & 1.100    & 0.929 \\
2000               & -0.058 & 1.980 & 1.035    & 0.941 & -0.072 & 2.421 & 1.093    & 0.928 \\
4000               &  0.011 & 1.887 & 1.000    & 0.946 &  0.050 & 2.185 & 0.961    & 0.960 \\ \bottomrule
\end{tabular}
\caption{Confirming theoretical properties of estimators: All nuisance parameters are consistently estimated at appropriate rates with the use of MoCo (with cross-fitting).}
\label{tab: sim1 cross}
\end{table}

\begin{table}[pt]
\label{sim2: multiple robust_cross}
\centering
\begin{tabular}{@{}llllll@{}}
\toprule
Setting & n & $\text{bias}_{\theta_{n, 0}^{\text{cf}}}$ & $\text{sd}_{\theta_{n, 0}^{\text{cf}}}$ & $\text{bias}_{\theta_{n, 1}^{\text{cf}}}$ & $\text{sd}_{\theta_{n, 1}^{\text{cf}}}$ \\ \midrule
{\multirow{5}{*}{(B2.1)}} & 200 & 0.0500 & 0.1289 & 0.0681 & 0.1455 \\
{} & 500 & 0.0341 & 0.1016 & 0.0558 & 0.0923 \\
{} & 1000 & 0.0103 & 0.0710 & 0.0470 & 0.0710 \\
{} & 2000 & -0.0007 & 0.0475 & 0.0380 & 0.0504 \\
{} & 4000 & -0.0026 & 0.0170 & 0.0331 & 0.0368 \\ \midrule
{\multirow{5}{*}{(B2.2)}} & 200 & -0.0575 & 0.1164 & -0.0510 & 0.1621 \\
{} & 500 & -0.0130 & 0.0865 & -0.0206 & 0.1125 \\
{} & 1000 & -0.0087 & 0.0594 & -0.0252 & 0.0800 \\
{} & 2000 & -0.0044 & 0.0405 & -0.0223 & 0.0555 \\
{} & 4000 & -0.0008 & 0.0264 & -0.0147 & 0.0400 \\ \midrule
\multirow{5}{*}{(B2.3)} & 200 & -0.0412 & 0.1199 & -0.0561 & 0.1275 \\
& 500 & -0.0021 & 0.0892 & -0.0073 & 0.1006 \\
& 1000 & -0.0019 & 0.0612 & -0.0043 & 0.0743 \\
& 2000 & -0.0011 & 0.0429 & -0.0027 & 0.0520 \\
& 4000 & 0.0008 & 0.0272 & 0.0019 & 0.0375 \\ \midrule
\multirow{5}{*}{(B2.4)} & 200 & -0.0585 & 0.1189 & -0.0488 & 0.1487 \\
& 500 & -0.0099 & 0.0915 & -0.0087 & 0.1035 \\
& 1000 & -0.0092 & 0.0608 & -0.0170 & 0.0756 \\
& 2000 & -0.0039 & 0.0429 & -0.0161 & 0.0532 \\
& 4000 & -0.0004 & 0.0272 & -0.0105 & 0.0383 \\ \midrule
\multirow{5}{*}{(B2.5)} & 200 & -0.0599 & 0.1217 & -0.0505 & 0.1556 \\
& 500 & -0.0118 & 0.0885 & -0.0143 & 0.1107 \\
& 1000 & -0.0107 & 0.0582 & -0.0248 & 0.0794 \\
& 2000 & -0.0054 & 0.0397 & -0.0230 & 0.0565 \\
& 4000 & -0.0008 & 0.0264 & -0.0148 & 0.0399 \\ \bottomrule
\end{tabular}
\caption{Bias and standard deviation(sd) of MoCo (with cross-fitting). The settings column indicates which nuisance parameters are consistently estimated based on assumption (B2) in Theorem 3.2 as outlined in the main manuscript.}\label{tab: sim2 multiple robust_cross}
\end{table}

\newpage
\section{Additional details of data analysis}

\subsection{MRI data and preprocessing}
We selected scans corresponding to 8 to 13-year-old children from the Kennedy Krieger Institute (KKI) and New York University (NYU) from ABIDE I and ABIDE II data releases. Resting-state fMRI scans were acquired using one of three protocols: 1) a 3T Philips Achieva scanner, 8-channel head coil, repetition time (TR)/echo time (TE)=2500/30 ms, flip angle 75$^\circ$, 3$\times$3$\times$3 mm voxels, SENSE phase reduction=3, 2 dummy scans, scan duration varied from 5m10s (one subject) to 6m45s (one subject) with most scans at either 5m20s (n=46) or 6m30s (n=99) (KKI); 2) the same but with a 32-channel head coil with most scans 6m30s (n=61) (KKI); 3) a 3T Siemens Allegra scanner, 8-channel head coil, TR/TE=2000/15 ms, flip angle=90$^\circ$, 3x3x4 mm voxels, scan duration=6:00 (NYU), and we removed the first two volumes. All children also had an anatomical T1 scan collected. 

\begin{figure}[!h]
	\centering
	\includegraphics[width=\textwidth]{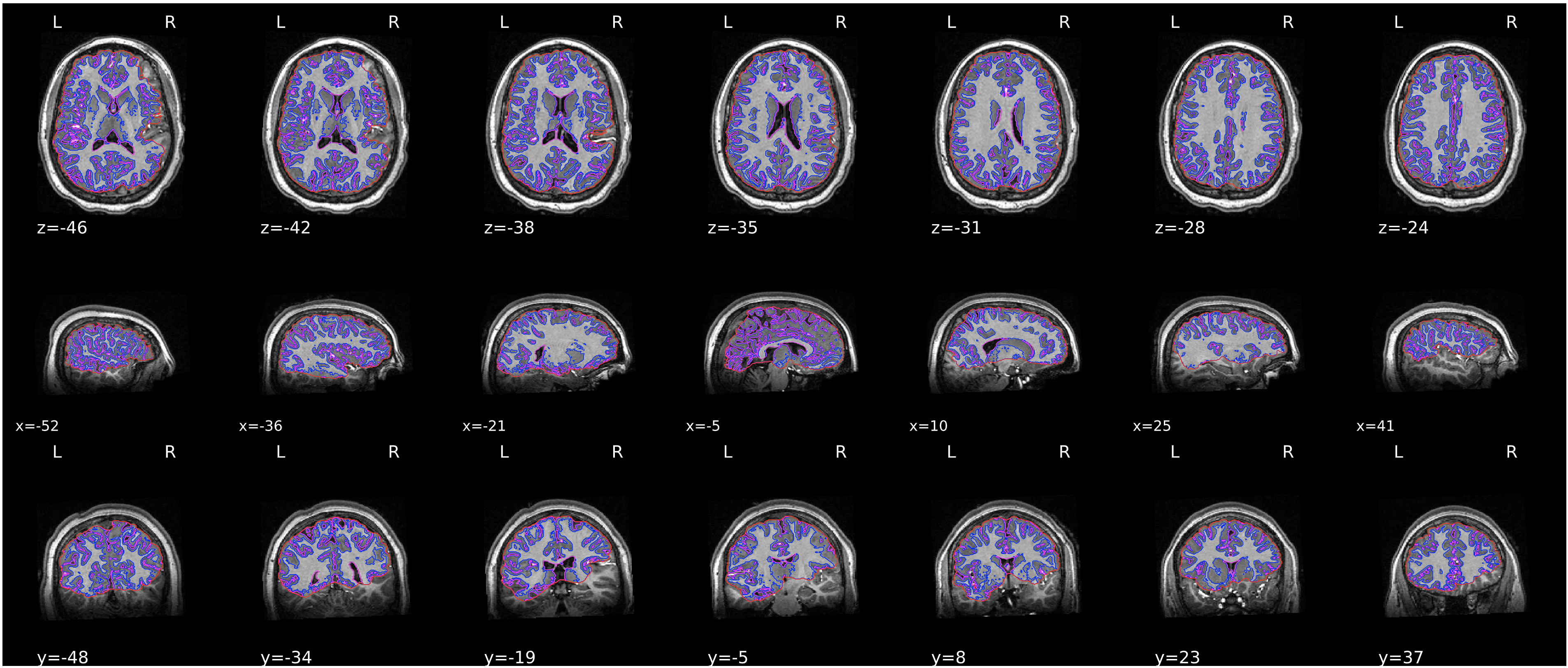}
	\caption{An example participant whose cortical segmentation failed in fMRIPrep. The template T1-weighted image is shown with contours outlining the detected brain mask and brain tissue segmentations. It is apparent from the middle and lower rows that large parts of the brain, including most of the temporal lobe and parts of the occipital lobe, were incorrectly excluded from the segmentation.}
	\label{fig:T1_failed_png}
\end{figure}

T1 anatomical and rs-fMRI data were processed with the cifti option for cortical surface registration using fMRIPrep, including anatomical tissue segmentation, surface construction, and surface registration, followed by fMRI motion correction, slice-time correction, boundary-based coregistration, and resampling to the fsaverage template  \citep{esteban2019fmriprep}. The detailed output from fMRIPrep is included in \cref{fmriprep_pre}. We visually inspected the accuracy of the cortical segmentation using the fMRIPrep quality control html files. We excluded 19 participants due to issues with the cortical segmentation. Issues with fMRIprep included image homogeneity issues, outliers in brain morphology, and motion during the T1 scan. An example participant that failed this step is included in \cref{fig:T1_failed_png}. 

Our final study sample was 377 children, with 245 non-autistic children and 132 ASD children. We defined the indicator of data usability $\Delta$ equal to one if a child had more than 5 minutes of data after removing frames with FD $>$ 0.2
mm \citep{power2014methods}. This resulted in $\Delta=0$ for 98/132 ASD (74.2\%) and 119/245 non-ASD children (48.6\%). 
Table \ref{tab:demo} displays characteristics of the analysis cohort by ASD diagnosis status. 
 
\begin{table}[h]
     \includegraphics[width = \textwidth]{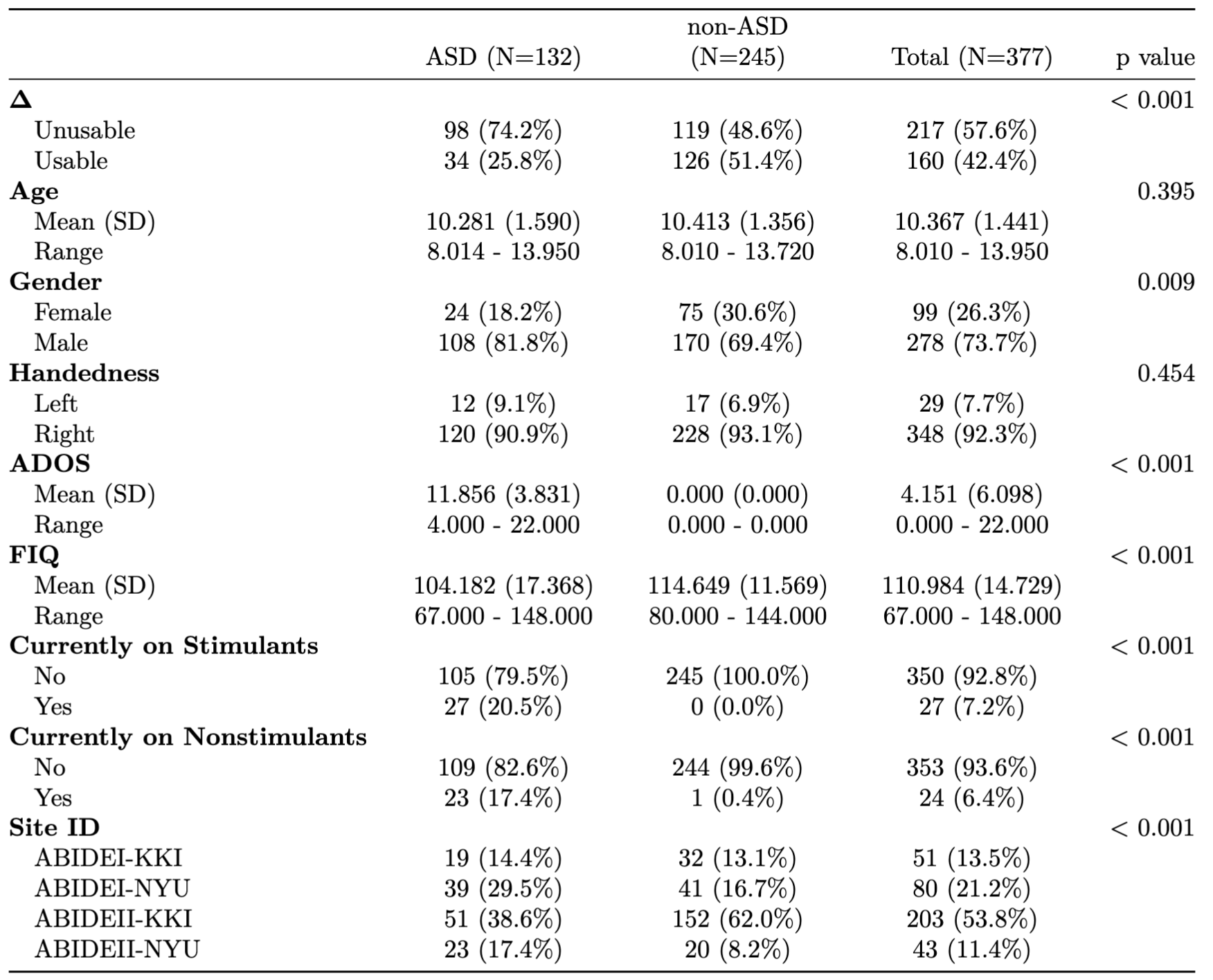}
      \caption{Demographic characteristics: Continuous variables are described using mean and standard deviation, and diagnostic groups are compared using the Kruskal-Wallis rank-sum test. Binary and categorical variables are reported as frequencies and percentages, and differences between diagnostic groups are assessed using either the Chi-square test or Fisher’s exact test.} \label{tab:demo}   
\end{table}

\subsection{Auto-generated text from fMRIprep data preprocessing}\label{fmriprep_pre}
Preprocessing was performed
using \emph{fMRIPrep} 21.0.2 (\citet{fmriprep1}; \citet{fmriprep2};
RRID:SCR\_016216), which is based on \emph{Nipype} 1.6.1
(\citet{nipype1}; \citet{nipype2}; RRID:SCR\_002502). The text below is automatically produced by fMRIprep.

\begin{description}
\item[Anatomical data preprocessing]
A total of 1 T1-weighted (T1w) images were found within the input BIDS
dataset. The T1-weighted (T1w) image was corrected for intensity
non-uniformity (INU) with \texttt{N4BiasFieldCorrection} \citep{n4},
distributed with ANTs 2.3.3 \citep[RRID:SCR\_004757]{ants}, and used as
T1w-reference throughout the workflow. The T1w-reference was then
skull-stripped with a \emph{Nipype} implementation of the
\texttt{antsBrainExtraction.sh} workflow (from ANTs), using OASIS30ANTs
as target template. Brain tissue segmentation of cerebrospinal fluid
(CSF), white-matter (WM) and gray-matter (GM) was performed on the
brain-extracted T1w using \texttt{fast} \citep[FSL 6.0.5.1:57b01774,
RRID:SCR\_002823,][]{fsl_fast}. Brain surfaces were reconstructed using
\texttt{recon-all} \citep[FreeSurfer 6.0.1,
RRID:SCR\_001847,][]{fs_reconall}, and the brain mask estimated
previously was refined with a custom variation of the method to
reconcile ANTs-derived and FreeSurfer-derived segmentations of the
cortical gray-matter of Mindboggle
\citep[RRID:SCR\_002438,][]{mindboggle}. Volume-based spatial
normalization to two standard spaces (MNI152 NLin6 Asym, MNI152 NLin 2009c Asym) was performed through nonlinear registration with
\texttt{antsRegistration} (ANTs 2.3.3), using brain-extracted versions
of both T1w reference and the T1w template. The following templates were
selected for spatial normalization: \emph{ICBM 152 Nonlinear
Asymmetrical template version 2009c} {[}\citet{mni152nlin2009casym},
RRID:SCR\_008796; TemplateFlow ID: MNI152NLin2009cAsym{]}, \emph{FSL's
MNI ICBM 152 non-linear 6th Generation Asymmetric Average Brain
Stereotaxic Registration Model} {[}\citet{mni152nlin6asym},
RRID:SCR\_002823; TemplateFlow ID: MNI152NLin6Asym{]}.
\item[Functional data preprocessing]
For each of the 1 BOLD runs found per subject (across all tasks and
sessions), the following preprocessing was performed. First, a reference
volume and its skull-stripped version were generated using a custom
methodology of \emph{fMRIPrep}. Head-motion parameters with respect to
the BOLD reference (transformation matrices, and six corresponding
rotation and translation parameters) are estimated before any
spatiotemporal filtering using \texttt{mcflirt} \citep[FSL
6.0.5.1:57b01774,][]{mcflirt}. BOLD runs were slice-time corrected to
1.22s (0.5 of slice acquisition range 0s-2.45s) using \texttt{3dTshift}
from AFNI \citep[RRID:SCR\_005927]{afni}. The BOLD time-series
(including slice-timing correction when applied) were resampled onto
their original, native space by applying the transforms to correct for
head-motion. These resampled BOLD time-series will be referred to as
\emph{preprocessed BOLD in original space}, or just \emph{preprocessed
BOLD}. The BOLD reference was then co-registered to the T1w reference
using \texttt{bbregister} (FreeSurfer) which implements boundary-based
registration \citep{bbr}. Co-registration was configured with six
degrees of freedom. Several confounding time-series were calculated
based on the \emph{preprocessed BOLD}: framewise displacement (FD) and three region-wise global signals. FD was computed following Power (absolute sum of relative motions,
\citet{power_fd_dvars})
calculated using the implementation in
\emph{Nipype} \citep[following the definitions by][]{power_fd_dvars}.
The three global signals were extracted within the CSF, the WM, and the
whole-brain masks. The BOLD time-series were resampled
into standard space, generating a \emph{preprocessed BOLD run in
MNI152NLin2009cAsym space}. First, a reference volume and its
skull-stripped version were generated using a custom methodology of
\emph{fMRIPrep}. The BOLD time-series were resampled onto the following
surfaces (FreeSurfer reconstruction nomenclature): \emph{fsaverage}.
\emph{Grayordinates} files \citep{hcppipelines} containing 91k samples
were also generated using the highest-resolution \texttt{fsaverage} as
intermediate standardized surface space. All resamplings can be
performed with \emph{a single interpolation step} by composing all the
pertinent transformations (i.e.~head-motion transform matrices and
co-registrations to anatomical and output spaces). Gridded (volumetric)
resamplings were performed using \texttt{antsApplyTransforms} (ANTs),
configured with Lanczos interpolation to minimize the smoothing effects
of other kernels \citep{lanczos}. Non-gridded (surface) resamplings were
performed using \texttt{mri\_vol2surf} (FreeSurfer).
\end{description}

Many internal operations of \emph{fMRIPrep} use \emph{Nilearn} 0.8.1
\citep[RRID:SCR\_001362]{nilearn}, mostly within the functional
processing workflow. For more details of the pipeline, see
\href{https://fmriprep.readthedocs.io/en/latest/workflows.html}{the
section corresponding to workflows in \emph{fMRIPrep}'s documentation}.

\hypertarget{copyright-waiver}{%
\subsubsection{Copyright Waiver}\label{copyright-waiver}}

The above boilerplate text was automatically generated by fMRIPrep with
the express intention that users should copy and paste this text into
their manuscripts \emph{unchanged}. It is released under the
\href{https://creativecommons.org/publicdomain/zero/1.0/}{CC0} license.

\subsection{Details of the application of MoCo to the ABIDE dataset}

The default mode network is a collection of brain regions that tend to co-activate during wakeful rest, including daydreaming or mind wandering. Hypoconnectivity between anterior and posterior parts of the default mode network was previously found in the Autism Brain Imaging Data Exchange (ABIDE) dataset \citep{di2014autism}. However, hypoconnectivity in the default mode network also arises from motion artifacts \citep{power2014methods}. 

We included covariates: diagnosis ($A)$; age, sex, and handedness ($X)$; Autism Diagnostic Observation Schedule (ADOS) score, Full-scale Intelligence Quotient (FIQ) score, stimulant medication status, and non-stimulant medication status ($Z$); and mean FD ($M$). The ADOS score is a standardized assessment tool used to diagnose ASD, with higher scores indicating greater social disability. 
We calculated the average time series for regions of interest defined using Schaefer's 400-node brain parcellation \citep{schaefer2018local}.
We then performed time-series level motion control by regressing the motion alignment parameters, the global signal, white matter, and cerebrospinal fluid calculated from fMRIPrep. We used COMBAT for site harmonization to account for three protocols \citep{yu2018statistical}. We calculated Fisher z-transformed correlations of every brain region with region 14 (`17networks\_LH\_DefaultA\_pCun\_1'), which is a hub of the posterior default mode network \citep{pham2022ciftitools}. 
 %Each node in this node system represents a distinct region of the cerebral cortex. These regions are based on anatomical and functional criteria and are designed to provide a spatial map of the different areas of the brain. We group and label the nodes based on the functional networks corresponding to Yeo's 7 major resting-state network system  \citep{yeo2011organization}. Specifically, the nodes are grouped into the visual network (Vis), somatomotor network (SMN), dorsal attention network (DAN), ventral attention network (VAN), limbic network (LN), the frontoparietal network (FP), and the default mode network (DMN). 

Although ASD is more prevalent in males than females, we treated sex as a confounder ($X$) rather than diagnosis-specific variable ($Z$) because sex-specific differences in functional connectivity have been previously documented \citep{shanmugan2022sex}, which could mask ASD-related differences in this cohort. Mean FD is an average of the frame-to-frame displacement calculated from the rigid body motion correction parameters used in quality control and motion correction \citep{esteban2019fmriprep,power2014methods}.

\subsection{Positivity assumptions in the ABIDE data}

For the dataset, the positivity assumption (A1.1) holds as no demographic characteristics $X$ can perfectly predict ASD. To assess the positivity assumptions (A1.2), we examine the values of four ratios present in the estimation process and the efficient influence function:

\begin{align}
    &\frac{p_{n, M \mid \Delta = 1, A, X}(M_i \mid 0, X_i)}{p_{n, M \mid \Delta = 1, A, X, Z}(M_i \mid A_i, X_i, Z_i)}  \text{ for } \Delta_i = 1  \\
    &\frac{p_{n, M \mid A, X}(M_i \mid A_i, X_i)}{p_{n, M \mid A, X, Z}(M_i \mid A_i, X_i, Z_i)}  \\
    &\frac{p_{n, M \mid \Delta = 1, A, X}(M_i \mid 0, X_i)}{p_{n, M \mid A, X, Z}(M_i \mid 0, X_i, Z_i)} \\
    &\frac{p_{n, M \mid \Delta = 1, A, X}(M_i \mid 0, X_i)}{p_{n, M \mid A, X, Z}(M_i \mid 1, X_i, Z_i)} .
\end{align}

\begin{figure}[h]
	\centering
	\includegraphics[width=0.8\textwidth]{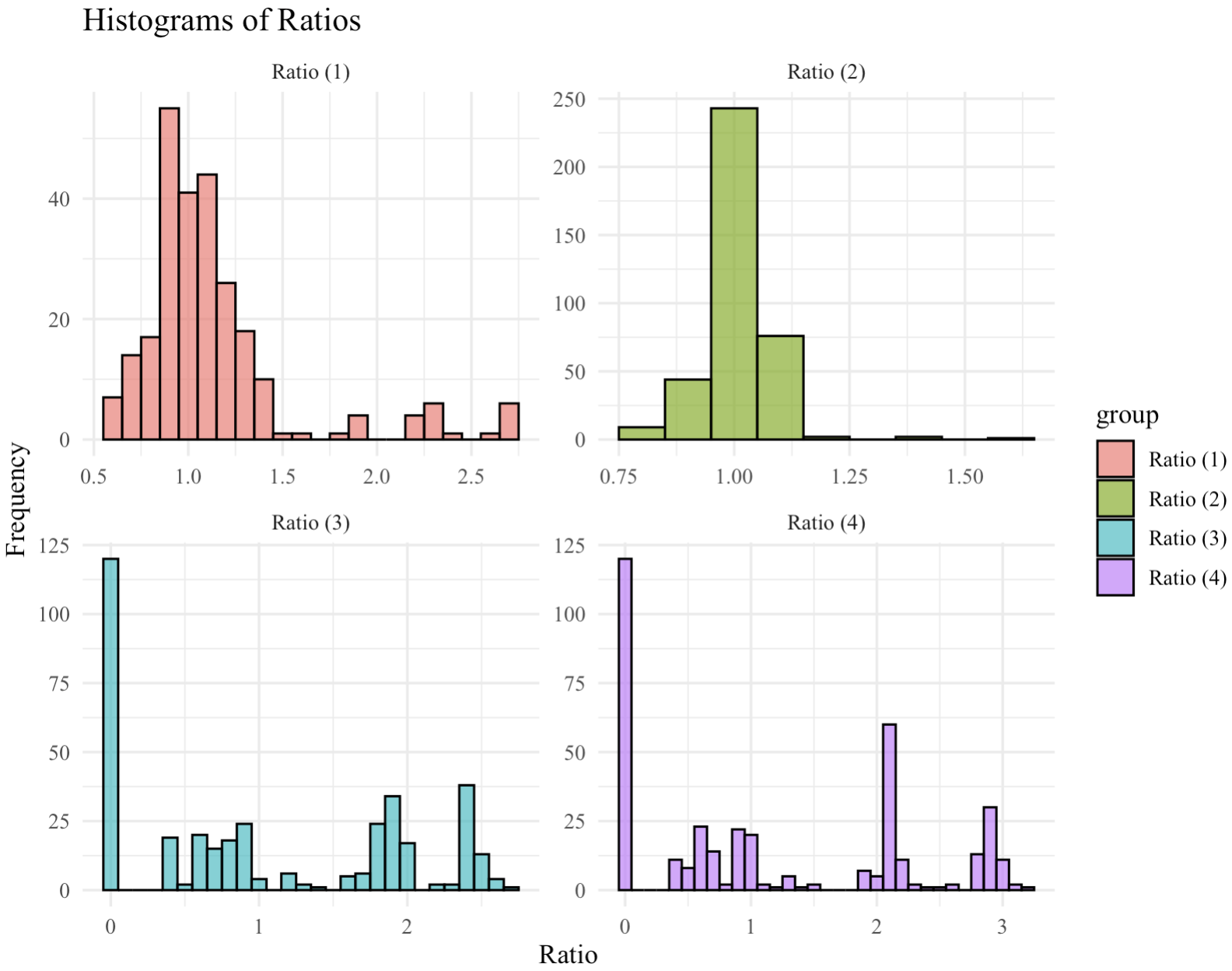}
	\caption{Histogram depicting the distribution of ratio values. Positivity assumptions appear reasonable since the ratios are not too large. Here, the ratios result in reasonable weights in the pseudo-regressions.} 
	\label{fig:positivity}
\end{figure}

There are no very large values for these ratios, which indicates the satisfaction of the positivity assumption. Figure \ref{fig:positivity} displays the histogram depicting the distribution of ratio values.

\subsection{Estimated functional connectivity}

\begin{figure}[h]
	\centering
	\includegraphics[width=0.88\textwidth]{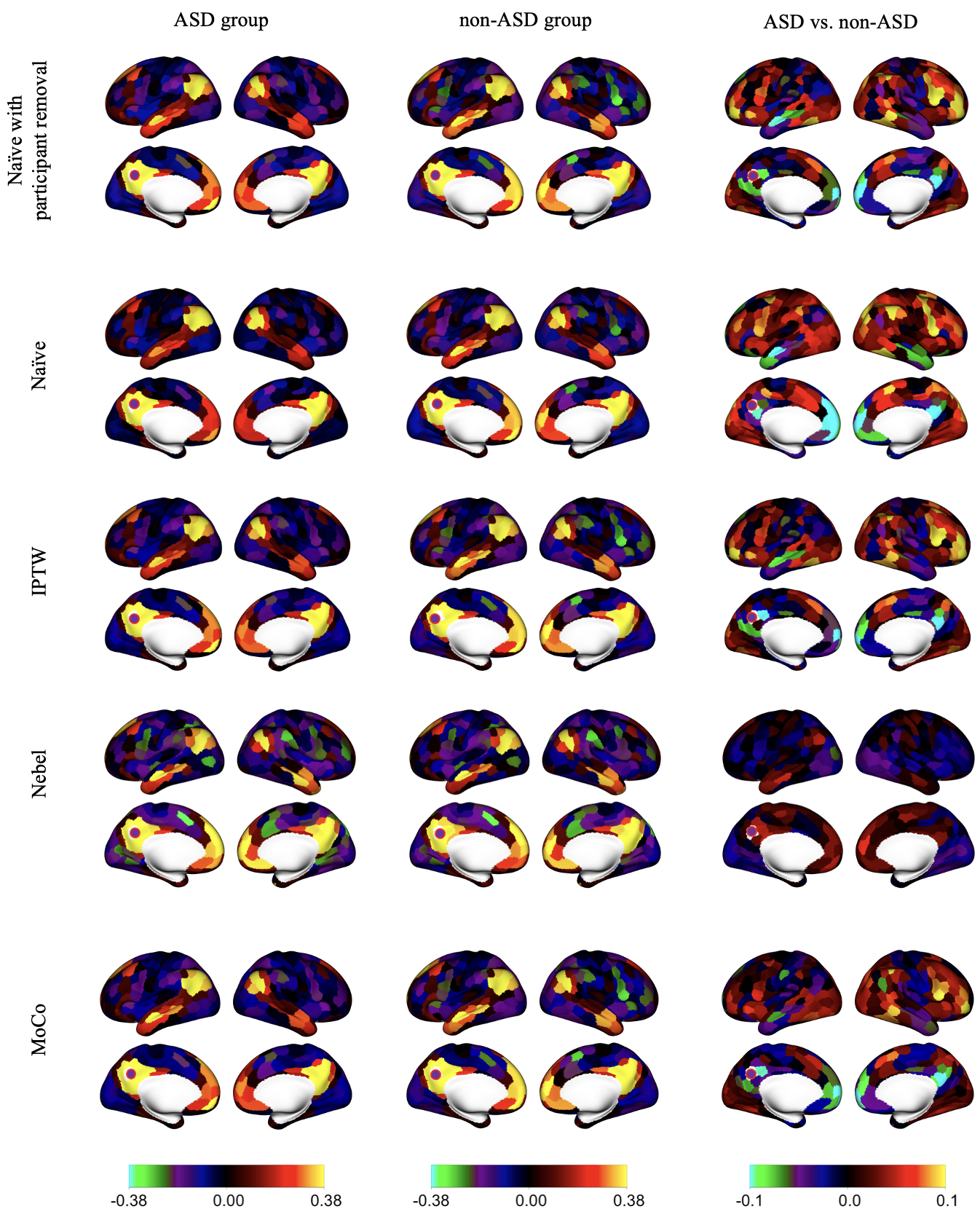}
	\caption{Estimated functional connectivity using the na{\"i}ve approach excluding high-motion participants, the na{\"i}ve approach, IPTW, Nebel's method, and MoCo for a seed region in the posterior default mode network (fuchsia point).} %Z-statistics thresholded using simultaneous confidence bands controlling for family-wise error rate at $\alpha=0.05$.} % increase to -0.09, 0.09
	\label{fig:est_functional connectivity}
\end{figure}

The estimated functional connectivity using the na{\"i}ve approach excluding high-motion participants, the na{\"i}ve approach including all participants, IPTW, Nebel's method, and MoCo are illustrated in Figure \ref{fig:est_functional connectivity}. The seed region in the posterior default mode network is defined by the fourteenth parcel but is represented by the fuchsia point for clarity. 

% \begin{figure}[h]
% 	\centering
% 	\includegraphics[width = 0.9\textwidth]{Figures/motion_distri.png}
%     \caption{Distributions of mean framewise displacement (FD) in the school-age children dataset. Panel A shows the distribution of mean FD over all children. %, which includes significant mass with unusable data, that is children with less than 5 minutes of data after removing frames with FD $>$ 0.2 mm. Our first hypothetical training program results in a reduction in motion to a tolerable level. 
%     Panel B shows the distribution of mean FD over children who meet the inclusion criteria. ASD children with usable data still have higher motion than non-ASD children, which motivates additional training in our hypothetical experiment such that the motion distribution in the ASD group (shown in green) would be identical to the motion distribution in the non-ASD group with $\Delta = 1$ (shown in yellow).}
% 	\label{fig:motion_distri} 
% \end{figure}

\newpage

\bibliographystyle{biom}
\bibliography{reference.bib}

\label{lastpage}

\end{document}